\documentclass[a4paper]{book}

\usepackage{setspace,comment,lipsum} 
\usepackage{graphicx}
\usepackage{dcolumn,float}
\usepackage{amsmath,braket}
\usepackage{bm,amsfonts,amssymb}
\usepackage{color} 
\usepackage{hyperref}   
\usepackage{wrapfig}
\usepackage{pdfpages,framed}

\newcommand{\rse}{\mathcal{R}}
\newcommand{\bes}[1]{j^{#1}_{L_{#1}}}
\newcommand{\han}[1]{h^{(1)#1}_{L_{#1}}}

\newcommand{\tmat}[2]{T_{#1#2}^{(L_{#1},L_{#2})}}
\newcommand{\ttpo}{$2\,{}^{3\!}P_1$}
\newcommand{\otpo}{$1\,{}^{3\!}P_1$}
\newcommand{\jpsi}{J\!/\!\psi}
\newcommand{\One}{1\!\!1}
\newcommand{\tso}{$^{3\!}S_1$}
\newcommand{\stso}{$2\,{}^{3\!}S_1$}
\newcommand{\ttso}{$3\,{}^{3\!}S_1$}
\newcommand{\ttpt}{$2\,{}^{3\!}P_2$}
\newcommand{\ttpz}{$2\,{}^{3\!}P_0$}
\newcommand{\ftso}{$4\,{}^{3\!}S_1$}
\newcommand{\tdo}{$^{3\!}D_1$}
\newcommand{\ftdo}{$1\,{}^{3\!}D_1$}
\newcommand{\stdo}{$2\,{}^{3\!}D_1$}
\newcommand{\ttdo}{$3\,{}^{3\!}D_1$}

\newcommand{\ds}{D_s}
\newcommand{\das}{D^\ast}

\newcommand{\dsas}{D_s^\ast}

\newcommand{\kas}{K^\ast}

\newcommand{\dsc}{$D_{s1}(2536)$}
\newcommand{\dsd}{$D_{s1}(2460)$}
\newcommand{\dc}{$D_{1}(2420)$}
\newcommand{\dd}{$D_{1}(2430)$}
\newcommand{\tpz}{${}^{3\!}P_0$}
\newcommand{\tpo}{${}^{3\!}P_1$}

\newcommand{\spo}{${}^{1\!}P_1$}

\newcommand{\be}{\begin{equation}}
\newcommand{\ee}{\end{equation}}
\newcommand{\bdm}{\begin{displaymath}}
\newcommand{\edm}{\end{displaymath}}
\newcommand{\upr}{\uparrow}
\newcommand{\dar}{\downarrow}
\newcommand{\barr}{\begin{array}}
\newcommand{\earr}{\end{array}}

\def\bea{\begin{eqnarray}}
\def\eea{\end{eqnarray}}

\newcommand{\tspo}{$2\,{}^{1\!}P_1$}

\newcommand{\otpz}{$1\,{}^{3\!}P_0$}

\newcommand{\osdt}{$1\,{}^{1\!}D_2$}

\DeclareMathAlphabet{\mathitbf}{OML}{cmm}{b}{it}

\def\ds{\displaystyle}

\def\upr{\uparrow}
\def\dar{\downarrow}

\title{Unquenched Meson Spectroscopy}
\author{Susana Coito}
\date{\today}

\begin{document}
\begin{titlepage}
\begin{minipage}[!t]{0.2\columnwidth}
\vspace*{-3cm}
\hspace*{-1cm}
\includegraphics[width=4cm]{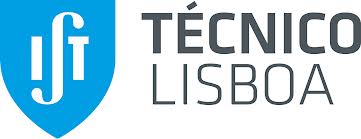}
\end{minipage}
\begin{minipage}[!t]{0.8\columnwidth}
\vspace*{-1.cm}
\begin{center}
{\bf\Large UNIVERSIDADE DE LISBOA}\\[3mm]
{\bf\Large INSTITUTO SUPERIOR T\'ECNICO}
\end{center}
\end{minipage}
\mbox{}\\[2.5cm]
\begin{center}
{\bf\Huge Unquenched Meson Spectroscopy}
\end{center}
\mbox{}\\[.5cm]
\begin{center}
{\Large Susana Patr\' icia Sim\~oes Coito }
\end{center}
\mbox{}\\
\begin{center}
\large
{\bf Supervisor:} Doctor Georges Rupp\\[.5cm]
{\bf Co-Supervisor:} Doctor Everardus Johannes Hubertus van Beveren
\end{center}
\mbox{}\\
\begin{center}
{\large\bf Thesis approved in public session to obtain the PhD Degree in Physics}\\[3mm]
{\large\bf Jury final classification: Pass with Merit}
\end{center}
\mbox{}\\
\begin{center}
\large
{\bf Jury}\\[.7cm]
{\bf Chairperson:} Chairman of the IST Scientific Board\\[.7cm]
{\bf Members of the Committee:}\\[.9cm]
\begin{tabular}{l}
 Doctor Brigitte Anabelle Vaz de Abreu Hiller\\[2mm]
 Doctor Alex Heinz Ladislaus Blin\\[2mm]
 Doctor Maria Paula Fraz\~ao Bordalo e S\'a\\[2mm]
 Doctor Everardus Johannes Hubertus van Beveren\\[2mm]
 Doctor Georges Rupp\\[2mm]
 Doctor Pedro Jos\'e de Almeida Bicudo\\[2mm]
\end{tabular}
\end{center}
\mbox{}\\
\begin{center}
\textbf{\large{2013}}
\end{center}
\clearpage
\thispagestyle{empty}
\end{titlepage}

\begin{titlepage}
\begin{minipage}[!t]{0.2\columnwidth}
\vspace*{-3cm}
\hspace*{-1cm}
\includegraphics[width=4cm]{logotipo}
\end{minipage}
\begin{minipage}[!t]{0.8\columnwidth}
\vspace*{-1.cm}
\begin{center}
{\bf\Large UNIVERSIDADE DE LISBOA}\\[3mm]
{\bf\Large INSTITUTO SUPERIOR T\'ECNICO}
\end{center}
\end{minipage}
\mbox{}\\[.9cm]
\begin{center}
{\bf\Huge Unquenched Meson Spectroscopy}
\end{center}
\mbox{}\\[-2mm]
\begin{center}
{\Large Susana Patr\' icia Sim\~oes Coito }
\end{center}
\mbox{}\\[-.5cm]
\begin{center}
\large
{\bf Supervisor:} Doctor Georges Rupp\\[.3cm]
{\bf Co-Supervisor:} Doctor Everardus Johannes Hubertus van Beveren
\end{center}
\mbox{}\\[-4mm]
\begin{center}
\large
{\bf Thesis approved in public session to obtain the PhD Degree in Physics}\\[3mm]
{\bf Jury final classification: Pass with Merit}\\[5mm]
{\bf Jury}\\[3mm]
{\bf Chairperson:} Chairman of the IST Scientific Board\\[.3cm]
{\bf Members of the Committee:}\\[.3cm]
\end{center}
\onehalfspacing
{\large
Doctor Brigitte Anabelle Vaz de Abreu Hiller, Investigadora Coordenadora (com Agrega\c{c}\~ao) da
Universidade de Coimbra\\[2mm]
Doctor Alex Heinz Ladislaus Blin, Investigador Coordenador (com Agrega\c{c}\~ao) da
Universidade de Coimbra\\[2mm]
Doctor Maria Paula Fraz\~ao Bordalo e S\'a, Professora Associada (com Agrega\c{c}\~ao) do
Instituto Superior T\'ecnico, da Universidade de Lisboa\\[2mm]
Doctor Everardus Johannes Hubertus van Beveren, Professor Associado,
Faculdade de Ci\^encias e Tecnologia, da Universidade de Coimbra\\[2mm]
Doctor Georges Rupp, Investigador Principal,
Centro de F\'isica das Interac\c{c}\~oes Fundamentais do Instituto Superior T\'ecnico, da Universidade de Lisboa\\[2mm]
Doctor Pedro Jos\'e de Almeida Bicudo, Professor Auxiliar (com Agrega\c{c}\~ao)
Instituto Superior T\'ecnico, da Universidade de Lisboa}
\mbox{}\\[-2mm]
\begin{center}
\large
{\bf Funding Institutions}\\[2mm]
Funda\c{c}\~ao para a Ci\^encia e Tecnologia\\
Centro de F\'isica das Interac\c{c}\~oes Fundamentais
\end{center}
\mbox{}\\[-1cm]
\begin{center}
\textbf{\large{2013}}
\end{center}
\clearpage
\thispagestyle{empty}
\end{titlepage}

\large
\onehalfspacing
\frontmatter
\chapter{Preface}
\thispagestyle{empty}
In October 2007, fifteen days after my graduation, I was very grateful to begin to work with George Rupp, who offered me a scholarship, in a research project entitled ``Non-Perturbative Hadron Spectroscopy". However, it was not until 2011 that I decided to go ahead towards a PhD, a delay which was mainly due to the lack of appropriate funding, after several failed applications to obtain a PhD fellowship.\\
At first, I was mainly motivated by the title. I knew that hadrons are related to the strong interactions and I was acquainted with the word ``spectroscopy" from atomic physics, but did not really know what it was in connection with the strong interaction, which made me curious. But mostly, I liked the exclusive expression ``non-perturbative". Non-perturbative, as opposed to perturbative, aims at a unitary description of reality which should be simpler, although eventually more complex, from the conceptual point of view.\\
The first six months of my studies were to familiarize myself with the model which is the basis of most of the research work I present in this thesis. Then I was interrupted by the lack of funding for five months, a time in which I worked as a secretary and as a guitar teacher. My uninterrupted research started finally by January 2009, until the fall of 2011, when I had to interrupt it again to fulfill the mandatory PhD courses. At last, I was forced to write the present thesis in only one month and a few days, due to an application to a Postdoc fellowship which requires the PhD completed until the end of this civil year. Therefore, I must ask apologies to the reader for the brevity of some explanations.\\    
My research work as well as the PhD courses were developed at \emph{Centro de F\'isica das Interac\c{c}\~ oes Fundamentais}, at {\it Instituto Superior T\'ecnico}, Lisboa, with Dr.\ George Rupp and always in straight collaboration with Prof.\ Eef van Beveren. It also included two stays at {\it Universidade de Coimbra}, one plus two weeks, where I had the opportunity to collaborate directly with Prof.\ Eef van Beveren. In the course of my research I was also supported to participate in six workshops concerning Hadronic Physics: ``Scadron70", IST, 2007; ``Excited QCD 2009", Zakopane (Poland); ``Excited QCD 2010", Stara Lesna (Slovakia); 33rd International School of Nuclear Physics, Erice (Italy), 2011;  IWHSS'12, Lisboa, 2012, and ``Excited QCD 2012", Peniche. In all the three ``Excited QCD" workshops, I presented a plenary talk.\\
All my research results, except for some extra details, and unfinished work, have been published in international journals subject to a scientific refereeing system. These papers are the basis of the writing of this thesis. In each paper where I am the first author I contributed in the development of the models, performed all the computational work, plotted all the figures and tables, organized the papers, wrote the first draft and followed all the subsequent improvements.\\

{\bf List of publications:}\\
\begin{enumerate}
\item S. Coito, G. Rupp, E. van Beveren, {\it The Nature of the $X(2175)$}, Acta Phys. Polon. Suppl. {\bf2}, 431 (2009). 

\item S. Coito, G. Rupp, E. van Beveren, {\it  Multichannel calculation of excited vector $\phi$ resonances and the $\phi(2170)$}, Phys. Rev. D {\bf80}, 094011 (2009). {\bf(3rd Chapter)}

\item Susana Coito, George Rupp, Eef van Beveren, {\it Delicate interplay between the $D^0-D^{*0},\ \rho_0-J/\psi,$ and  $\omega-J/\psi$ channels in the $X(3872)$ resonance}, Eur. Phys. J. C {\bf71}, 1762 (2011). {\bf(5th Chapter)}

\item S. Coito, G. Rupp, E. van Beveren, {\it Coupled-channel analysis of the $X(3872)$}, Acta Phys. Polon. Suppl. {\bf3}, 983 (2010).

\item Susana Coito, George Rupp, Eef van Beveren, {\it Quasi-bound states in the continuum: a dynamical coupled-channel calculation of axial-vector charmed mesons}, Phys. Rev. D {\bf84}, 094020 (2011). {\bf(4th Chapter)}

\item S. Coito, G. Rupp, E. van Beveren, {\it Is the X(3872) a molecule?}, Acta Phys. Polon. Suppl. {\bf 5}, 1015 (2012).

\item Susana Coito, George Rupp, Eef van Beveren, \emph{\it $X(3872)$ is not a true molecule}, Eur. Phys. J. C {\bf73}, 2531 (2013). {\bf(6th Chapter)}
\item G. Rupp, S. Coito, E. van Beveren, {\it Light and Not So Light Scalar Mesons}, Acta Phys. Polon. Suppl. {\bf 2}, 437 (2009).

\item G. Rupp, S. Coito, E. van Beveren, {\it Complex masses in the S-matrix}, Acta Phys. Polon. Suppl. {\bf3}, 931 (2010).

\item G. Rupp, S. Coito, E. van Beveren, {\it Towards Meson Spectroscopy Instead of Bump Hunting}, Prog. in Part. and Nucl. Phys. {\bf67}, 449 (2012).

\item G. Rupp, S. Coito, E. van Beveren, {\it Meson Spectroscopy: Too Much Excitement and Too Few Excitations}, Acta Phys. Polon. Suppl. {\bf5}, 1007 (2012).

\item Eef van Beveren, Susana Coito, George Rupp, {\it Substructures from weak interactions in light of possible threshold signals at LEP and LHC}, arXiv:1304.7711 [hep-ph].
\end{enumerate}\mbox{}\\
{\bf Financial support:}\\
\begin{itemize}
\item 9 Oct.\ 2007 $-$ 8 Apr.\ 2008: {\it Non-Perturbative Hadron Spectroscopy},\\ \hspace*{5.5cm}{ PDCT/FP/63907/2005}.
\item 1 Aug.\ 2008 $-$ 31 Oct.\ 2008: {\it Non-Perturbative Methods and Phenomena in the}\\ \hspace*{5.5cm} {\it Standard Model},  {POCI/FP/81913/2007}. 
\item 1 Jan.\ 2009 $-$ 31 Jul.\ 2009: {\it Hadron Production, Scattering and Spectroscopy},\\ \hspace*{5.5cm}{CERN/FP/83502/2008}.  
\item 1 Aug.\ 2009 $-$ 31 Jan.\ 2010: {SFA-2-91/CFIF}.
\item 3 Feb.\ 2010 $-$ 2 Oct.\ 2010: {\it Hadronic Resonances and Confinement},\\ \hspace*{5.5cm}{CERN/FP/109307/2009}.
\item 3 Oct.\ 2010 $-$ 2 Feb.\ 2011: {SFA-2-91/CFIF}. 
\item 3 Feb.\ 2011 $-$ 2 Nov.\ 2011: {\it Hadronic Resonances and Non-Resonances},\\ \hspace*{5.5cm}{CERN/FP/116333/2010}.
\item 3 Nov.\ 2011 $-$ 2 May 2012: {SFA-2-91/CFIF}.
\item 3 May 2012 $-$ 31 Dec.\ 2013: {\it Genuine Hadron Spectroscopy instead of Mere}\\ \hspace*{5.5cm}{\it Bump Hunting}, { CERN/FP/123576/2011}.
\end{itemize}
\clearpage
\thispagestyle{empty}

\chapter{Agradecimentos}
\thispagestyle{empty}
Em primeiro lugar agrade\c{c}o ao Senhor, a Sabedoria Infinita, por me ter conduzido por este caminho, e por ter providenciado sempre com tudo quanto necessitei. Em particular pelo dom da intelig\^encia e ci\^encia, e pelos frutos da paci\^encia e perseveran\c{c}a, por entre todas as dificuldades, interiores e exteriores.\\ 
Em segundo lugar agrade\c{c}o ao meu ``boss", George Rupp, que foi a pedra angular de todo este meu percurso, pelo seu apoio em todas as circunst\^ancias. O George foi para mim um orientador sempre presente, sempre optimista e motivador. Acreditou em mim, mesmo tendo em conta a minha modesta m\'edia de licenciatura, e incentivou-me a realizar este doutoramento. Por outro lado, encontrou sempre maneira de me financiar com bolsas de investiga\c{c}\~ao, sem as quais n\~{a}o teria sido poss\'ivel concluir esta etapa. Para al\'em disso, ajudou-me sempre com todas as burocracias e problemas v\'arios que foram surgindo ao longo do percurso. Finalmente, agrade\c{c}o-lhe tamb\'em o especial facto de nunca me ter mandado trabalhar!\\ 
Em terceiro lugar agrade\c{c}o ao meu co-orientador, Eef van Beveren, o mestre. Por um lado, por me ter recebido sempre muito bem em Coimbra. Por outro lado, pela sua originalidade e excentricidade, e capacidade de aten\c{c}\~ao aos detalhes.\\
Agrade\c{c}o tamb\'em aos membros do Centro de F\'isica das Interac\c{c}\~ oes Fundamentais por me terem apoiado na concess\~ao das bolsas e das instala\c{c}\~oes, e ao pessoal da secretaria, pela sua simpatia e disponibilidade.\\
Aos meus colegas das pausas, especialmente ao Nuno Cardoso, Dmitri Antonov e Carlos Zarro, tamb\'em pela sua amizade, e aos outros que me ajudaram e apoiaram como colegas em diversas circunst\^ancias. Agrade\c{c}o tamb\'em \`a Cristina, respons\'avel pela limpeza do meu gabinete, pela sua amizade e compreens\~ao. E aos meus colegas de corredor, porque me deixaram sempre tocar guitarra sem se queixarem.\\
Aos meus amigos que foram o meu b\'aculo nos momentos mais complicados, e que partilharam a alegria dos momentos bons, em particular a Ana Rita Seita, Elsa Silva, Ab\'ilio Pequeno, Firmino Batista, In\^es Costa, Marcos e Elena Borges de Pinho. E a outros que, vivendo presentes no meu cora\c{c}\~ao, n\~ao puderam acompanhar-me mais de perto durante este tempo.\\
Agrade\c{c}o ainda, de cora\c{c}\~ao, a toda a Congrega\c{c}\~ao das Servas de Nossa Senhora de F\'atima, que me acompanhou sempre no meu crescimento humano e espiritual, e, em particular, \`a comunidade da Casa Geral, que me recebeu na sua casa durante este \'ultimo m\^es e meio, a fim de que eu pudesse escrever a tese na melhor paz. E de modo muito especial \`a irm\~a Mafalda Leit\~ao, que me acompanhou ao longo de todo este tempo, mostrando-se sempre dispon\'ivel e de cora\c{c}\~ao aberto.\\
Por \'ultimo, agrade\c{c}o \`a minha fam\'ilia, aos meus pais e irm\~aos, por o amor ser sempre mais forte do que todas as incompreens\~oes. E tamb\'em aos meus av\'os, que j\'a partiram deste mundo, que viveram do trabalho das suas m\~aos atrav\'es do culto da terra, por serem os alicerces.


\clearpage
\thispagestyle{empty}
\vspace*{6cm}
\begin{quote}
\hspace*{1cm}{\it ``A ci\^encia humana mais aguda \'e ignor\^ancia cega ante a divina."}\\[4mm]
{\normalsize \hspace*{2cm} in Liturgia das Horas, do Hino ``Que salmos ou que versos"}\\
{\normalsize \hspace*{2cm} in {\it V\'arias Rimas ao Bom Jesus}, Diogo Bernardes}
\end{quote} 
\clearpage
\thispagestyle{empty}
\chapter{Abstract}
\thispagestyle{empty}
Quantum chromodynamics (QCD), the quantum field theory of strong interactions, is highly nonpertubative in the low-energy sector, where confinement dominates and resonance phenomena are observed. Therefore, phenomenological unquenched models based on the old ideas of the $\mathcal{S}$-matrix theory give a fundamental contribution to understand the complex pattern of masses, widths and shapes of experimentally observed meson resonances.\\
In the present thesis we employ the Resonance-Spectrum-Expansion coupled-channel model to study two enigmatic meson states, the isoscalar vector $\phi(2170)$ and the charmonium-like axial-vector $X(3872)$. The same model is applied to describe the peculiar pattern of masses and widths of the open-charm axial-vectors -- pseudovectors $D_1(2420)$ and $D_1(2430)$, and $D_{s1}(2460)$ and $D_{s1}(2536)$. Furthermore, a simplified Schr\" odinger model is used to study the dominant wave-function components of $X(3872)$ near its resonance mass.\\
Both models successfully describe the whole variety of special features observed in experiment, which are not so easily explained in QCD-inspired quenched models. 

\vfill
{\bf Keywords:} meson spectroscopy, unquenched, unitarity, coupled-channels, confinement, $\mathcal{S}$-matrix poles, Schr\" odinger models, dynamical resonances, hidden and open charm, axial-vectors.
\clearpage
\thispagestyle{empty}
\chapter{Resumo}
\thispagestyle{empty}
A teoria de campo para as interac\c{c}\~oes fortes, conhecida como Cromodin\^amica Qu\^antica (QCD), \'e altamente n\~ao perturbativa no regime de baixas energias, onde domina o confinamento de quarks, e onde s\~ao observados os fen\'omenos de resson\^ancia de hadr\~oes. Por conseguinte, os modelos fenomenol\'ogicos baseados nas ideias antigas da teoria da matriz $\mathcal{S}$ podem contribuir, de modo fundamental, para a compreens\~ao do padr\~ao complexo de massas, larguras e formas de sec\c{c}\~ao eficaz, das resson\^ancias mes\~ao observadas experimentalmente.\\        
Na presente tese aplica-se um modelo ``unquenched" de canais acoplados, denominado ``Expans\~ao do Espectro de Resson\^ancias", para estudar dois estados mes\~ao enigm\'aticos, o vector isoscalar $\phi(2170)$ e o vector-axial do tipo charm\'onio $X(3872)$. O mesmo modelo \'e utilizado para descrever o padr\~ao peculiar de massas e larguras dos vectores-axiais -- pseudovectores com charm aberto $D_1(2420)$ e $D_1(2430)$, e $D_{s1}(2460)$ e $D_{s1}(2536)$. Desenvolve-se ainda um modelo simplificado de Schr\" odinger, para estudar as componentes dominantes da fun\c{c}\~ao de onda de $X(3872)$, junto ao limiar da sua massa de resson\^ancia.\\
Ambos os modelos descrevem com sucesso toda a variedade de tra\c{c}os especiais destes mes\~oes, observados na experi\^encia, e que n\~ao s\~ao explicados t\~ao facilmente atrav\'es dos modelos ``quenched" inspirados pela QCD.

\vfill
{\bf Palavras-chave:} espectroscopia de mes\~oes, ``unquenched", unitariedade, canais acoplados, confinamento, polos da matriz $\mathcal{S}$, modelos de Schr\" odinger, resson\^ancias din\^amicas, charm escondido e aberto, vectores-axiais.
\clearpage
\thispagestyle{empty}
\tableofcontents
\clearpage
\thispagestyle{empty}
\listoffigures
\clearpage
\thispagestyle{empty}
\listoftables
\clearpage
\thispagestyle{empty}
\mainmatter
\chapter{Introduction}
\thispagestyle{empty}
\section{Historical Overview}
In 1935 Hideki Yukawa proposed a field that should be responsible for the short-range force between the neutron and the proton, where the quanta would be bosons (integer spin) with charge $\pm e$ and mass about 200 times the mass of the electron \cite{PPMSJ17p48}. Homi Bhabha introduced a similar particle but uncharged to account for the interaction between two protons \cite{N141p117}. This new particle, proposed as the mediator of the field,  was named {\it meson}, for having a mass intermediate between the electron and the proton. It was discovered from cosmic rays in 1947, by Cecil Powell in collaboration with Giuseppe Occhialini \cite{N159p186}. In the same year, also with C\'esar Lattes, the same authors identified another ``meson", the already discovered {\it muon}, which does not interact strongly, and distinguished it from the former meson which was called {\it pion} \cite{N160p453}. This was followed by the discovery of a panoply of strongly interacting particles, both baryons, with mass above the 
proton mass and half-integer spin (fermions), and mesons. The term ``meson" was then redefined to designate solely particles with intermediate mass which interact strongly, thus excluding the $muon$.\\
In 1961 Murray Gell-Mann proposed ``The Eightfold Way", a scheme of classification of baryons and mesons based on symmetries and on the already discovered states, where the eight known baryons were grouped in a supermultiplet, and the mesons in two octets of pseudoscalars and vectors, plus two singlets \cite{CRCTSL20}. A similar representation was independently proposed by Yuval Ne'eman in the same year \cite{NP26p222}. Inquiring about the origin of the internal symmetries of isospin or hypercharge, the mechanism of {\it bootstrap} was introduced, which states that the internal symmetries can be expressed as equalities among certain masses and couplings, whose values are not put in by hand but emerge from self-consistency \cite{PR132p1831}. However, a different scheme was proposed in 1964 by Gell-Mann, which postulates that the baryons and mesons are composite systems of {\it quarks} $q$ and {\it antiquarks} $\bar{q}$, where baryons are combinations $(qqq),\ (qqqq\bar{q})$, etc., and mesons $(q\bar{q}),\ (qq\bar{q}\bar{q})$, etc., within $SU(3)$ flavor symmetry \cite{PL8p214}. A similar picture was proposed independently by George Zweig, which used the term {\it aces} instead of quarks \cite{CERN-TH-401}. Evidence of a composite proton came out in 1969, in experiments of deep inelastic scattering performed by E.\ Bloom {\it et al.} and by M.\ Breidenbach {\it et al.} \cite{PRL23p930}. James Bjorken and Sidney Drell contributed significantly to this discovery by proposing experimental methods to study the proton pointlike constituents, the ``partons", after Feynman, relating further {\it partons} to quarks \cite{PR185p1975}.\\
As mentioned above, there were two known internal symmetries of the mesons and baryons, viz.\ {\it isospin} and the {\it hypercharge}, a sum over {\it baryon number} and {\it strangeness}. A new internal symmetry was revealed after the discovery of the meson $J/\psi$ in 1974 \cite{PRL33p1404}, namely {\it charm}, which adds to hypercharge.\\
Finally, a new quantum group associated with the strong interaction was postulated in 1973 by David Gross and Frank Wilczek, the {\it color} gauge group, in the context of non-Abelian field theories \cite{PRD8p3633}.

\section{\label{cqm}The constituent quark model}
The constituent quark model is based on all the discoveries mentioned above, in particular the ``eightfold way", and some further developments. It offers the most complete classification scheme of mesons and baryons. Generically, we define {\it hadrons} as all particles and states that are subject to the strong interaction. Within the quark model, a meson is a hadron composed of a quark-antiquark pair $q\bar{q}$, while the baryon is a hadron composed of three quarks $qqq$. No other hadrons were contemplated in the original model. The name ``constituent" is used, because the postulated elementary quarks have never been observed in isolation, which implies that their mass must be estimated from the composite hadrons, as an effective, constituent mass. As far as we know, from experiment and Standard Model predictions, there are six different quark {\it flavors}, namely $up-u$ and $down-d$, associated with isospin symmetry, $strange-s$ associated with the strangeness quantum number, $charm-c$, $bottom-b$ 
and $top-t$.\\
The symmetries considered in the quark model are the flavor group $SU(3)$, or $SU(4)$, the spin group $SU(2)$, electric charge, parity, $C$-parity and $G$-parity. The flavor group is not an exact symmetry due to differences among the masses of the quarks. Yet $SU(3)$, with $q=u,d,s$, is a reasonably good symmetry, whereas $SU(4)$, with $q=u,d,s,c$, is badly broken due to the considerably larger charm mass. Quarks are fermions with spin $1/2$ and SU(2) is an exact symmetry. Baryons may have spin $S=1/2,3/2$ while mesons have $S=0,1$. Henceforth only mesons will be described, as they are the subject of this thesis. Parity is defined as $P=(-1)^{L+1}$, where $L$ is the orbital angular momentum and the additional factor $-1$ is due to the intrinsic parity of the fermion-antifermion pair $q\bar{q}$. Charge-conjugation or $C$-parity is an operator given by $C=(-1)^{L+S}$, being applicable only to particles which are their own antiparticles. Also, the $G$-parity operator is defined by $G=(-1)^IC$, where $I$ is the 
isospin. Parity, $C$-parity, $G$-parity and $J$ are conserved under strong interactions. In Table \ref{qnumbers} the lowest angular excitations of the quark-model mesons are summarized.
\begin{table}
\centering
\begin{tabular}{c|c|c|ccc|c}
&&&&&&\\[-3mm]
   &&$J^{PC}$&$J$&$L$&$S$&$^{2S+1}L_J$\\
\hline
&&&&&&\\[-3mm]
Pseudoscalar &P  &$0^{-+}$&0&0&0&$^{1}S_0$\\[1mm]
Vector       &V  &$1^{--}$&1&0, 2&1&$^{3}S_1$, $^{3}D_1$\\[1mm]
Scalar       &S  &$0^{++}$&0&1&1&$^{3}P_0$\\[1mm]
Axial Vector &A$^+$  &$1^{++}$&1&1&1&$^{3}P_1$\\[1mm]
Pseudovector &A$^-$  &$1^{+-}$&1&1&0&$^{1}P_1$\\[1mm]
Tensor       &T  &$2^{++}$&2&1, 3&1&$^{3}P_2$, $^3F_2$\\[1mm]
Tensor       &T  &$2^{-+}$&2&2&0&$^{1}D_2$\\[1mm]
Tensor       &T  &$2^{--}$&2&2&1&$^{3}D_2$\\[1mm]
\end{tabular}
\vspace*{2mm}
\caption{\label{qnumbers}Quantum Numbers of $q\bar{q}$. 5th column: spectroscopic notation.}
\end{table}
The flavor multiplets are constructed as follows:
\begin{description}
\item $SU(3)\to3\otimes\bar{3}=8\oplus1$
\item Isovectors: $I=1,\ I_3=1,0,-1\to\Big(u\bar{d},\frac{1}{\sqrt{2}}(u\bar{u}-d\bar{d}),d\bar{u}\Big)$
\item Isodoublets: $I=1/2$, $I_3=1/2,-1/2\to\Big(u\bar{s},s\bar{d},d\bar{s},s\bar{u}\Big)$ 
\item Isoscalars: $I=0\to \ket{8}=\frac{1}{\sqrt{6}}(u\bar{u}+d\bar{d}-2s\bar{s})$, $\ket{1}=\frac{1}{\sqrt{3}}(u\bar{u}+d\bar{d}+s\bar{s})$
\end{description}
Here, all states within the nonet are orthogonal. In addition,
\begin{description}
\item $SU(4)\to4\otimes\bar{4}=15\oplus1$
\item Open charm: $C=1,\ C_3=1,-1 \to\Big(c\bar{u},c\bar{d},c\bar{s},u\bar{c},d\bar{c},s\bar{c}\Big)$
\item Unflavored: $\ket{15}=\frac{1}{\sqrt{12}}(u\bar{u}+d\bar{d}+s\bar{s}-3c\bar{c})$, $\ket{1}_{SU(4)}=\frac{1}{\sqrt{4}}(u\bar{u}+d\bar{d}+s\bar{s}+c\bar{c})$
\end{description}
However, the $c\bar{c}$ component decouples from the $\ket{15}$ and $\ket{1}_{SU(4)}$ states and in practice the unflavored states of the broken group will be $\ket{8}$, $\ket{1}$, and $\ket{c\bar{c}}$, i.e., {\it charmonium}. Within $SU(3)$ there is some mixing between the {\it strangeonium} $\ket{s\bar{s}}$ and the $\ket{n\bar{n}}:=\frac{1}{\sqrt{2}}(u\bar{u}+d\bar{d})$ components. Table \ref{pv} lists the light-quark nonets, as well as the heavy-quark mesons with charm, for pseudoscalars and vectors. The octet and singlet states are mixtures of physical states which can be written as:
\be
\label{pmixa}
\begin{split}
&\ket{\psi}=\cos\theta\ket{8}-\sin\theta\ket{1},\ \ket{\psi}=\cos\phi\ket{n\bar{n}}-\sin\phi\ket{s\bar{s}},\\
&\ket{\psi'}=\sin\theta\ket{8}+\cos\theta\ket{1},\ \ket{\psi'}=\sin\phi\ket{n\bar{n}}+\cos\phi\ket{s\bar{s}},
\end{split}
\ee
where $\phi=\theta+54.736^o$.
The mixing angle, which is determined empirically, is only relevant for the pseudoscalars $\eta,\ \eta'$. Finally, the $q\bar{q}$ states within the quark model are the whole set of angular plus radial excitations, the latter associated with the quantum number $n$.  
\begin{table}
\centering
\begin{tabular}{c|c|c}
$q\bar{q}$&P&V\\[2mm]
\hline
&&\\[-3mm]
$u\bar{d},\frac{1}{\sqrt{2}}(u\bar{u}-d\bar{d}),d\bar{u}$&$\pi^+,\pi^0,\pi^-$&$\rho^+,\rho^0,\rho^-$\\[2mm]
$u\bar{s},s\bar{d},d\bar{s},s\bar{u}$&$K^+,\bar{K}^0,K^0,K^-$&$K^{*+},\bar{K}^{*0},K^{*0},K^{*-}$\\[2mm]
$\ket{8}$&$\cos\theta_P\ket{\eta}+\sin\theta_P\ket{\eta'}$&$\cos\theta_V\ket{\omega}+\sin\theta_V\ket{\phi}$\\[2mm]
$\ket{1}$&$-\sin\theta_P\ket{\eta}+\cos\theta_P\ket{\eta'}$&$-\sin\theta_V\ket{\omega}+\cos\theta_V\ket{\phi}$\\[2mm]
\hline
&&\\[-3mm]
$c\bar{u},c\bar{d},c\bar{s},u\bar{c},d\bar{c},s\bar{c}$&$D^0,D^+,D_s^+,\bar{D}^0,D^-,D_s^-$&$D^{*0},D^{*+},D_s^{*+},\bar{D}^{*0},D^{*-},D_s^{*-}$\\[2mm]
$c\bar{c}$&$\eta_c$&$J/\psi$\\
\end{tabular}
\vspace*{2mm}
\caption{\label{pv} Pseudoscalar and Vector Mesons}
\end{table}
 
\section{Quantum Chromodynamics}
Although the constituent quark ``model", as outlined above, is a good starting point for spectroscopy work, it does not yet include any dynamics and so cannot explain all hadronic states nor resonance phenomena. Even within its quality of a classification scheme, it is a simplification of reality. In a physical quantum system, there are continuous creation and annihilation phenomena, which can be described through the method of second quantization. Therefore, a mesonic system should be a composition of a ``permanent" $q\bar{q}$ component, the {\it valence} quarks, and a fluctuation component composed of $q\bar{q}$ pairs that are continuously being created and annihilated, the {\it sea} quarks. This is still a simplified scheme.\\
{\it Quantum Chromodynamics} (QCD) is a gauge-field theory developed upon the concept of {\it color} and based on the successful theory of quantum electrodynamics, which can be treated perturbatively \cite{PRD8p3633,qcd}. QCD introduces a new fundamental degree of freedom, which is the mediator of the strong interaction, called {\it gluon}, a massless particle of spin 1, electrically neutral, but color charged and with the property of being self-interacting. The theory only allows colorless compositions of quarks in the final states, like the common mesons and baryons, but additionally $qq\bar{q}\bar{q}$ or $q\bar{q}q\bar{q}$ tetraquarks, $qqqq\bar{q}$ pentaquarks, etc., and also combinations between quarks and gluons as well, the {\it hybrids}, and compositions of gluons only, the {\it glueballs}. All these configurations are labeled {\it exotics}. The theory also predicts another internal degree of freedom without external ``legs", the {\it ghost} fields, to account for gauge invariance.\\
Empirical observations of the strong interaction reveal two fundamental properties of the force, namely {\it confinement} and {\it asymptotic freedom}, both related to the spatial or, equivalently, momentum scale of the interaction. That is, quarks and gluons are always bound in colorless singlets at the femtometer scale, and behave as quasi-free below 1 fm, a region which can be probed at high momentum transfer. This is explained within QCD via the properties of the renormalized {\it coupling} strength of the interaction, which grows with the separation distance between the elementary quarks and gluons.\\
Although QCD might be an attractive theory for its conceptual simplicity, due to the highly {\it nonperturbative} character of the strong interaction this simplicity is deceiving. In practice the perturbation expansion can be performed only at very high energies, outside the physical regime of interest here, where confinement dominates and resonance phemonena are observed. Anyhow, the machinery of calculus is always very extensive, involving many details and techniques.\\
In what concerns the subject of hadron spectroscopy, QCD may contribute in two ways. The most direct approach is via {\it lattice QCD}, i.e., the heavy computational calculus of the full nonperturbative theory on a finite and discrete lattice. The indirect contribution is via models that are QCD inspired.\\
Finally, there is an empirical rule applicable to mesons, related to the dynamics of the interaction, viz.\ the {\it Okubo-Zweig-Iizuka} (OZI) rule \cite{CERN-TH-401,PL5p165}, which states that meson decay channels which require the creation of a new $q\bar{q}$ pair besides the original pair, by breaking the latter's {\it string}, are strongly favored - OZI-{\it allowed}, whereas the decay channels that involve a pair creation from the vacuum without breaking the string of the original pair are highly suppressed - OZI-{\it suppressed}. In the latter type, intermediate gluons must carry energy enough to transform into a $q\bar{q}$ pair.

\section{\label{smatrixtheory}$\mathcal{S}$-matrix theory}
There is another way to approach a physical problem without going into all details of a microscopic description, which is by reducing the parameters of the system to only a few, and then to calculate the dynamics fully. From a conceptual point of view it may be harder to understand, but the touchstone conferring reliability to these approaches is the comparison with experimental data. This is the spirit in which the {\it Scattering matrix - $\mathcal{S}$-matrix - theory} of strong interactions was built and developed. One of the reasons why field theories are often preferred is the claim of being built upon first principles. Henry Stapp discussed \cite{RMP34p390} the physical relevance of such principles on one side, and the possibility of constructing an $\mathcal{S}$-matrix theory over axioms on the other side. In what concerns field theories, he pointed out: (i) it is not known whether the principles admit any rigorous solutions, except for trivial ones; (ii) the 
principles depart from hypothetical space-time points, which are not observables; (iii) the specific principles of positive definiteness, nondegeneracy of the vacuum, completeness, locality, and energy spectrum are restrictive and arbitrary. In addition, he understood there is a disconnection between field theory and practical calculations, adding: 
\begin{quote}
\it ``Practical calculations are the heart of physics, and it is the job of physical axioms to specify a connection between experience and a well-defined mathematical scheme in which practical calculations have place."
\end{quote}
Then, he proposed seven postulates for $\mathcal{S}$-matrix theory:
\begin{enumerate}
\item There is a linear relation between the probabilities and the squares of amplitudes, according to basic quantum theory. This postulate leads to unitarity.
\item Certain sets of experiments are complete, e.g., the measurement of momentum, spin and particle type of all particles present are a complete set. Interference effects are observable.
\item The connection of momentum functions to space-time coordinates is given by a Fourier transformation.
\item Relativistic invariance.
\item The physical interpretation of the quantities of the theory be such that translational and rotational invariance imply conservation of energy-momentum and angular momentum.
\item From the above postulates one can construct a set of scattering functions which satisfy unitarity, and are analytic in the interior and on the boundary of their physical sheet, except for singularities required by unitarity, due to phase space.
\item All physical-type points of the physical sheet correspond to processes actually occurring in nature.
\end{enumerate}
Geoffrey Chew added \cite[a.]{RMP34p394} to these postulates three assumptions: (i) maximal smoothness, i.e., maximal analyticity; (ii) maximal strength, i.e., saturation of the unitary condition; and (iii) isospin, strangeness, baryon number, etc., conservation.\\
In a different paper \cite[b.]{RMP34p394}, Chew and Steven Frautschi stressed the importance of the definition of ``pure potential scattering" instead of scattering of ``independent" particles , and they stated, after observing several experimental tests:
\begin{quote}
\it ``It is plausible that none of the strongly interacting particles are completely independent but that each is a dynamical consequence of interactions between others."
\end{quote}
The theory of the scattering matrix was constructed on the basis of the principles of quantum mechanics and the pioneering work of John Wheeler \cite{PR52p1107}. He proposed the ``method of resonating group structure" to nuclear physics, which he contrasted with the Hartree-Fock procedure, i.e., the method of building up a wave function for a whole problem out of partial wave functions that describe the close interactions within the individual groups, instead of directly building up a wave function for a system of many particles. He obtained the scattering matrix as a unitary relation which {\it ``connects the asymptotic behavior of an arbitrary particular solution with that of solutions possessing a standard asymptotic form"}, cf. Eq. (48), which corresponds to $S_J=e^{i2\delta_J}$, where $\delta_J$ is the phase shift.\\
Another important precursor work \cite{NC14p951} was due to Tullio Regge, who extended the analyticity of the radial Schr\" odinger equation to complex orbital momenta, relating potentials with scattering amplitudes, and further deriving pole positions from the transmitted momentum. These {\it Regge poles} were associated with bound states and resonances that can be viewed in the complex energy $E$ plane for fixed angular momentum $J$ or vice versa. The trajectory of a single pole in the $J$ plane as $E$ changes would corresponds to a family of ``particles" of different mass, thus defining a {\it Regge trajectory} \cite[c.]{RMP34p394}.\\
Some classical reading about $S$-matrix theory may be found in Ref. \cite{Chew}.

\section{Phenomenological Models}    
Bound states calculated within the quark model essentially describe the valence-quark contribution, i.e., the {\it quenched} spectrum.
Phenomenological models which treat the resonant spectrum of bare states without considering any other relevant hadron degrees of freedom are thus considered {\it quenched} approaches. This is the case of, e.g., the mainstream model of Stephan Godgrey and Nathan Isgur (GI), a QCD inspired model \cite{PRD32p189}, where the meson spectrum is built up over a Coulomb-plus-linear ``funnel" potential. Here, the authors distinguish between ``soft" QCD - quenched, and ``true QCD" - {\it unquenched}, which includes the whole QCD action. In the same spirit, also the exotic states should have a ``quenched" spectrum which, in combination with the regular $q\bar{q}$ mesons, or $qqq$ baryons, would give rise to the complex structures 
observed in the experiments.\\
Instead, unquenched descriptions state that, besides the important valence-quark contribution, a meson or baryon is dressed with other relevant hadron components which must be included in an appropriate description of the experimental data. This ``dressing" comes from the strongly coupled nonlinear character of the interaction. In Ref. \cite{EPJA35p253} the authors suggested two opposite unquenching methods, by either dressing quark-model states and comparing the outcome to the experiments, or by taking into account self-energy contributions, implicitly included in the measured scattering-matrix poles, through an ``undressing" procedure and comparing the outcome to the quark model. This undressing may be performed through a coupled-channel model involving hadronic mass shifts, related to off-shell effects.\\
Nonperturbative microscopical approaches are employed to study the confinement problem. Within QCD, quarks and gluons interact via effective {\it strings} with a potential that grows linearly with the separation distance between two color particles, at the scale $1/\Lambda_{QCD}\sim 5$ GeV$^{-1}$, where $\Lambda_{QCD}$ is the QCD scale parameter, and the slope of the potential is given by the string tension, $\propto \Lambda_{QCD}^2$.\\
Another important nonperturbative phenomenon is the spontaneous breaking of {\it chiral} symmetry, which leads to the appearance of pseudoscalar Goldstone bosons, a role played by the pion. This symmetry corresponds to the $SU_L(N_f)\times SU_R(N_f)$ symmetry group, with $L,R$ for left- and right-handed quark fields, respectively, and $N_f$ the number of flavors of light quarks. A textbook concerning nonperturbative methods in gauge theories can be found in Ref.~\cite{Dmitri}. Effective approaches to spontaneous chiral-symmetry breaking are given by the nonrenormalizable Nambu-Jona-Lasinio model \cite{PR122p345} and the linear sigma model \cite{NC16p705}. A nonperturbative version of the latter model, at the quark instead of nuclear level, was constructed by Delbourgo and Scadron, both for $N_f=2$ and $N_f=3$ \cite{MPLA10p251}, leading to accurate predictions of a host of low-energy observables. Another approach to dynamical chiral-symmetry breaking is by employing a current quark model with a chirally 
symmetric confining potential \cite{PRD42p1611}. Related with chiral symmetry is the concept of {\it Adler zeros}, i.e., zero-mass pions which are emitted or absorbed in a strong interaction or first-order electromagnetic process \cite{PR139pB1638}.\\
A distinction is usually made among quark-mass assignments. The constituent quark masses, used in hadron spectroscopy, the dynamical quark masses, generated through chiral noninvariance of the QCD vacuum, and current quark masses, which correspond to the bare quarks and are associated with current divergences and higher momentum transfers \cite{PRD40p3670}.\\    
In constituent quark models the meson spectrum is generally obtained by solving the Schr\" odinger equation using some effective potential, which includes a $q\bar{q}$ confining part and also spin-dependent components, such as spin-orbit and color hyperfine interactions. The free parameters are then tuned to give agreement with experiment \cite{RMP71p1411}. Some models include relativistic effects \cite{PRD32p189} that should be comparable to the orbital splittings in the light-quark systems. The standard confining potential is the already mentioned ``funnel" potential given by
\be
V(\vec{r}_{ij})=-\frac{4}{3}\frac{\alpha_s(r)}{r}+br+C,
\ee  
where $\alpha_s(r)$ is the running coupling constant of the strong interactions and $C$ is an integration constant. The Coulomb part is dominant for smaller radii, typically for heavy systems, while the linear component dominates at larger radii, characteristic of light-quark systems. Linear Regge trajectories are usually seen as a consequence of the linear part of the potential.\\
Meson decays involve creation of $q\bar{q}$ pairs. In some models a pair is formed through intermediate gluons in a $^3S_1$ state, with $J^{PC}=1^{--}$. The most common assumption is pair creation with vacuum quantum numbers, i.e., in a $^3P_0$ state, with $J^{PC}=0^{++}$. This can be formulated in terms of a harmonic-oscillator spatial $SU(6)$ basis \cite{PRD8p2223}. Comparison with experimental decays highly favors the $^3P_0$ model \cite{PRD50p6855}. It is possible that the mechanism of $q\bar{q}$ pair creation is strongly related to the OZI rule \cite{PRD64p094507}.

\section{Experimental Data}
\begin{figure}[h]
\centering
\resizebox{!}{161pt}{\includegraphics{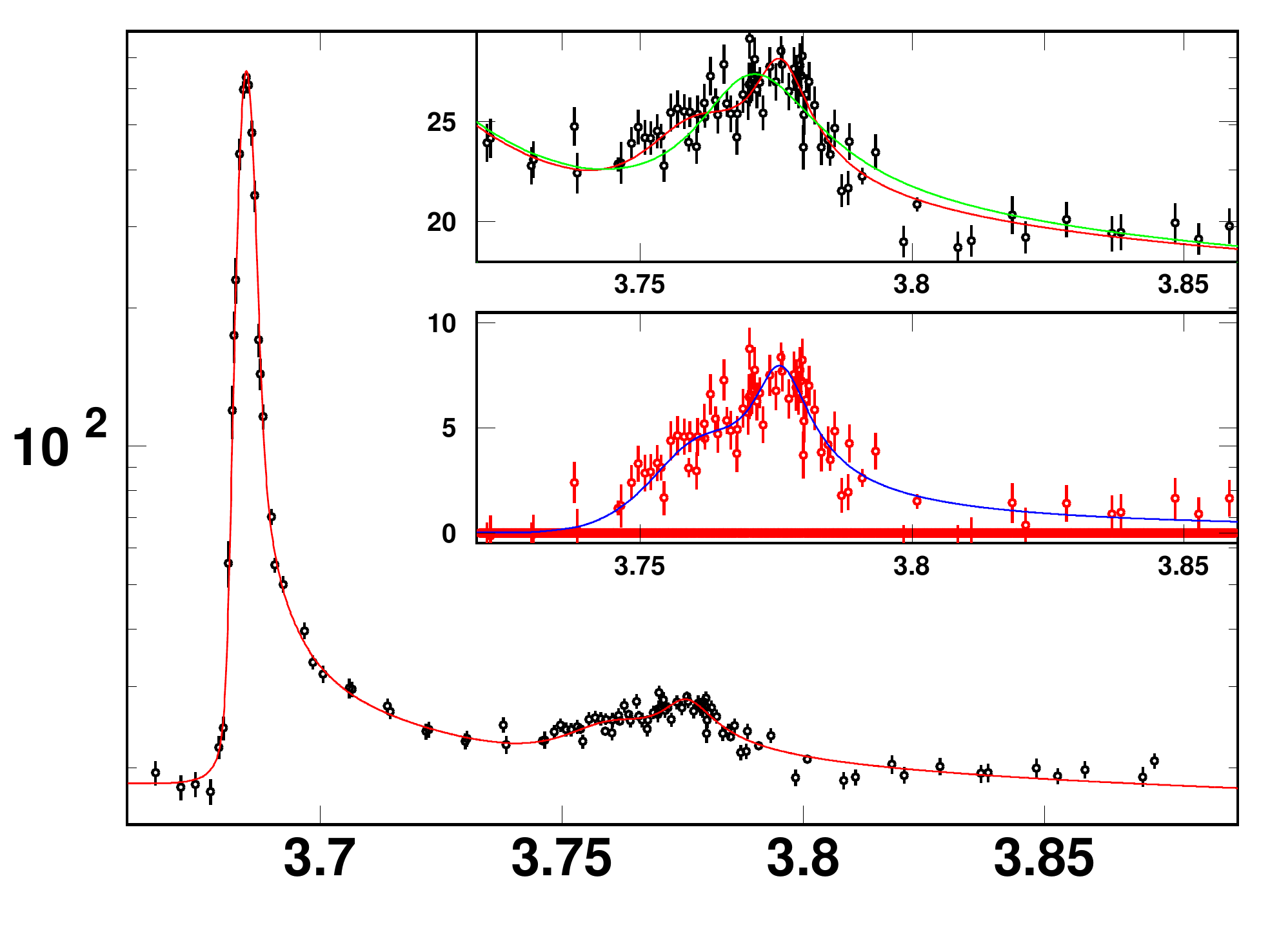}}
\put(-130,-5){\large $E_{\rm cm}$ (GeV)}
\put(-230,75){\rotatebox{90}{\large {$\sigma$ (nb)}}} 
\caption{$\psi(3770)$, BES Collab. \cite{0807.0494}.}
\label{psi3770}
\end{figure}

Scattering-matrix poles may be studied via partial-wave analyses, typically by measuring experimental phase shifts, amplitudes, and cross sections. These quantities are always unquenched, by definition. A typical resonant cross section is usually described by a simple {\it Breit-Wigner} (BW) formula, original Ref.~\cite{PR49p519}, for a process $1+2\to1'+2'$:
\be
\sigma=\frac{4\pi}{q^2}\frac{2J+1}{(2s_1+1)(2s_2+1)}|A|^2,
\ee
where $s_1$, $s_2$ are the spins of the incoming particles, $J$ is the spin of the resonance, $q$ is the c.m. momentum, and the BW scattering amplitude $A$ is given by
\be
A=\frac{x\Gamma/2}{M-\sqrt{s}-i\Gamma/2}. 
\ee
where $x$ is the branching fraction of a given decay channel, $\sqrt{s}$ is the c.m. energy, $M$ is the resonance mass, and $\Gamma$ its total decay width. These kinds of bell-shaped structures usually define the mass and width of a resonance. However, resonance peaks may be very broad and the line shapes very deformed, in which cases the definition of both parameters is not clear at all. Figure \ref{psi3770}, Ref.~\cite{0807.0494}, illustrates one of these ``deformed" signals viz., at the $\psi(3770)$ resonance.
Although the BW mass is assumed by some to be the proper physical property, the real part of the corresponding pole position in the complex energy plane may be also a good definition of resonance mass. These two definitions of mass are only equivalent for very narrow resonances. In a recent work \cite{PRL111p112004} a modified BW formula, with a new phase parameter, adjusts to different resonance shapes in a model-independent way.\\
Thresholds play a key role in understanding resonances. Some mesons appear very close to a specific decay channel, such as the $X(3872)$, which we will discuss further on, but also the scalars $f_0(980)$, $a_0(980)$, $K_0(1430)$, and the tensor $f_2(1565)$. This kind of resonances often display a {\it cusp} like line shape. However, not all cusps are necessarily generated by resonant states, but may instead be produced by some inelastic mechanism, as has been discussed by David Bugg \cite{JPGNPP37p055002}.\\
Other enhancements with resonant shapes observed in the decay channels may not correspond to true resonances, as suggested in \cite{PRL105p102001} for the case of the $X(4260)$. Furthermore, a resonant state may produce a dip instead of a bump, due to the opening of competing decay channels.\\
Resonant phenomena associated with hadrons are not trivial to resolve from the experimental point of view. Signal distortions may be due to the superposition of partial waves, or inelasticities due to competing decay channels and nearby thresholds. Also contributing to the complexity of data analysis are the large widths of some resonances, and the opening of many decay channels, which diminishes the branching fraction of each channel, especially in the case of radial excitations. A careful look at the listings of the Particle Data Group \cite{PRD86p010001} allows us to understand that many states predicted by the quark model are missing. In particular, the first and second radial excitations are not well established for all angular excitations in any flavor sector. Experimental research in the low-energy resonance region is far from being complete, for besides the need of a deeper understanding of the enhancements, which often requires more statistics, many more states are predicted than those that are 
observed. 
Moreover, the real existence of 
exotic hadrons is still not unequivocal. Phenomenological models, while trying to explain resonances and nonresonant structures from a theoretical point of view, depend entirely on the experimental data to be tested. In conclusion, more efforts are needed, especially experimental ones, in this very topical issue of hadron spectroscopy. For further reading, see Ref.~\cite{PR397p257}.
\chapter{The Models}
\thispagestyle{empty}
In this thesis two coupled-channel models are employed to study three types of mesonic resonances: the strangeonium vector $\phi(2170)$, the pseudovectors$-$axial-vectors $D_1(2420)$, $D_1(2430)$, $D_{s1}(2536)$ and $D_{s1}(2460)$, and the charmonium-type axial-vector $X(3872)$. All resonances are studied within the {\it Resonance-Spectrum-Expansion} (RSE) model. The $X(3872)$, alias $\chi_{c1}(2P)$, is additionally studied in a simple two-channel model where the wave function is fully determined. Both models are unquenched, are based on the simplified constituent quark model presented in Sec.~\ref{cqm}, and rely on the spirit of $\mathcal{S}$-matrix theory discussed in Sec.~\ref{smatrixtheory}.

\section{Some scattering formalism}
At instant $t=0$ a freely moving wave packet, at distance $r_0$ from a target with radius $a$, where $r_0\gg a$, is given by
\be
\psi(\vec{r},0)=\frac{1}{(2\pi)^{3/2}}\int d^3k\ \varphi(\vec{k})e^{i\vec{k}\cdot(\vec{r}-\vec{r_0})}.
\ee
This plane wave may be replaced by  
\be
\psi(\vec{r},0)=\int d^3k\ \varphi(\vec{k})e^{-i\vec{k}\cdot\vec{r_0}}\psi^{(+)}(\vec{k},\vec{r}),
\ee 
where the retarded wave function $\psi^{(+)}$ is a solution of the stationary Schr\"odinger equation \eqref{sch} and an eigenfunction of the Hamiltonian with a short-range central potential 
\be
\label{sch}
\big[-\nabla^2+2mV(\vec{r})\big]\psi^{(+)}(\vec{k},\vec{r})=k^2\psi^{(+)}(\vec{k},\vec{r}).
\ee
The Green's function $G(\vec{r},\vec{r'})$ is a propagator function defined as a solution of the differential equation
\be
\label{dgf}
\frac{1}{2m}\big(\nabla^2+k^2\big)G(\vec{r},\vec{r'})=\delta^{(3)}(\vec{r}-\vec{r'}).
\ee
One particular solution is the retarded Green's function
\be
\label{retgf}
G_+(\vec{r},\vec{r'})=-\frac{m}{2\pi}\frac{e^{ik|\vec{r}-\vec{r'}|}}{|\vec{r}-\vec{r'}|}.
\ee
The {\it Lippman-Schwinger} (LS) equation is then written as
\be
\label{ls}
\psi^{(+)}(\vec{k},\vec{r})=\frac{e^{i\vec{k}\cdot\vec{r}}}{(2\pi)^{3/2}}+\int d^3r'G_+(\vec{r},\vec{r'})V(\vec{r'})\psi^{(+)}(\vec{k},\vec{r'}).
\ee
This integral equation, which includes the boundary conditions, is a formal solution of \eqref{sch}.\\
Substituting \eqref{retgf} in \eqref{ls}, the asymptotic expression for $\psi^{(+)}$ becomes
\be
\psi^{(+)}(\vec{k},\vec{r})\approx\frac{1}{(2\pi)^{3/2}}\Big(e^{i\vec{k}\cdot\vec{r}}+\frac{e^{ikr}}{r}f(\vec{k},\hat{r})\Big)\ \ (r\ \mathrm{large}),
\ee
where
\be
\label{amp2}
f(\vec{k},\hat{r})=-(2\pi)^{1/2}m\int d^3r'\ e^{-ik\hat{r}\cdot\vec{r'}}V(\vec{r'})\psi^{(+)}(\vec{k},\vec{r'}).
\ee
With the help of the M\o ller operator 
\be
\Omega^{(+)}=\lim_{t\to\infty}e^{iHt}e^{-iH_0t},
\ee
the definition $\ket{\psi^{(+)}(\vec{k})}=\Omega^{(+)}\ket{\vec{k}}$, and
\be
G(z)=(z-H_0-V)^{-1}=(z-H)^{-1},
\ee
where $z$ is an arbitrary complex variable, we get
\be
\ket{\psi^{(+)}(\vec{k})}=\big[1+\lim_{\epsilon\to0}\ G(E(k')+i\epsilon)V\big]\ket{\vec{k}}.
\ee
The transition operator is defined by
\be
\label{t}
T(z)=V+VG(z)V,
\ee
so that
\be
\braket{\vec{k'}|V|\psi^{(+)}(\vec{k})}=\lim_{\epsilon\to0}\ \braket{\vec{k'}|T(E(k)+i\epsilon)|\vec{k}}.
\ee
In case of spherical symmetry let us assume
\be
\label{above}
\braket{\vec{k}|T(k^2+i\epsilon)|\vec{k'}}=\sum_{l=0}^{\infty}(2l+1)P_l(\hat{k}\cdot\hat{k'})T_l(k,k'),
\ee
where $P_l$ are Legendre polynomials.
Using $\psi^{(+)}(\vec{k},\vec{r})=\braket{\vec{r}|\psi^{(+)}(\vec{k})}$, Eq.~\eqref{above}, and also a plane-wave expansion in spherical harmonics, the integration of Eq. \eqref{amp2} yields
\be
f(\vec{k},\hat{r})=-4\pi^2m\sum_{l=0}^{\infty}(2l+1)P_l(\hat{k}\cdot\hat{r})T_l(k).
\ee
The scattering operator is defined by
\be
S(\vec{k'},\vec{k})=\braket{\psi^{(-)}(\vec{k'})|\psi^{(+)}(\vec{k})}.
\ee
It can be proved that its relation to $T$ is given by
\be
\label{st}
S(\vec{k'},\vec{k})=\delta^{(3)}(\vec{k'}-\vec{k})-2\pi i\delta(E(k')-E(k))\braket{\vec{k'}|T(E(k)+i\epsilon)|\vec{k}},
\ee
which equation explicitly shows that the matrix elements of $S$ are {\it on-shell} while those of $T$ may be {\it off-shell}.\\
As an example, consider the spherically symmetric potential
\be
V(\vec{r},\vec{r'})=\delta^{(3)}(\vec{r}-\vec{r'})V(r),
\ee
which in momentum space reads
\be
\label{fofv}
V(\vec{k},\vec{k'})=\frac{1}{2\pi^2}\sum_{l=0}^\infty(2l+1)P_l(\hat{k}\cdot\hat{k'})\int_0^\infty r^2drV(r)j_l(kr)j_l(k'r),
\ee
where $j_l$ is a spherical Bessel function. If we define
\be
\label{tpot}
V(r)=\frac{\lambda}{2\mu a}\delta(r-a),
\ee
Eq.~\eqref{fofv} reduces to
\be
V(\vec{k},\vec{k'})=\frac{\lambda a}{4\pi^2\mu}\sum_{l=0}^\infty(2l+1)P_l(\hat{k}\cdot\hat{k'})j_l(ka)j_l(k'a).
\ee
Finally, using Eq.~\eqref{t} we get
\be
\label{tborn}
\braket{\vec{p}|T|\vec{p'}}=\braket{\vec{p}|V|\vec{p'}}+\int d^3k'\int d^3k\ \braket{\vec{p}|V|\vec{k'}}\braket{\vec{k'}|G_0(z)|\vec{k}}\braket{\vec{k}|T(z)|\vec{p'}},
\ee
\be
G_0(\vec{k'},\vec{k};z)=\braket{\vec{k'}|(z-H_0)^{-1}|\vec{k}}=\frac{2\mu}{2\mu z-k'^2}\braket{\vec{k'}|\vec{k}}.
\ee
The first term of the Born expansion of Eq.~\eqref{tborn} gives
\be
T_l^{(1)}(p,p')=\frac{\lambda a}{4\pi^2\mu}j_l(pa)j_l(p'a),
\ee
the second order term
\be
T_l^{(2)}(p,p')=-\frac{i\lambda^2 p a^2}{4\pi^2\mu}j_l^2(pa)h_l^{(1)}(pa)j_l(p'a),
\ee
and the whole $T$ operator becomes
\be
\label{tm}
T(\vec{p},\vec{p'})=\frac{\lambda a}{4\pi^2\mu}\sum_{l=0}^\infty(2l+1)P_l(\hat{p}\cdot\hat{p'})\frac{j_l(pa)j_l(p'a)}{1+i\lambda paj_l(pa)h_l^{(1)}(pa)}.
\ee
\section{The Resonance-Spectrum-Expansion (RSE) model \label{RSE}}
\begin{figure}[h]
\centering
\begin{tabular}{cc}
\resizebox{!}{50pt}{\includegraphics{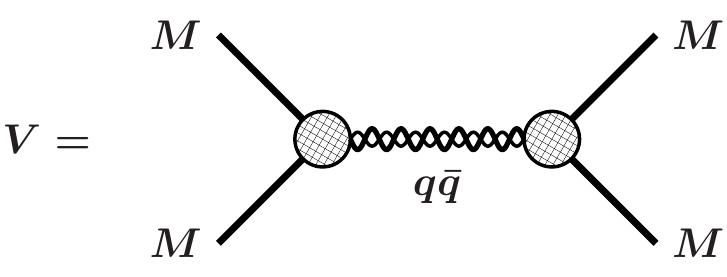}}&\resizebox{!}{50pt}{\includegraphics{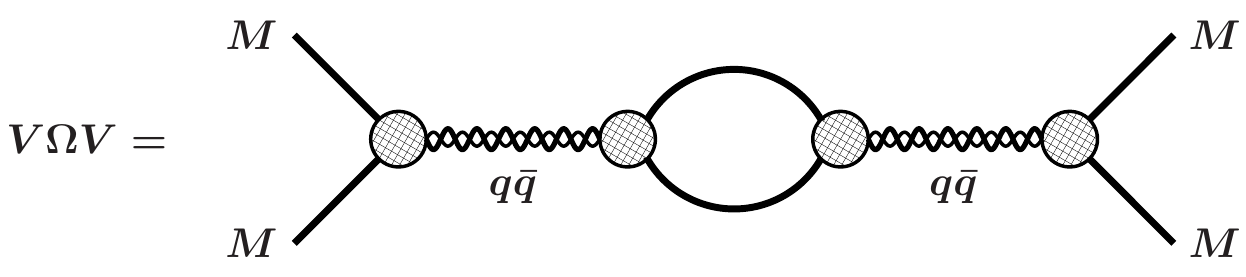}}\\
\end{tabular}
\caption{\label{be}Born Expansion, ref.~\cite{AP324p1620}.}
\end{figure}

The RSE coupled-channel model describes elastic scattering of the form $AB \rightarrow CD$, where $A, B, C$, and $D$ may be in principle any hadrons. In all applications here, they are non-exotic mesons $M$. The transition operator, Eq.~\eqref{tborn}, is described by a matrix, where each row or column represents a different channel. Its Born expansion may be represented by the diagrams in Fig.~\ref{be}.   
The effective meson-meson potential consists of an intermediate-state $s$-channel $q\bar{q}$ propagator between two $q\bar{q}$-meson-meson vertex functions for the initial and final state,
reading \cite{AP324p1620}
\be
\label{effpot}
V_{ij}^{(L_i,L_j)}(p_i,p_j;E)=\lambda^2j_{L_i}^i(p_ia)\mathcal{R}j_{L_j}^j(p_j'a)
\ee
with
\be
\label{rsep}
\mathcal{R}_{ij}=\sum_{l_c,S}\sum_{n=0}^\infty\frac{g^i_{nl_cS}\ g^j_{nl_cS}}{E-E_n^{(l_c)}}
\ee
where the {\it RSE} propagator contains an infinite tower of $s$-channel bare $q\bar{q}$ states, corresponding to the spectrum of an, in principle, arbitrary confining potential. Here, $E_n^{(l_c)}$ is the discrete energy of the $n$-th recurrency in the bare $q\bar{q}$ channel with orbital angular momentum $l_c$ and spin $S$, and $g^i_{nl_cS}$ is the corresponding coupling to the $i$-th meson-meson channel. Furthermore, in Eq.~(\ref{effpot}), $\lambda$ is an overall coupling, and $\bes{i}(p_i)$ and $p_i$ are the $L_i$-th order spherical Bessel function and the (relativistically defined) off-shell relative momentum in meson-meson channel $i$, respectively. The spherical Bessel function originates in our string-breaking picture of OZI-allowed decay, being just the Fourier transform of a spherical delta function of radius $a$; see Eqs.~\eqref{fofv} and \eqref{tpot}. Together with the overall coupling constant $\lambda$, the radius $a$ is a freely adjustable parameter here, though its range of allowed values 
turns 
out to be quite limited in practice.
Because of the separable form of the effective meson-meson interaction in Eq.~(\ref{effpot}), the fully off-shell $T$-matrix can be solved in closed form with straightforward algebra, resulting in the expression
\be
\label{tmatp}
\hspace*{-4mm}T_{ij}^{L_i,L_j}(p_i,p_j';E)=-2a\lambda^2\sqrt{\mu_ip_i}j_{L_i}^i(p_ia)\sum_{m=1}^N\mathcal{R}_{im}\{[\One-\Omega\mathcal{R}]^{-1}\}_{mj}j_{L_j}^j(p'_ja)\sqrt{\mu_jp'_j},\ 
\ee
with
\be
\Omega_{ij}(k_j)=-2ia\lambda^2\mu_jk_j\,\bes{j}(k_ja)\,\han{j}(k_ja)\,
\delta_{ij}\;, 
\label{omegap}
\ee
where $\han{j}(k_ja)$ is the spherical Hankel function of the first kind, 
$k_j$ and $\mu_j$ are the on-shell relative momentum and reduced mass in
meson-meson channel $j$, respectively, and the matrix $\mathcal{R}(E)$ is
given by Eq.~\eqref{rsep}.\\
Although in principle any confinement potential can be employed for the spectrum of the $q\bar{q}$ states, in practical applications of RSE, a harmonic oscillator (HO) with constant frequency has been used, with excellent results. For more details and further references, see Refs.\ \cite{AP324p1620}-\cite{PRL97p202001}.
Therefore, it is used in all cases here as well. The HO spectrum is given by
\be
\label{hop}
E_n=m_q+m_{\bar{q}}+\omega(2n+l_c+3/2),\ n=0,1,2,...,
\ee
where a constant frequency of $\omega$ is used, a value fixed long ago \cite{PRD27p1527}. Both $\omega$ and the constituent quark masses, ibid., are defined in Eq.~\eqref{ctepar}, where $n=u,d$.
\be
\label{ctepar}
\begin{split}
&\omega=190\ \mathrm{MeV},\\
&m_n=406\ \mathrm{MeV},\\
&m_s=508\ \mathrm{MeV},\\
&m_c=1562\ \mathrm{MeV}.
\end{split}
\ee

\subsubsection{Relative Couplings}
The relative couplings $g_{nl_cS}$ in Eq.~\eqref{rsep} are computed in accordance with the $^3P_0$ model for quark-pair creation and the OZI rule, using the formalism of Ref.~\cite{ZPC21p291}, based on overlaps of HO wave functions. These values were computed by Eef van Beveren in the $SU(3)$ flavor basis and angular momentum basis, with three-meson vertices, and are listed in Ref.~\cite{EPJC11p717} for several cases. As the $n$-dependence may be written as $g_n^2=r_n^2/4^n$, where $r_n$ is a polynomial with degree below 4 for the lowest angular excitations, in practice convergence of the series is achieved by truncating it after 20 terms.  

\subsection{Redefining the $S$-matrix\label{redsm}}
\label{unitarization}
It is straightforward to show that the $S$-matrix $S(E)\equiv\One+2iT(E)$ (cf.\ Eq.\ \eqref{st}),
where $T(E)$ is the on-energy-shell restriction of the multichannel $T$-matrix
in Eqs. \eqref{rsep}--\eqref{omegap}, is unitary and symmetric, when limited to open
channels and real energies. However, it is also easy to see 
that, for complex masses and so complex relative momenta, the unitarity of $S$
is lost, though not its symmetry. The latter property can be used to redefine
the physical $S$-matrix.\\
Since $S$ is always a symmetrix matrix, it can
be decomposed, via Takagi \cite{JJM1p82} factorization, as
\be
S \; = \; VDV^{T} \; ,
\label{takagip}
\ee
where $V$ is unitary and $D$ is a real nonnegative diagonal matrix.
Then we get
\be
S^\dag S = (V^T)^\dag DV^\dag VDV^T = (V^T)^\dag D^2V^T
= U^\dag D^2U ,
\label{sdagsp}
\ee
where we have defined $U\equiv V^T$, also unitary. So
the diagonal elements of $D=\sqrt{US^\dag SU^\dag}$  are the square
roots of the eigenvalues of the positive Hermitian matrix $S^\dag S$,
which are all real and nonnegative. Moreover, since $S=\One+2iT$ is
manifestly nonsingular, the eigenvalues of $S^\dag S$ are even all
nonzero and $U$ is unique. Thus, we may define
\be
S^\prime \; \equiv \; SU^\dag D^{-1}U \; .
\label{sprimep}
\ee
Then, using Eq.~(\ref{takagip}) and $V=U^T$, we have
\be
S^\prime \; = \; U^TDUU^\dag D^{-1}U \; = \; U^TU \; ,
\label{sprimesymp}
\ee
which is obviously symmetric and,
as
\be
(U^TU)^\dag=U^\dag(U^\dag)^T=U^{-1}(U^{-1})^T=(U^TU)^{-1}\;,
\label{sprimeunitp}
\ee
also unitary. So $S^\prime$ has the required properties to be defined as
the $S$-matrix for a scattering process with complex masses in the asymptotic
states.

\section{Two-coupled-channel Schr\" odinger model\label{CSSM}}
The RSE approach does not allow to obtain wave functions in a straightforward fashion. Here, we resort to the equivalent \cite{IJTPGTNO11p179} coordinate-space coupled-channel formalism of Ref.~\cite{ZPC19p275}, which was used to study the influence of strong decay channels on hadronic spectra and wave functions, besides several more specific phenomenological applications.\\
Consider now a system composed of a confined $q\bar{q}$ channel coupled to a meson-meson channel $M_1M_2$. Confinement is still described by Eq.~\eqref{hop}. In the scattering channel, no direct interactions between the two mesons are considered, with $\mu_f$
and $l_f$ being the reduced two-meson mass and orbital angular momenta
in the free channel, respectively. Transitions between the two channels are modeled via an off-diagonal delta-shell potential with strength $g$, which mimics string breaking at a well-defined distance $a$. The corresponding Hamiltonian, transition potential, and $2\times2$ matrix Schr\"odinger equation are given in Eqs.~(\ref{hcp})--(\ref{schrp}), with the usual definition $u(r)=rR(r)$, where
$R(r)$ is the radial wave function.
\be
\label{hcp}
h_c=\frac{1}{2\mu_c}\bigg(-\frac{d^ 2}{dr^ 2}+\frac{l_c(l_c+1)}
{r^2}\bigg)+\frac{\mu_c\omega^2r^2}{2}+m_q+m_{\bar{q}} \; ,
\ee
\be
\label{hf}
h_f=\frac{1}{2\mu_f}\bigg(-\frac{d^ 2}
{dr^ 2}+\frac{l_f(l_f+1)}{r^2}\bigg)+m_{M_1}+m_{M_2} \; ,
\ee
\be
\label{pot}
V=\frac{g}{2\mu_ca}\delta(r-a) \; ,\ \mathrm{cf.}\eqref{tpot}
\ee
\be
\label{schrp}
\left(\barr{cc}
h_c & V\\
V & h_f
\earr\right)
\left(\barr{c}
u_c\\
u_f
\earr\right)=
E\left(\barr{c}
u_c\\
u_f
\earr\right) \; .
\ee\\
Once the $1\times1$ $S$ matrix (cf.\ Eq.~(\ref{scotanp})) has been
constructed from the wave function, possible bound or virtual states as well
as resonances can be searched for.

\subsection{\label{appAp}Solving the coupled-channel Schr\"odinger equation}
We twice integrate the Schr\"odinger equation~\eqref{schrp} in order to get
two sets of boundary conditions, viz.\ Eqs.~(\ref{bc1p}) and (\ref{bc2p}):
\be
\label{bc1p}
\begin{split}
&u_c'(r\upr a)-u_c'(r\dar a)+\frac{g}{a}u_f(a)=0 \; ,\\
&u_f'(r\upr a)-u_f'(r\dar a)+\frac{g\mu_f}{a\mu_c}u_c(a)=0 \; ;
\end{split}
\ee
\be
\begin{split}
\label{bc2p}
&u_c(r\upr a)=u_c(r\dar a) \; , \\
&u_f(r\upr a)=u_f(r\dar a) \; .
\end{split}
\ee\mbox{} \\[2mm]
A general solution to this problem is the two-component wave function given by
Eqs.~(\ref{wfcp}) and \eqref{fwfp}, for the confined and meson-meson channel, 
respectively:
\be
\label{wfcp}
u_c(r)=
\left\lbrace\barr{lc}
A_cF_c(r) & r<a \; , \\[5pt]
B_cG_c(r) & r>a \; ;
\earr\right.
\ee
\be
\label{fwfp}
u_f(r)=\left\lbrace\barr{lc}
A_f J_{l_f}(kr)\; , & r<a \; , \\[5pt]
B_f\Big\lbrack J_{l_f}(kr)k^{2l_f+1}\cot\delta_{l_f}(E)-N_{l_f}(kr)\Big\rbrack \; , &r>a\;.
\earr\right.
\ee
In Eq.~(\ref{wfcp}), the function $F_c(r)$ vanishes at the origin, whereas
$G_c(r)$ falls off exponentially for $r\to\infty$, their explicit expressions
being
\begin{eqnarray}
\label{ffc}
\hspace*{-20pt} F(r)=&\displaystyle\:\frac{1}{\Gamma(l+3/2)}\,z^{(l+1)/2}\,
e^{-z/2}\,\Phi(-\nu,l+3/2,z) \; , & \\
\displaystyle
\label{fgc}
\hspace*{-20pt} G(r)=&\displaystyle-\frac{1}{2\sqrt{\mu\omega}}\Gamma(-\nu)\,z^{(l+1)/2}\,e^{-z/2}
\,\Psi(-\nu,l+3/2,z) \; ,  &
\end{eqnarray}
where $\Phi$ and $\Psi$ are the confluent hypergeometric functions of first and
second kind (see Sec.\ \ref{appBp}), respectively, $\Gamma(-\nu)$ is the
complex gamma function, $\nu$ is given by
\be
\label{evep}
\nu(E)=\frac{E-m_q-m_{\bar{q}}}{2\omega}-\frac{l_c+3/2}{2},\ \ \mathrm{cf.}\eqref{hop}
\ee 
and $z=\mu\omega r^2$. Note that only in the case of integer $\nu$, i.e., for 
$g=0$, do $\Phi$ and $\Psi$ reduce to the usual Laguerre polynomials for the
three-dimensional HO potential.
Furthermore, the functions $J$ and $N$ in Eq.~(\ref{fwfp}) are simple
redefinitions of the standard spherical Bessel and Neumann functions, i.e.,
$J_l(kr)=k^{-l}rj_l(kr)$ and $N_l(kr)=k^{l+1}rn_l(kr)$.
From the boundary conditions (\ref{bc1p}) and (\ref{bc2p}), as well as the
wave-function expressions (\ref{wfcp}) and (\ref{fwfp}), we get, with the
definition $\kappa=ka$,
\begin{eqnarray}
\label{bc1b}
\displaystyle
G_c'(r)F_c(a)-F_c'(a)G_c(a)&=&
\frac{g}{a}J_{l_f}(\kappa)F_c(a)\frac{A_f}{B_c} \; , \nonumber \\
\mbox{ } \\
\displaystyle
J_{l_f}'(\kappa)N_{l_f}(\kappa)-J_{l_f}(\kappa)N_{l_f}'(\kappa)&=&
\frac{g}{a}\frac{\mu_f}{\mu_c}J_{l_f}(\kappa)F_c(a)\frac{A_c}{B_f} \; .
\nonumber
\end{eqnarray}
Using now the Wronskian relations
\begin{eqnarray}
\label{bc2b}
\hspace*{-10pt} W(F_c(a),G_c(a)) \equiv F_c(a)G_c'(a)-F_c'(a)G_c(a)=1\;,
 \nonumber \\ \mbox{ } \hspace{-20pt} \\
W(N_{l_f}(\kappa),J_{l_f}(\kappa)) \equiv N_{l_f}(\kappa)J_{l_f}'(\kappa) -
N_{l_f}'(\kappa)J_{l_f}(\kappa)=-1 \; , \hspace{-15pt} \nonumber
\end{eqnarray}
and continuity of the wave function at
$r\!=\!a$ (cf.\ Eq.~(\ref{bc2p})), we can solve for three of the four unknowns
$A_c$, $B_c$, $A_f$, and $B_f$. Note that Eqs.~(\ref{bc1p}) and (\ref{bc2p})
are not entirely linearly independent, so that solving all four constants
is not possible. This is logical, as the overall wave-function normalization
does not follow from the Schr\"{o}dinger equation. Expressing all in terms of
$A_c$ then yields
\be
\label{ampp}
\begin{array}{cc}
A_c \; , &
\displaystyle A_f=-\left[\frac{g}{a}J_{l_f}(\kappa)G_c(a)\right]^{-1}\!A_c\;,
\\[10pt] \displaystyle
B_c=\frac{F_c(a)}{G_c(a)}\,A_c \; , &
\displaystyle B_f=\frac{g}{a}\frac{\mu_f}{\mu_c}J_{l_f}(\kappa)F_c(a)\,A_c\;.
\end{array}
\ee
Note that, in order to obtain the $D^0D^{*0}$ wave function in the outer
region, we must substitute $\cot\delta_{l_f}(E)=i$ in
Eq.~(\ref{fwfp}) (also see below). Finally, the normalization constant
$\mathcal{N}$ of the total wave function is determined by computing
\be
\label{np}
\int_0^\infty \!dr\ |u(r)|^2=
\int_0^\infty \!dr\ \left(u_c^2(r)+u_f^2(r)\right)=\mathcal{N}^2 \; .
\ee
Then, we can also calculate the root-mean-square radius
$\bar{r}=\sqrt{\langle r^2\rangle}$ of the two-component system by
\be
\label{rmsp}
\langle r^2\rangle = \frac{1}{\mathcal{N}^2}\int_0^\infty dr\,r^2
\left(u_c^2(r)+u_f^2(r)\right) \; .
\ee
As for the $S$-matrix poles corresponding to resonances, bound
states, or virtual bound states, $\cot\delta_{l_f}(E)$ can be
solved from continuity of $u_f(r)$ at $r\!=\!a$ in Eq.~(\ref{fwfp}),
resulting in the expression
\be
\label{cotanp}
\cot\delta_{l_f}(E)=
-\left[g^2\frac{\mu_f}{\mu_c}kj_{l_f}^2(\kappa) F_c(a)G_c(a)\right]^{-1}+
\frac{n_{l_f}(\kappa)}{j_{l_f}(\kappa)} \; ,
\ee
with the $1\times1$ $S$-matrix simply given by
\be
\label{scotanp}
S_{l_f}(E)=\frac{\cot\delta_{l_f}(E)+i}
{\cot\delta_{l_f}(E)-i} \; .
\ee
Real or complex poles can then be searched for numerically, by using
Newton's method to find the energies for which
$\cot\delta_{l_f}(E)=i$, on the appropriate Riemann sheet.

\section{\label{appBp}Special functions, numerical methods, and kinematics}
The confluent hypergeometric functions $\Phi$ and $\Psi$ introduced in
Subs.~\ref{appAp} are defined in Ref.~\cite{B53}, Eqs.~(6.1.1) and
(6.5.7), respectively. Thus, the function $\Phi$ is easily programmed as 
a rapidly converging power series, while the definition (6.5.7) of $\Psi$
in terms of $\Phi$ and the gamma function $\Gamma$ then also allows
straightforward computation, by employing Gauss's multiplication formula
for $\Gamma(-\nu)$ (see Ref.~\cite{AS70}, Eq.~(6.1.20)) so as to map the
argument $-\nu$ to lying well inside the unit circle in the complex plane,
whereafter a very fast converging power-series expansion of
$1/\Gamma(-\nu)$ (see Ref.~\cite{AS70}, Eq.~(6.1.34)) can be applied.\\
The integrals for wave-function normalization and computation of r.m.s.\
radii are carried out by simple Gauss integration, choosing increasing
numbers of points on a finite interval for the $c\bar{c}$ channel, and an
infinite one for $D^0D^{*0}$. Note that, in the former case, the wave
function falls off fast enough to allow convergence for a finite cutoff,
whereas in the latter a suitable logarithmic mapping is used. In both
cases though, because of the wave-function cusp at $r\!=\!a$ and in order
to avoid numerical instabilities, the domain of integration is split into
two pieces, with up to 16 Gauss points in the inner region and 64 in the
outer one, thus resulting in a very high precision of the results.\\
Although the $X(3872)$ bound state can reasonably be considered a
nonrelativistic system, we still use relativistic kinematics in the
$D^0D^{*0}$ channel, since parts of the resonance-pole trajectories
involve relatively large (complex) momenta. For consistency, the same
is done for all energies. The manifest unitarity of the $S$
matrix is not affected by this choice. Thus, the relative $D^0D^{*0}$
momentum reads
\be
\label{momp}
k(E)=\frac{E}{2}\left\lbrace\left[1-\left(\frac{T}{E}\right)^2\right]
\left[1-\left(\frac{P}{E}\right)^2\right]\right\rbrace^{\frac{1}{2}} \; ,
\ee
where $T$ and $P$ are the threshold ($m_{D^{*0}}+m_{D^0}$) and
pseudothreshold ($m_{D^{*0}}-m_{D^0}$) energies, respectively.
The corresponding relativistic reduced mass is defined as
\be
\label{redmp}
\mu_f(E)\equiv\frac{1}{2}\frac{dk^2}{dE}=
\frac{E}{4}\left[1-\left(\frac{TP}{E^2}\right)^2\right] \; .
\ee
Note that in the $c\bar{c}$ channel the reduced mass is defined in the
usual way, i.e., $\mu_c=m_c/2$, owing to the inherently nonrelativistic
nature of the HO potential and the ensuing wave function.

\chapter{Strangeonium: $\phi$'s and the $\phi(2170)$\label{chp3}}
\thispagestyle{empty}
\vspace*{-1.2cm}{\normalsize S. Coito, G. Rupp, and E. van Beveren, {\it PRD} {\bf80}, 094011 (2009).}\\[1cm]
In 2006, the BABAR Collaboration announced \cite{PRD74p091103} the discovery of a new vector-meson resonance, called $X(2175)$, in the initial-state-radiation process $e^+e^-\to K^+K^-\pi\pi\gamma$, observed in the channel $\phi(1020)f_0(980)$, with the $\phi$ meson decaying to $K^+K^-$ and the $f_0(980)$ to $\pi^+\pi^-$ or $\pi^0\pi^0$. Two years later, the BES Collaboration confirmed \cite{PRL100p102003} this resonance, then denoted $Y(2175)$, in the decay $J/\psi\to\eta[\to\!\gamma\gamma]\ \phi[\to\!K^+K^-]\,f_0(980)[\to\!\pi^+\pi^-]$. At present, the new state is included in the PDG listings as
the $\phi(2170)$ \cite{PRD86p010001} with average mass $M=(2175\pm15)$~MeV and width $\Gamma=(61\pm18)$~MeV. However, these resonance parameters are strongly challenged by the Belle \cite{PRD80p031101} results on the $Y(2175)$, alias $\phi(2170)$, observed in the process $e^+e^-\to\phi\,\pi^+\pi^-$, yielding $M=(2079\pm13^{+79}_{-28})$~MeV and $\Gamma=(192\pm23^{+25}_{-61})$~MeV. The observation of this highly excited $\phi$-type resonance with (probably) modest width, besides the peculiar, seemingly preferential, decay mode $\phi f_0(980)$, triggered a variety of model explications, most of which proposing exotic solutions. Let us mention first a strangeonium-hybrid ($s\bar{s}g$) assignment, in the flux-tube as well as the constituent-gluon model \cite{PLB650p390}, and a perturbative comparison of $\phi(2170)$ decays in these exotic ansatzes with a standard $2\,{}^{3\!}D_1$ $s\bar{s}$ description from both the flux-tube and the ${}^{3\!}P_0$ model, by the same 
authors \cite{PLB657p49}. Other approaches in terms of exotics, with QCD sum rules, are an $ss\bar{s}\bar{s}$ tetraquark assignment \cite{NPA791p106}, and an analysis \cite{PRD78p034012} exploring both $ss\bar{s}\bar{s}$ and $s\bar{s}s\bar{s}$ configurations. In an effective description based on Resonance Chiral Perturbation Theory \cite{PRD76p074012}, the bulk of the experimental data is reproduced except for the $\phi(2170)$ peak. This then led to a 3-body Faddeev calculation \cite{PRD78p074031}, with the pair interactions taken from the chiral unitary approach. Indeed, a resonance with parameters reasonably close to those of the $\phi(2170)$ is thus generated, though a little bit too narrow. Finally, a review on several puzzling hadron states \cite{IJMPE17p283} mentions the possibility that the $\phi(2170)$  arises from $S$-wave threshold effects. In this chapter, we shall study the possibility that the $\phi(2170)$ is a normal excited $\phi$ meson, by coupling a complete confinement spectrum of $s\bar{s}
$ states to a variety of $S$- and $P$-wave two-meson channels, composed of pairs of ground-state pseudoscalar (P), vector (V), scalar (S), and axial-vector (A) mesons. The employed formalism is a multichannel generalization of the Resonance-Spectrum Expansion (RSE), Sec.~\ref{RSE}, which allows for an arbitrary number of confined and scattering channels \cite{0905.3308}. 

\section{The RSE applied to $\phi$ recurrencies}
In the present investigation of strangeonium vector mesons, both the \tso\ and \tdo\ $s\bar{s}$ confinement channels are included. We could in principle also consider deviations from ideal mixing, by coupling the corresponding two $(u\bar{u}+d\bar{d})/\sqrt{2}$ channels as well, but such fine corrections will be left for possible future studies. For the meson-meson channels, we consider the most relevant combinations of ground-state P, V, S, and A mesons that have nonvanishing coupling to either of the two confinement channels in accordance with the $^{3\!}P_0$ model and the OZI rule. The resulting 17 channels are listed, with all their relevant quantum numbers, in Table~\ref{MMp}. For the channels containing an $\eta$ or $\eta'$ meson, we assume a pseudoscalar mixing angle of $37.3^\circ$, in the flavor basis, though our results are not very sensitive to the precise value. Also note that channels with the same particles but different relative orbital angular momentum $L$ or total spin $S$ are considered 
different. This is only strictly necessary for different $L$, because of the corresponding wave functions, but is also done when $S$ is different, for the purpose of clarity. All relative couplings are given in Table~\ref{MMp} for the lowest recurrencies ($n=0$). As a matter of fact, we list their squares, which are rational numbers, but given as rounded floating-point numbers in the table, also for clarity's sake. For higher $n$ values, the couplings fall off very rapidly. Their $n$ dependence, for the various sets of decay channels, is presented in Table~\ref{gnp}. The threshold values in Table~\ref{MMp} are obtained by taking the meson masses given in the PDG 2008 tables or listings \cite{PLB667p1}, with the exception of the $K_0^*(800)$ (alias $\kappa$), for which we choose the real part of the pole position from Ref.~\cite{PLB641p265}, as it lies closer to the world average of $\kappa$ masses. Note that we take sharp thresholds, even when (broad) resonances are involved. We shall come back to this point 
in 
Sec.~\ref{concphi}. Finally, we should notice that a number of channels that also couple to $s\bar{s}$ vector states according to the scheme of Ref.~\cite{ZPC21p291}, viz.\ $P$-wave channels involving axial-vector mesons as well as some channels with tensor mesons, have not been included in the final calculations presented here. However, their influence has been tested and turned out to be very modest, due to the corresponding small couplings.\\
Now we evaluate the on-shell components of the $T$-matrix defined in Eqs.~\eqref{rsep},\eqref{tmatp} for the channels given in Tables~\ref{MMp}--\ref{HOpp}.
\begin{table}[t]
\centering
\begin{tabular}{c|c c|c c | c c|c}
 & $g^2_{(l_c=0)}$ & $g^2_{(l_c=2)}$ & &  &  &  & Threshold \\[1mm]
Channel & $\times10^{-3}$ & $\times10^{-3}$ & $l_1$ & $l_2$ & $L$ & $S$ &
(MeV) \\ 
\hline & & & & & & & \\[-3mm]
$KK$             & $27.8$  & $9.26$  & $0$&$0$&$1$&$0$ & $987$\\[3mm]
$KK^*$           & $111$   & $9.26$  & $0$&$0$&$1$&$1$ & $1388$\\[1mm] 
$\eta\phi$       & $40.8$  & $3.40$  & $0$&$0$&$1$&$1$ & $1567$\\[1mm] 
$\eta'\phi$      & $70.3$  & $5.86$  & $0$&$0$&$1$&$1$ & $1977$\\[3mm]
$K^*K^*$         & $9.26$  & $3.09$  & $0$&$0$&$1$&$0$ & $1788$\\[1mm] 
$K^*K^*$         & $185$   & $0.62$  & $0$&$0$&$1$&$2$ & $1788$\\[3mm]
$\phi(1020)f_0(980)$  & $83.3$  & $0$     & $0$&$1$&$0$&$1$ & $1999$\\[1mm]  
$K^*K_0^*(800)$  & $83.3$  & $0$     & $0$&$1$&$0$&$1$ & $1639$\\[1mm] 
$\phi(1020)f_0(980)$  & $0$     & $14.7$  & $0$&$1$&$2$&$1$ & $1999$\\[1mm]  
$K^*K_0^*(800)$  & $0$     & $14.7$  & $0$&$1$&$2$&$1$ & $1639$\\[3mm]
$\eta h_1(1380)$ & $10.2$  & $5.67$  & $0$&$1$&$0$&$1$ & $1928$\\[1mm]  
$\eta' h_1(1380)$& $17.6$  & $9.76$  & $0$&$1$&$0$&$1$ & $2338$\\[1mm]  
$KK_1(1270)$     & $83.3$  & $20.6$  & $0$&$1$&$0$&$1$ & $1764$\\[1mm] 
$KK_1(1400)$     & $0$     & $2.57$  & $0$&$1$&$0$&$1$ & $1894$\\[3mm]
$K^*K_1(1270)$   & $167$   & $10.3$  & $0$&$1$&$0$&$1$ & $2164$\\[1mm] 
$K^*K_1(1400)$   & $0$     & $1.29$  & $0$&$1$&$0$&$1$ & $2294$\\[1mm] 
$\phi f_1(1420)$ & $111$   & $3.86$  & $0$&$1$&$0$&$1$ & $2439$\\[3mm]
\end{tabular}
\caption{\label{MMp}Included two-meson channels, their internal and relative angular
momenta and spins, couplings squared for $n=0$, and thresholds. See
Ref.~\cite{PLB667p1} for properties of listed mesons, except for the
$K_0^*(800)$, discussed in the text.}
\end{table}
\begin{figure}[h]
\centering
\resizebox{!}{420pt}{\includegraphics{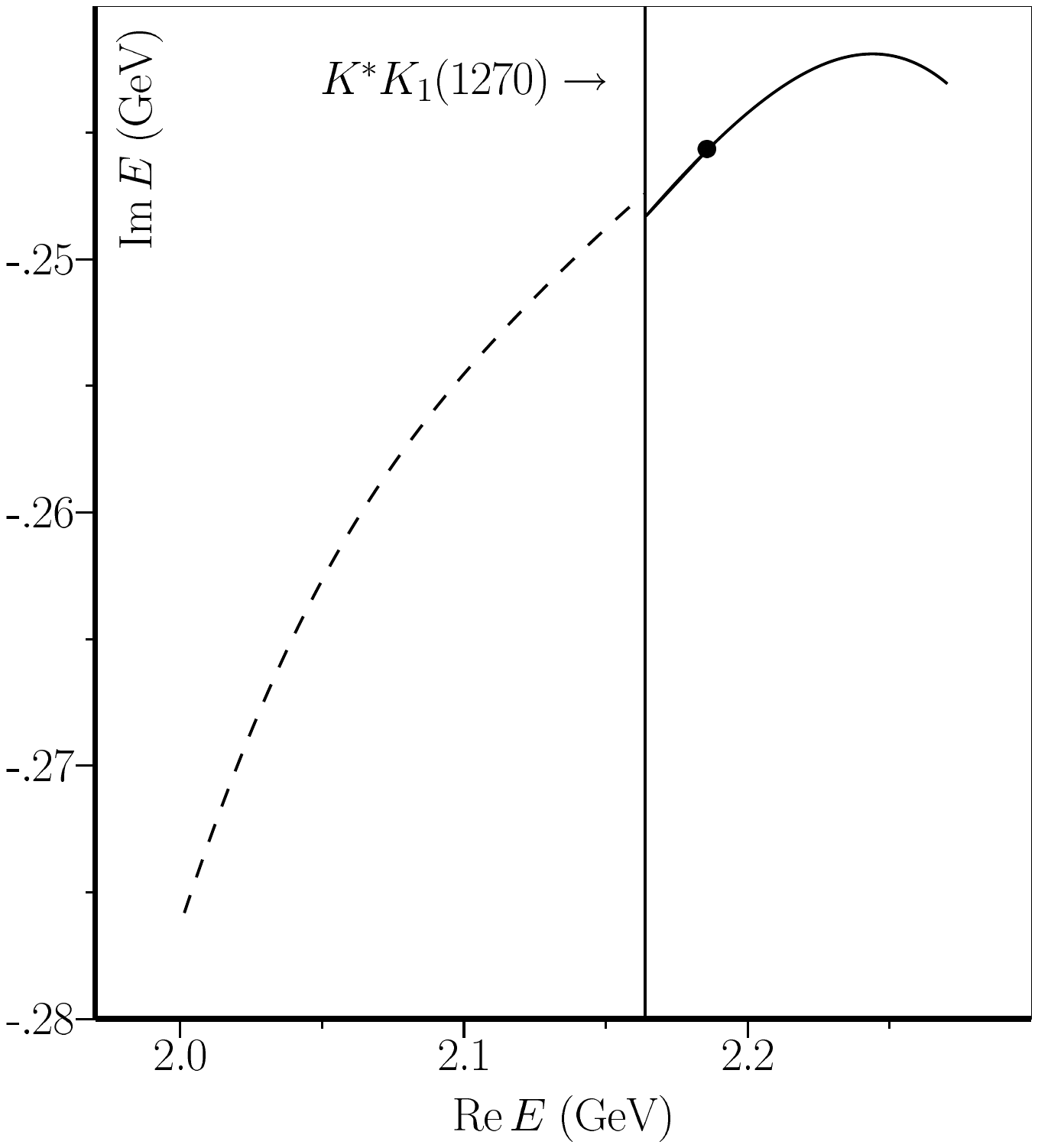}}
\mbox{}\\[-5.5cm]
\caption{Trajectory of first continuum pole, for $2.26\leq\lambda\leq5.99$
(GeV$^{-3/2}$), from left to right. Bullet represents $\lambda=4$~GeV$^{-3/2}$,
while dashed line indicates unphysical Riemann sheet.}
\label{2170}
\end{figure} 
\begin{figure}[h]
\centering
\resizebox{!}{420pt}{\includegraphics{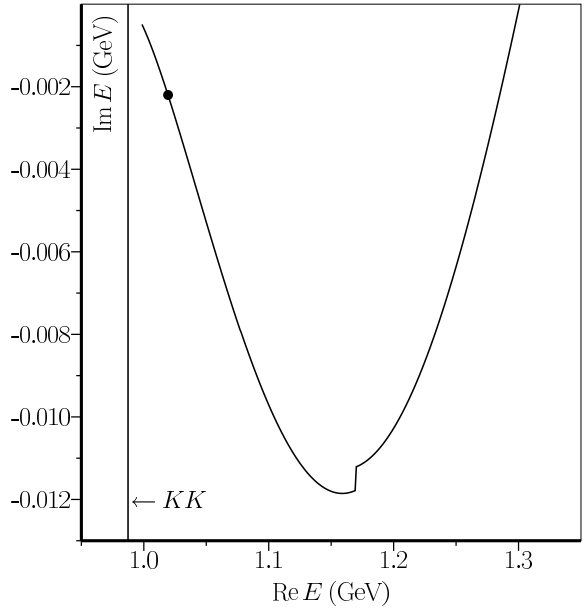}}
\mbox{}\\[-5.5cm]
\caption{$1\,{}^{3\!}S_1$ confinement pole for $4.31\geq\lambda\geq0$
(GeV$^{-3/2}$). Bullet represents $\lambda=4$~GeV$^{-3/2}$.}
\label{zerothconf} 
\end{figure}
\begin{figure}[h]
\centering
\resizebox{!}{420pt}{\includegraphics{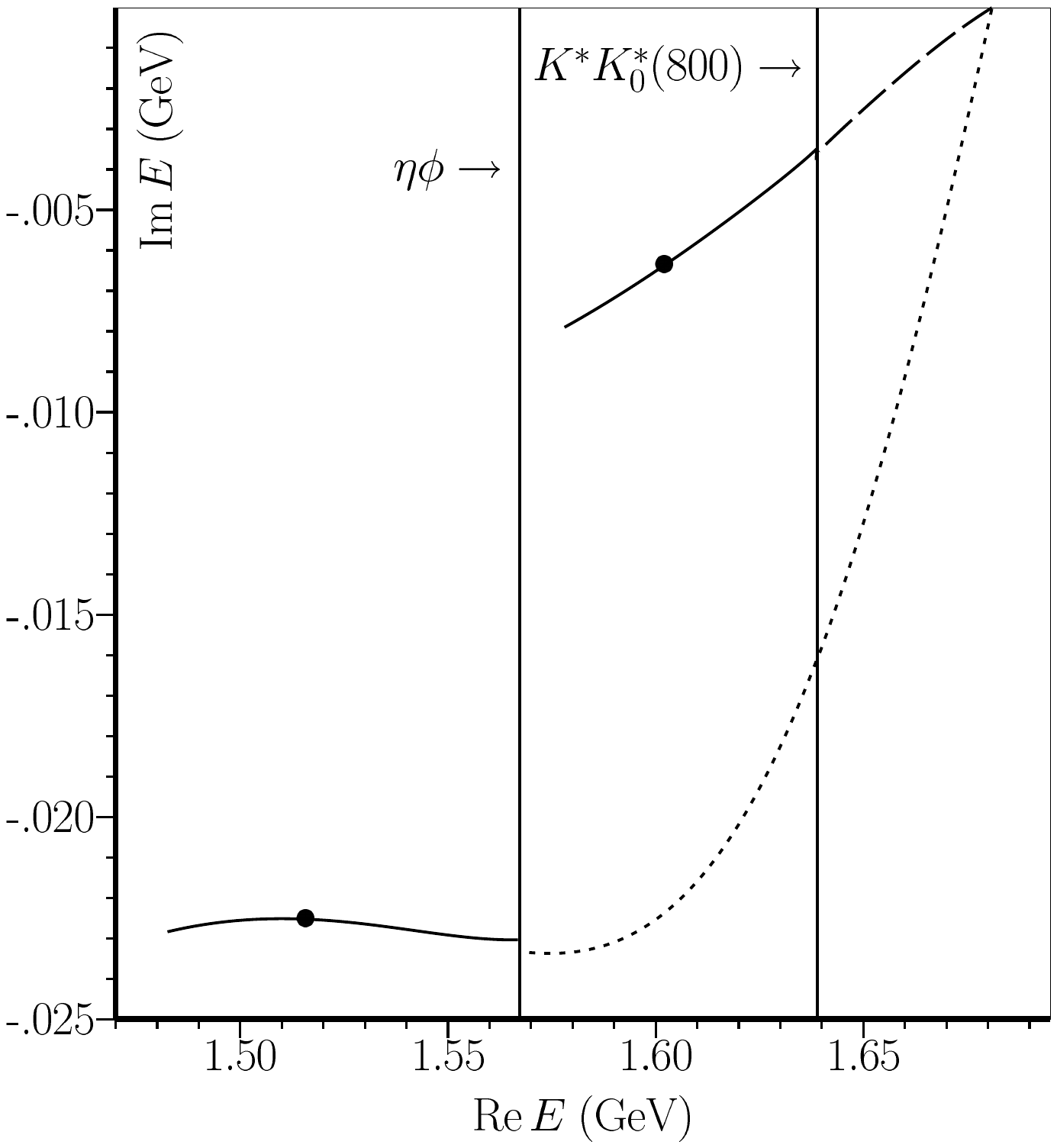}}
\mbox{}\\[-5.5cm]
\caption{\stso\ (lower) and \ftdo\ (upper) confinement poles for
$5.0\geq\lambda\geq0$ (GeV$^{-3/2}$) and $4.76\geq\lambda\geq0$ (GeV$^{-3/2}$),
respectively. Bullets represent $\lambda=4$~GeV$^{-3/2}$, while dotted and
dashed lines indicate unphysical Riemann sheets.}
\label{firstconf} 
\end{figure}
\begin{table}[h]
\centering
\begin{tabular}{c|c|c}
Channel & $\tilde{g}^2_{(l_c=0,n)}\times4^n$ &
          $\tilde{g}^2_{(l_c=2,n)}\times4^n$ \\ 
\hline & & \\[-3mm]
PP & $(2n+3)/3$ & $n+1$ \\[1mm]
PV & $(2n+3)/3$ & $n+1$ \\[1mm]
VV & $(2n+3)/3$ & $n+1$ \\[1mm]
SV & $(2n-3)^2/9$ & $(n+1)(2n+5)/5$ \\[1mm]
PA & $(2n-3)^2/9$ & $(n+1)(2n+5)/5$ \\[1mm]
VA & $(2n-3)^2/9$ & $(n+1)^2$ \\[1mm]
\end{tabular}
\caption{\label{gnp}Dependence of couplings squared on recurrency $n$.}
\end{table}
\begin{table}
\centering
\begin{tabular}{c|c|c}
$n$ & $l_c=0$ & $l_c=2 $\\ [0.5ex]
\hline & & \\[-3mm]
0 & 1301 & 1681\\[1mm]
1 & 1681 & 2061\\[1mm]
2 & 2061 & 2441\\[1mm]
3 & 2441 & 2821\\[1mm]
\end{tabular}
\caption{\label{HOpp}Masses of bare $s\bar{s}$ states in MeV, for HO potential with
$\omega=190$~MeV and $m_s=508$ MeV (see Eqs.~\eqref{hop} and \eqref{ctepar}).}
\end{table}

\section{Experimental status of $\phi$ states}
Before adjusting our two free parameters $\lambda$ and $a$ from Eq.~(\ref{tmatp}), let us first have a look at the experimental status of vector  $\phi$ resonances. According to the 2012 PDG listings \cite{PRD86p010001}, there are only 3 observed states, viz. the $\phi(1020)$, $\phi(1680)$, and $\phi(2170)$. Their PDG masses and widths are given in Table~\ref{phis}. 
\begin{table}[h]
\centering
\begin{tabular}{c|c|c}
  & $M$ (MeV) & $\Gamma$ (MeV) \\ 
\hline & & \\[-3mm]
$\phi(1020)$ & 1019.455$\pm$0.020 & 4.26$\pm$0.04\\[1mm]
$\phi(1680)$ & 1680$\pm$20 & 150$\pm$50\\[1mm]
$\phi(2170)$ & {\boldmath$2175\pm15$} & {\boldmath $61\pm18$}\\[1mm]
\end{tabular}
\caption{\label{phis}Listed $J^{PC}=1^{--}$ $\phi$ resonances, with masses and widths 
\cite{PLB667p1} (values for $\phi(1680)$ are estimates \cite{PLB667p1}).}
\end{table}
Clearly, this is a very poor status, as several additional states must exist in the energy range 1--2~GeV according to the quark model, and also if we compare with e.g.\ observed $\rho$ resonances \cite{PLB667p1} in the same energy interval. Moreover, the $\phi(1680)$ can hardly be the first radial excitation of the $\phi(1020)$, in view of the well established $K^*(1410)$, which is almost 300~MeV lighter, and a typical mass difference of 100--150 MeV between the strange and nonstrange ($u,d$) constituent quarks \cite{PRD27p1527,IJMPA13p657}. This conclusion is further supported if indeed the $\rho(1250)$ is confirmed as the first radial recurrence of the $\rho(770)$ \cite{NPA807p145,PRD27p1527}. So the $\phi(1680)$ is more likely to be the $1\,{}^{3\!}D_1$ state, with a hitherto undetected $2\,{}^{3\!}S_1$ state somehwere in the mass range 1.5--1.6~GeV. As a matter of fact, in Ref.~\cite{PRD57p4334} a vector $\phi$ resonance was reported at roughly 1.5~GeV, though this observation is, surprisingly, included 
under the $\phi(1680)$ entry \cite{PLB667p1}. Even more oddly, another $\phi$-like state, at $\sim\!1.9$~GeV and reported in the same paper \cite{PRD57p4334}, is {\em also} \/included under the $\phi(1680)$ \cite{PLB667p1}. However, a resonance at about 1.9~GeV should be a good candidate for the next radial $s\bar{s}$ recurrency, if we take the observed $\rho$ resonances in
Ref.~\cite{NPA807p145} for granted.

\section{Hunting after poles}
\begin{figure}[h]
\centering
\resizebox{!}{420pt}{\includegraphics{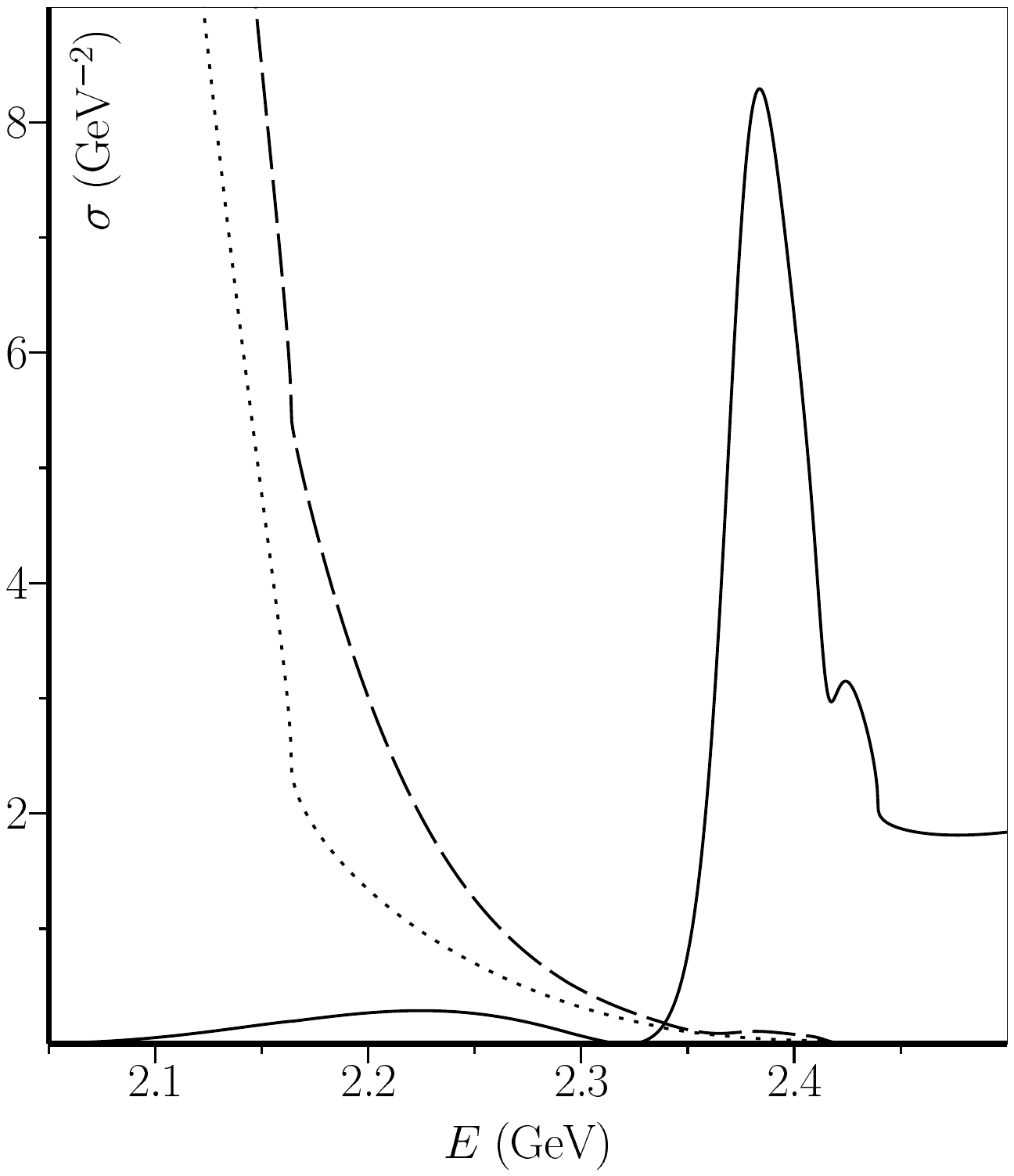}}
\mbox{}\\[-5.5cm]
\caption{Elastic $D$-wave (solid line) and $S$-wave (dashed line)
$\phi(1020)f_0(980)$ cross section. Dotted line: $S$-wave cross section for
\tso\ channel only.}
\label{f0phi}
\end{figure}

In view of the poor status of excited $\phi$ states, let us adjust our parameters $\lambda$ and $a$ to the mass and width of the $\phi(1020)$. Here, we should mention that an additional phenomenological ingredient of our model is an extra suppression of subthreshold contributions, using a  form factor, on top of the natural damping due to the spherical Bessel and Hankel functions in Eq.~(\ref{rsep}). Such a procedure is common practice in multichannel phase-shift analyses. Thus, for closed meson-meson channels we make the substitution
\begin{equation}
\left(g^i_{(l_c,n)}\right)^2\;\to\; \left(g^i_{(l_c,n)}\right)^2
e^{\alpha k_{i}^{2}}
\;\;\;\;\mbox{for}\;\;\;\;
\Re\mbox{e}\, k_{i}^{2}<0
\; .
\label{damping}
\end{equation}
The parameter $\alpha$ is chosen at exactly the same value as in previous work \cite{PLB641p265,PRL97p202001}, viz.\ $\alpha=4$~GeV$^{-2}$.\\
Choosing now $\lambda=4$ and $a=4$~GeV$^{-1}$, we manage to
reproduce mass and width of the $\phi(1020)$ with remarkable accuracy, namely
$M_\phi=1019.5$ MeV and $\Gamma_\phi=4.4$ MeV. Note that these values of
$\lambda$ and $a$ are of the same order of magnitude as in the work mentioned
before \cite{PLB641p265,PRL97p202001}, which dealt with scalar mesons.\\
In Table~\ref{polesp} we collect all resonance poles encountered on the
respective physical Riemann sheets, which correspond to $\Im\mbox{m}\,k_i>0$
for closed channels and $\Im\mbox{m}\,k_i<0$ for open ones. When the latter
conditions are not fulfilled, we call the corresponding Riemann sheets
unphysical. Moreover, we also show here the pole positions obtained
by taking only the \tso\ $s\bar{s}$ channel and switching off the \tdo, for
fixed $\lambda$ and $a$. Focusing for the moment on those poles that originate
\begin{table}[h]
\centering
\begin{tabular}{c|cc|cc|c}
&\multicolumn{2}{c|}{${}^{3\!}S_1$ only}&
\multicolumn{2}{c|}{${}^{3\!}S_1+{}^{3\!}D_1$}&\\
\hline & & & & & \\[-3mm]
Pole&$\Re$e & $\Im$m & $\Re$e & $\Im$m & Type of Pole \\ 
\hline & & & & & \\[-3mm]
1&$1027.5$ &$-2.7$& $1019.5$ &$-2.2$& conf., $n=0$, $1\,{}^{3\!}S_1$\\[1mm]
2&$1537$  &$-13$ & $1516$  &$-23$ & conf., $n=1$, $2\,{}^{3\!}S_1$\\[1mm]
3&   -   &    -  & $1602$  &$ -6$& conf., $n=0$, $1\,{}^{3\!}D_1$\\[1mm]
4&$1998$  & $-16$& $1932$  &$-24$ & conf.\, $n=2$, $3\,{}^{3\!}S_1$\\[1mm]
5&   -   &    -  & $1996$  &$-14$ & conf.\, $n=1$, $\,{}2^{3\!}D_1$\\[1mm]
6&$2397$  & $-214$& {\boldmath$2186$} & {\boldmath $-246$}&continuum \\[1mm]
7&$2415$  & $-6$& $2371$  & $-29$  & conf., $n=3$, $4\,{}^{3\!}S_1$\\[1mm]		
8&-      & -     & $2415$  & $-8$  & conf., $n=2$, $3\,{}^{3\!}D_1$ \\[1mm]
9&$2.501$  & $-236$& $2551$  & $-193$  & continuum \\[1mm]
\end{tabular}
\mbox{} \\[1mm]
\caption{\label{polesp}Complex-energy poles in MeV, for ${}^{3\!}S_1$ $s\bar{s}$ channel
only, and for both ${}^{3\!}S_1$ and ${}^{3\!}D_1$. See text for further
details.}
\end{table}
in the states of the confinement spectrum (indicated by ``conf.'' in the
table), we see good candidates for the resonances at $\sim\!1.5$~GeV and
$\sim\!1.9$~GeV reported in Ref.~\cite{PRD57p4334}, and possibly also for the
$\phi(1680)$, though our $1\,{}^{3\!}D_1$ state seems somewhat too light. Note,
however, that under the $\phi(1680)$ entry \cite{PLB667p1} in the PDG listings
there is a relatively recent observation \cite{PLB551p27} with a mass of
$(1623\pm20)$~MeV, which is compatible with our pole at 1602~MeV.
Furthermore, the imaginary parts of the confinement poles are generally too
small, except for the $\phi(1020)$. We shall come back to this point in the
conclusions below. Besides the latter poles, also two so-called {\em
continuum} \/poles are found, often designated as {\em dynamical} \/poles,
the most conspicuous of which is the one at $(2186-i246)$~MeV, as the real part
is very close to the mass of the $\phi(2170)$ as measured by BABAR
\cite{PRD74p091103} and BES \cite{PRL100p102003}. However, in view of the much
too large width, even as compared to the Belle \cite{PRD80p031101} value,
considerable caution is urged. Also this point will be further discussed in the
conclusions.\\
Some words are in place here about our identification of the \tso\ and \tdo\
confinement poles in Table~\ref{polesp}. The point is that, rigorously speaking,
these designations only make sense for pure confinement states and, moreover,
without any \tso\,/\,\tdo\ mixing. Now, in our approach, the very mixing is
provided by the coupling to common decay channels. So for any nonvanishing
value of the overall coupling $\lambda$ there are no longer pure \tso\ and
\tdo\ states, while for the physical value of $\lambda$ the mixing is probably
considerable. Moreover, there is no obvious way to tell which pole of a pair
originating in a degenerate confinement state stems from either \tso\ or \tdo.
Therefore, our identification is partly based on the couplings in
Table~\ref{MMp}, which on the whole suggest larger shifts for \tso\ than for
\tdo, partly on a comparison with a perturbative approach employed in
Ref.~\cite{PRD21p772} to find poles for small $\lambda$.\\
The designation {\em continuum} \/pole becomes clear when plotting a
corresponding trajectory as a function of the overall coupling $\lambda$. In
Fig.~\ref{2170}, the first such pole is shown to have an {\em increasingly} 
\/large imaginary part for {\em decreasing} \/$\lambda$, eventually
disappearing in the continuum for $\lambda\to0$. 
\begin{figure}[h]
\centering
\resizebox{!}{420pt}{\includegraphics{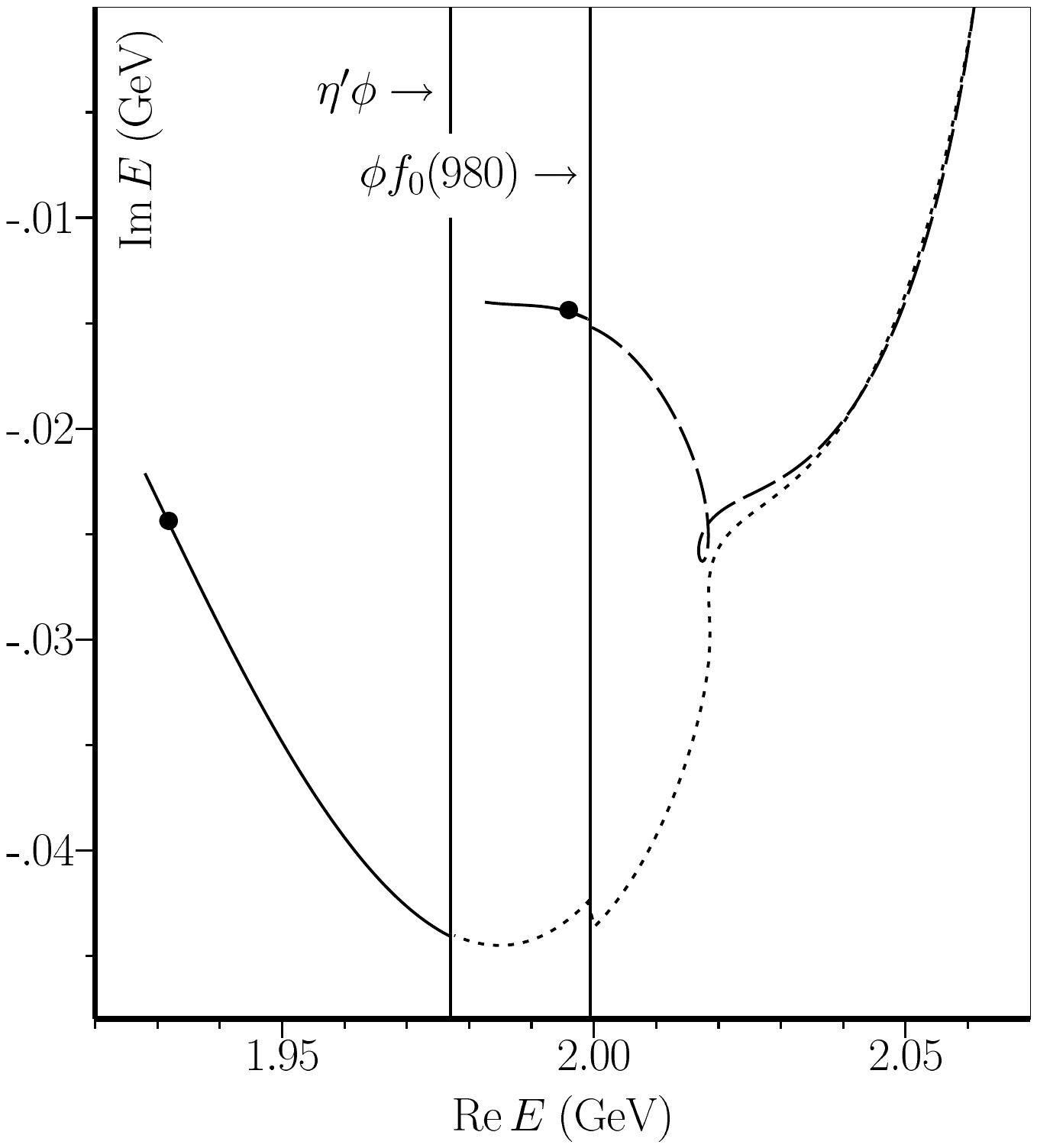}}
\mbox{}\\[-5.5cm]
\caption{\ttso\ (lower) and \stdo\ (upper) confinement poles for
$4.2\geq\lambda\geq0$ (GeV$^{-3/2}$) and $5.99\geq\lambda\geq0$ (GeV$^{-3/2}$),
respectively. Bullets represent $\lambda=4$~GeV$^{-3/2}$, while dotted and
dashed lines indicate unphysical Riemann sheets.}
\label{secondconf}
\end{figure}
\begin{figure}[h]
\centering
\resizebox{!}{420pt}{\includegraphics{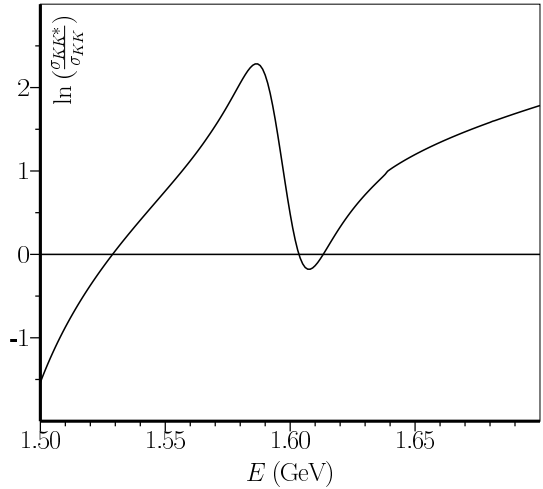}}
\mbox{}\\[-6.5cm]
\caption{Natural logarithm of the ratio of the elastic $KK^*$ and $KK$
cross sections.}
\label{KKSKK}
\end{figure} 
Turning now to the $\phi(2170)$ energy region, we show in Fig.~\ref{f0phi}
the elastic $S$- and $D$-wave
\begin{figure}[h]
\centering
\resizebox{!}{420pt}{\includegraphics{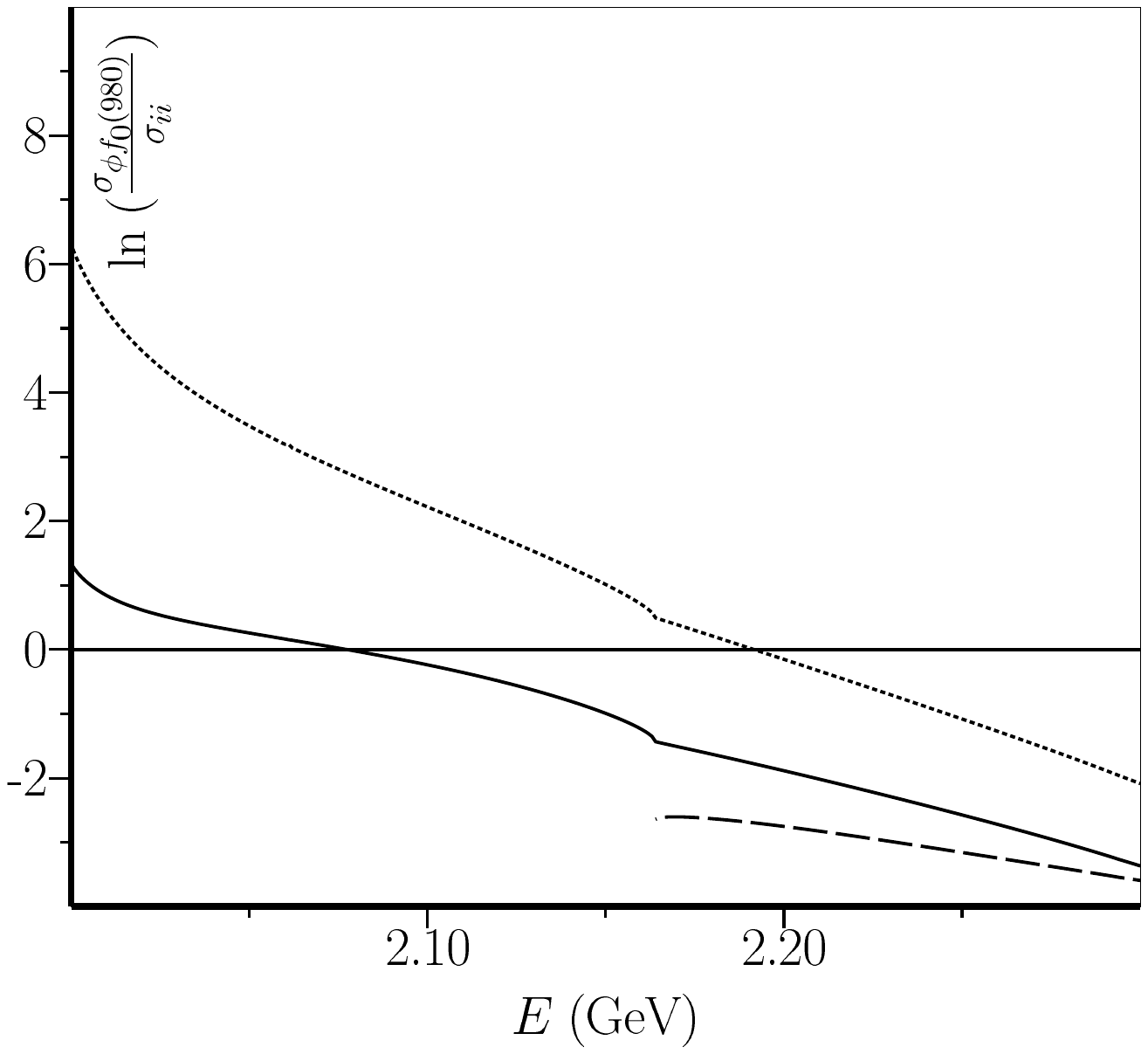}}
\mbox{}\\[-6.5cm]
\caption{Natural logarithm of the ratios of the elastic $S$-wave
$\phi(1020)f_0(980)$ cross section and the elastic $K^*K^*$ (solid line),
$\phi(1020)\eta'$ (dotted line), and $K^*K_1(1270)$ (dashed line) cross
sections.}
\label{f0phiKK}
\end{figure}
Note that the small jump
at the important $S$-wave $K^*K_1(1270)$ threshold is due to a minor
threshold discontinuity of the damping function in Eq.~(\ref{damping}) for
complex momenta.
Figure~\ref{zerothconf} shows a similar trajectory, but now for the lowest
confinement pole, which ends up as the $\phi(1020)$ resonance. Notice the
large negative mass shift ($\approx\!280$~MeV), as well as the way the pole
approaches the $K\bar{K}$ threshold, which is typical for $P$-wave decay
channels. Also note that the tiny jump in the trajectory is due to the way 
relativistic reduced mass is defined below threshold, which in the case
of closed channels with highly unequal masses ($KK_1(1270)$ here) requires 
an intervention to prevent the reduced mass from becoming negative.
In Fig.~\ref{firstconf}, we depict the trajectories of the \stso\ and
\ftdo\ confinement poles. Note that the coupling to decay channels lifts
the original degeneracy of the \stso\ and \ftdo\ HO states.
The trajectories of the next pair of confinement poles, i.e., \ttso\ and \stdo,
are drawn in Fig.~\ref{secondconf}. Note the highly nonlinear behavior of the
poles, showing the unreliability of perturbative methods to estimate
coupled-channel effects.

\section{Cross sections}
Now we shall show, as mere illustrations, some of the cross sections related to
the resonance poles found in the preceding section.
\begin{figure}[h]
\centering
\resizebox{!}{420pt}{\includegraphics{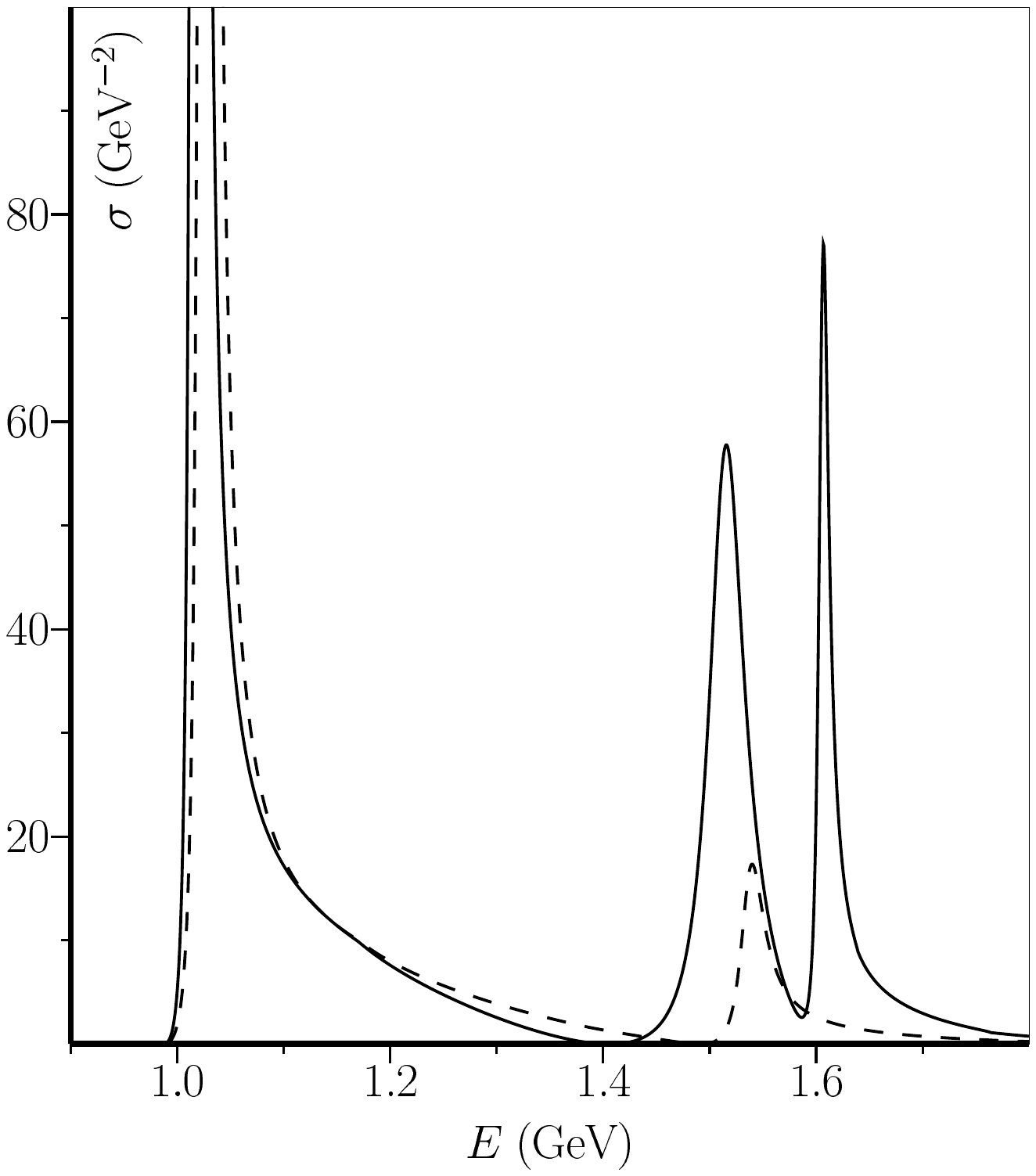}}
\mbox{}\\[-5.5cm]
\caption{Elastic $P$-wave $KK$ cross section. Full line: both \tso\ and \tdo\
$s\bar{s}$ channels included; dashed line: only \tso.}
\label{KK}
\end{figure}
In Fig.~\ref{KK}, the elastic $P$-wave $KK$ cross section is depicted in
the energy region covering the $\phi(1020)$ as well as the \stso\ and \ftdo\
resonances. We see that including the \tdo\ $s\bar{s}$ channel has the effect
of lowering the \stso\ state, besides the generation of an additional
resonance, of course. This ``repulsion'' between the \tso\ and \tdo\ poles is
also noticed for the \ttso\ and \stdo\ states.\\
Figure~\ref{KKSKK} shows the relative importance of the $KK$ and $KK^*$
channels in the energy interval 1.5--1.7~GeV, which should be relevant for
the $\phi(1680)$. The plotted quantity is the logarithm of the ratio of
the elastic $KK^*$ and $KK$ cross sections, which shows that the $KK^*$
channel is strongly dominant, except at low energies, because of phase
space, and close to the pole at $\sim\!1.6$~GeV, where the two cross sections
are comparable. Dominance of the $KK^*$ decay mode is reported under the
$\phi(1680)$ PDG entry \cite{PLB667p1}.
$\phi(1020)f_0(980)$ cross sections. The effect of the continuum pole at
$(2186-i246)$~MeV is noticeable as a small and very broad enhancement in the
$D$-wave cross section. In the $S$-wave case, its effect is completely
overwhelmed by the huge cross section at threshold, partly due to the
\ttso\ pole not far below. Also quite conspicuous are the here predicted
\ftso\ and \ttdo\ resonances (see Table~\ref{polesp} for the respective pole
positions). Of course, all these model {\em elastic} \/cross
sections have little direct bearing upon the experimentally observed 
{\em production} \/cross sections. The production process of the
$\phi(2170)$ may be studied with the RSE production formalism 
\cite{AOP323p1215}, but that lies outside the scope of the present
investigation, which focused on the possibility of generating a
$\phi(2170)$ resonance pole through coupled channels.\\
Finally, in Fig.~\ref{f0phiKK} we plot the logarithm of the ratios
of the elastic $S$-wave $\phi(1020)f_0(980)$ cross section and the 
elastic $K^*K^*$, $\phi(1020)\eta'$, and $K^*K_1(1270)$ cross sections,
in the energy interval 2.0--2.3~GeV. We see that the $S$-wave $\phi(1020)f_0(980)$ cross section dominates
up to about 2.08~GeV, but getting overwhelmed first by the 
($P$-wave) $K^*K^*$ channel, and then even more so by the $S$-wave
$K^*K_1(1270)$ channel, right from its threshold at $\approx\!2.16$~GeV
upwards. Also the $\phi(1020)\eta'$ channel is becoming more important
here. As for the $K^*K^*$ channel, it gives rise to a final state with
two kaons and two pions, i.e., the same as that for which the
$\phi(2170)$ was observed. So the experimental status of the
$\phi(2170)$ might be improved if one succeeded in identifying and
isolating the $K^*[\to K\pi]K^*[\to K\pi]$ decay mode, which should be
quite important.
 
\section{\label{concphi}Summary and conclusions}
In this chapter, we have applied the RSE formalism for non-exotic multichannel
meson-meson scattering to calculate the resonance spectrum of excited vector
$\phi$ mesons, and to find out whether this way the $\phi(2170)$ can be
generated. The inclusion of all relevant two-meson channels that couple to
the bare \tso\ and \tdo\ $s\bar{s}$ states should guarantee a reasonable
description.
Thus, several vector $\phi$ resonances are predicted, some of which are good
candidates for observed states, while others may correspond to others,
undetected so far, but quite plausible in view of observed partner states in 
the excited $\rho$ spectrum. Finally, a very broad $\phi$-like resonance pole
of a dynamical origin is found, with real part very close to that of the
$\phi(2170)$, but a much too large imaginary part, so that its interpretation
remains uncertain. On the other hand, the calculated resonances originating in
the confinement spectrum are generally too narrow.\\
These considerations bring us to the main problem of our description,
namely the inclusion of sharp thresholds only. The point is that many
of the channels in Table~\ref{MMp} involve highly unstable particles, several
of which are broad to very broad resonances themselves. Treating the
corresponding thresholds as sharp is clearly an approximation. In particular,
the $f_0(980)$ meson included in the $\phi(1020)f_0(980)$ channels is a
very pronounced resonance in the coupled $\pi\pi$-$KK$ system. This feature
is crucial in the three-body calculation of the $\phi(2170)$ in
Ref.~\cite{PRD78p074031}, which indeed produces a clear resonance signal
at almost the right energy, and even with a somewhat too {\em small} \/width.
We believe that in our approach, too, a narrower $\phi(2170)$
might be generated, if we could account for the physical width of the
$f_0(980)$ meson, and also for the widths of the $K^*$ and
$K_1(1270)$ resonances in the here included $K^*K_1(1270)$ channel. The
reason is that the widths effectively cause these channels to act already
below their central thresholds, which will strongly influence poles just
underneath. Especially the width of very strongly coupling $K^*K1(1270)$
channel, whose threshold lies only some 25~MeV below the real part of the
continuum pole at $(2186-i\times246)$~MeV, will surely have a very significant
effect on this pole's trajectory. Because of the typical behavior of continuum
poles, with decreasing width for increasing coupling, we expect that the width
of our $\phi(2170)$ candidate may thus be reduced.  Conversely, including the
widths of final-state resonances will probably {\em increase} \/the
widths of the now too narrow excited $\phi$ resonances stemming from the
confinement spectrum.\\
To account for the nonvanishing widths of mesons in the
coupled channels is a very difficult problem, since the simple 
substitution of the here used real masses by the true complex masses
will destroy the manifest unitarity of the $S$-matrix.\\

\chapter{Axials with open-charm\label{chp4}}
\thispagestyle{empty}
\vspace*{-1.2cm}{\normalsize S. Coito, G. Rupp, and E. van Beveren, {\it PRD} {\bf84}, 094020 (2011).}\\[1cm]
The axial-vector (AV) charmed mesons \dc\ and \dsc\ \cite{PRD86p010001} have the
puzzling feature that their decay widths are much smaller than one would
expect on the basis of their principal $S$-wave decay modes. Namely, the
\dc\ decays to $D^\ast\pi$ (possibly also in a $D$ wave), with a
phase space of more than 270 MeV, but has a total width of only 20--30
MeV \cite{PRD86p010001,NPB866p229}. On the other hand, the \dsc\ decays to $D^\ast K$
in $S$ and $D$ wave with a phase space of about 30 MeV, resulting in
an unknown tiny width $<\!\!2.3$ MeV, limited by the experimental 
resolution \cite{PRD86p010001}. The discovery of the missing two AV charmed
mesons, namely the very narrow \dsd\ and the very broad
\dd, first observed by CLEO \cite{PRD68p032002} and Belle
\cite{PRD69p112002}, respectively, completed an even more confusing
picture. While the tiny width of the \dsd\ can be easily understood, since
this meson lies underneath its lowest Okubo-Zweig-Iizuka--allowed (OZIA) and
isospin-conserving decay threshold, the huge \dd\ width, in $D^\ast\pi$, is
in sharp contrast with that of the \dc. Moreover, the \dsd\ lies 76 MeV below
the \dsc, whereas the \dc\ and \dd\ are almost degenerate in mass, if one
takes the central value of the latter resonance.\\
Quark potential models, with standard spin-orbit splittings, fail dramatically
in reproducing this pattern of masses. For instance, in the relativized quark
model \cite{PRD32p189} the $c\bar{s}$ state that is mainly \tpo\ comes 
out at 2.57 GeV, assuming the already then well-established \dsc\ to be
mostly \spo, though with a very large mixing between \tpo\ and \spo. Reference
\cite{PRD32p189} similarly predicted a too high mass for the dominantly \tpo\
state in the $c\bar{q}$ ($q=u,d$)
sector, viz. 2.49~GeV. In the chiral quark model for heavy-light systems of
Ref.~\cite{PRD64p114004}, the result for the mainly \tpo\
$c\bar{q}$ state is also 2.49~GeV, while the discrepancy
is even worse in the $c\bar{s}$ sector, with a prediction of 2.605~GeV for the
mostly \tpo\ state, now with a small mixing in both sectors.\\
More recently and after the discovery of the \dsd\ (and \dd), chiral Lagrangians
for heavy-light systems (see e.g.\
Refs.~\cite{PRD68p054024,PLB599p55,MPLA19p2083,PRD72p034006}) have
been employed in order to understand the masses of the AV charmed mesons, in
particular the mass splittings with respect to the vector (V) mesons with charm
$D_s^\ast$ and $D^\ast$, respectively. Reference~\cite{PLB599p55} analyzed in
detail the curious experimental \cite{JPG37p075021} observation that the AV-V mass
difference is considerably larger in the charm-nonstrange sector than in the
charm-strange one, which is not predicted by typical quark potential models
\cite{PRD32p189,PRD64p114004}. The same discrepancy applies to the
scalar-pseudoscalar mass difference in either sector \cite{JPG37p075021,PLB599p55}.
In Ref.~\cite{PLB599p55}, the problem was tackled by calculating chiral loop
corrections, but the result turned out to be exactly the opposite of what is
needed to remove or alleviate the discrepancy.\\
An alternative approach to the AV charmed mesons is by trying to generate them
as dynamical resonances in chiral unitary theory \cite{EPJA33p119}. Indeed, in
the latter paper, describing AV mesons in other flavor sectors as well,
several charmed resonances were predicted, including the \dc, \dd, \dsc, and
\dsd, with reasonable results, though the $c\bar{q}$ states
came out about 100 MeV off. However, dynamical generation of mesonic resonances,
including the ones that are commonly thought to be of a normal 
quark-antiquark
type, may give
rise to interpretational difficulties, besides predicting several genuinely
exotic and so far unobserved states \cite{EPJA33p119}.
Dynamically generated AV charmed as well as bottom mesons can be found in
Ref.~\cite{PLB647p133},
too.\\
Finally, in
Ref.~\cite{PRD77p074017} a coupled-channel calculation of positive-parity
$c\bar{s}$ and $b\bar{s}$ was carried out in a chiral quark model, similar
to our approach in its philosophy, and with results for the \dsc\ and \dsd\
close to the present ones.

\section{\label{ozi}OZI-allowed channels for AV charmed mesons}
In order to account for the two possible spectroscopic channels \tpo\ and \spo\ contributing to a $J^P=1^+$ state with undefined $C$-parity, we couple both $q\bar{q}$ channels to the most important meson-meson channels. Now we describe the physical AV charmed resonances by coupling bare \tpo\ and \spo\ $c\bar{n}$, $c\bar{s}$ channels to all OZI-allowed ground-state pseudoscalar-vector (PV) and vector-vector (VV) channels. It is true that there are also relevant pseudoscalar-scalar (PS) channels (in $P$-wave), most notably $Df_0(600)$ and $D_0^\ast(2400)\pi$ \cite{JPG37p075021} in the AV $c\bar{q}$ case, and $DK_0^\ast(800)$ for $c\bar{s}$. These will contribute to the observed \cite{JPG37p075021} $D\pi\pi$ and $D\pi K$ decay modes, respectively. Although we have developed, Subec.~\ref{redsm} an algebraic procedure to deal with resonances in asymptotic states whilst preserving unitarity, the huge widths of the $D_0^\ast(2400)$, $f_0(600)$, and $K_0^\ast(800)$ resonances may lead to fine sensitivities that 
will tend to obscure the point we want to make, apart from the fact that there will also be nonresonant contributions to the $D\pi\pi$ and $D\pi K$ final states. So we restrict ourselves to the open and closed PV and VV channels in the present investigation, but we shall further discuss this issue below. The here included channels for $c\bar{q}$ and $c\bar{s}$ are given in Tables~\ref{cn} and \ref{cs}, respectively, together with the corresponding orbital angular momenta, threshold energies, and ground-state couplings squared {$(\tilde{g}^i_{(S=1(0),n=0})^2$, where $S=1(0)$ refers to the \tpo\ (\spo) quark-antiquark component.\\
\begin{table}[h]
\centering
\begin{tabular}{c|c|c|c|c}
Channel & $\left(\tilde{g}^i_{(S=1,n=0)}\right)^2$ &
$\left(\tilde{g}^i_{(S=0,n=0)}\right)^2$ & $L$ & Threshold \\ [0.5ex]
\hline&&&& \\[-9pt]
$\das\pi$    & 0.02778  &   0.01389  &    0 & 2146\\[1mm]
$\das\pi$    & 0.03472  &   0.06944  &    2 & 2146\\[1mm]
$\das\eta$   & 0.00524  &   0.00262  &    0 & 2556\\[1mm]
$\das\eta$   & 0.00655  &   0.01310  &    2 & 2556 \\[1mm]
$\dsas K$    & 0.01852  &   0.00926  &    0 & 2608\\[1mm]
$\dsas K$    & 0.02315  &   0.04630  &    2 & 2608\\[1mm]
$D\rho$      & 0.02778  &   0.01389  &    0 & 2643\\[1mm]
$D\rho$	     & 0.03472  &   0.06944  &    2 & 2643\\[1mm]
$D\omega$    & 0.00926  &   0.00463  &    0 & 2650\\[1mm]
$D\omega$    & 0.01157  &   0.02315  &    2 & 2650\\[1mm]
$\das\rho$   & 0        &   0.01389  &    0 & 2784\\[1mm]
$\das\rho$   & 0.01042  &   0.06944  &    2 & 2784\\[1mm]
$\das\omega$ & 0        &   0.00463  &    0 & 2791\\[1mm]
$\das\omega$ & 0.03472  &   0.02315  &    2 & 2791\\[1mm]
$\ds\kas$    & 0.01852  &   0.00926  &    0 & 2862\\[1mm]
$\ds\kas$    & 0.02315  &   0.04630  &    2 & 2862\\[1mm]
$\das\eta'$  & 0.00402  &   0.00201  &    0 & 2996\\[1mm]
$\das\eta'$  & 0.00502  &   0.01004  &    2 & 2996\\[1mm]
$\dsas\kas$  & 0        &   0.00926  &    0 & 3006\\[1mm]
$\dsas\kas$  & 0.06944  &   0.04630  &    2 & 3006\\[1mm]
\end{tabular}
\caption{\label{cn}Included meson-meson channels for \dc\ and \dd, with ground-state couplings squared, Sec.~\ref{RSE}, orbital angular momenta, and thresholds in MeV. For $\eta$ and $\eta'$, a pseudoscalar mixing angle of $37.3^\circ$ is used, as in Chapter.~\ref{chp3}.}
\end{table}
\begin{table}[h]
\centering
\begin{tabular}{c|c|c|c|c}
Channel & $\left({\tilde{g}}^i_{(S={1},n=0)}\right)^2$ &
 $\left({\tilde{g}}^i_{(S=0,n=0)}\right)^2$
& L & Threshold \\[1mm]
\hline &&&& \\[-9pt]
$\das K$     & 0.03704 & 0.01852 & 0 & 2504\\[1mm]
$\das K$     & 0.04630 & 0.09259 & 2 & 2504\\[1mm]
$\dsas\eta$  & 0.00803 & 0.00402 & 0 & 2660\\[1mm]
$\dsas\eta$  & 0.01004 & 0.02009 & 2 & 2660\\[1mm]
$D\kas$	     & 0.03704 & 0.01852 & 0 & 2761\\[1mm]
$D\kas$	     & 0.04630 & 0.09259 & 2 & 2761\\[1mm]
$\das\kas$   & 0       & 0.01852 & 0 & 2902\\[1mm]
$\das\kas$   & 0.01389 & 0.09259 & 2 & 2902\\[1mm]
$\ds\phi$    & 0.01852 & 0.00926 & 0 & 2988\\[1mm]
$\ds\phi$    & 0.02315 & 0.04630 & 2 & 2988\\[1mm]
$\dsas\eta'$ & 0.01048 & 0.00524 & 0 & 3069\\[1mm]
$\dsas\eta'$ & 0.01310 & 0.02621 & 2 & 3069\\[1mm]
$\dsas\phi$  & 0       & 0.00926 & 0 & 3132\\[1mm]
$\dsas\phi$  & 0.06944 & 0.04630 & 2 & 3132\\[1mm]
\end{tabular}
\caption{\label{cs}As Table~\ref{cn}, but now for  \dsc\ and \dsd.}
\end{table}
In Sec.~\ref{couplings}, we show in more detail how the ground-state coupling constants in Tables~\ref{cn} and \ref{cs} depend on the isospin and $J^{PC}$ quantum numbers of the various meson-meson channels.
The latter squared couplings must be multiplied
by $(n+1)/4^n$ for $L=0$ and by $(2n/5+1)/4^n$ for $L=2$, so as to obtain the couplings for the radial recurrences $n$ in the RSE sum of Eq.~(\ref{rsep}). A subthreshold suppression of closed channels is used just as in Eq.~\eqref{damping}.\\
The energies of the bare AV $c\bar{n}$ and $c\bar{s}$ states we determine from Eqs.~\eqref{hop} and \eqref{ctepar}. This yields masses of $2443$ MeV and $2545$ MeV for the bare AV $c\bar{n}$ and $c\bar{s}$ states, respectively, which are very close to values found in typical single-channel quark models \cite{PRD32p189,PRD64p114004}. 

\section{\label{poles}Quasi-bound states in the continuum and other poles}
Next we search for poles in the $S$ matrix. Starting with the $c\bar{n}$ case, we choose $r$ in the range 3.2--{3.5}~GeV$^{-1}$ (0.64--{0.70}~fm), which is in between the values of 2.0~GeV$^{-1}$ for an AV $c\bar{c}$ system, Chapter \ref{chp5}, and 4.0~GeV$^{-1}$ for vector $s\bar{s}$ states, Chapter \ref{chp3}. In Fig.~\ref{donesoft}, we plot several pole trajectories in the complex E plane as a function of the overall coupling $\lambda$.
\begin{figure}[h]
\centering
\resizebox{!}{400pt}{\includegraphics{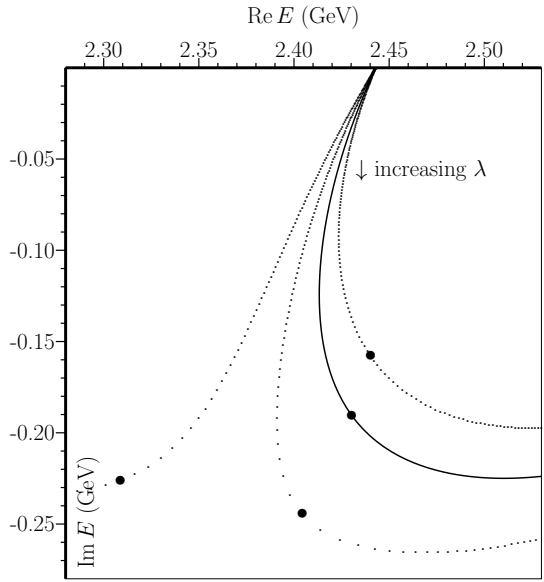}}
\mbox{}\\[-5.5cm]
\caption{\label{donesoft}\dd\ pole trajectories as a function of $\lambda$, for $r_0=$ 3.2--3.5 GeV$^{-1}$ (left to right). Solid curve and bullets correspond to $r_0=$ 3.40~GeV$^{-1}$ and $\lambda=1.30$, respectively.}
\end{figure}
We see that this pole rapidly acquires a large imaginary part, whereas the real part changes considerably less, especially in the range {$r_0=3.3$--$3.5$} GeV$^{-1}$, making it a good candidate for the broad \dd\ resonance. For $\lambda={1.30}$ and ${r_0=3.40}$ GeV$^{-1}$, the pole comes out at $({2430}-i\times{191})$~MeV, being thus fine-tuned to the experimental mass and width \cite{JPG37p075021}. However, there should be another pole in the $S$ matrix, since there are 2 quark-antiquark channels and more than 2 MM channels. From the structure of the $T$-matrix in Eqs.~(\ref{rsep}--\ref{omegap}), one can algebraically show that the number of poles for each bare state is equal to min$(N_{q\bar{q}},N_{MM})$, besides possible poles of a purely dynamical nature. Indeed, another pole originating from the bare $c\bar{q}$ state is encountered, with its trajectories depicted in Fig.~\ref{donehard}. Quite remarkably, this pole moves very little, acquiring an imaginary part that is a factor {55} smaller than in the \
dd\ case,
 for the values $\lambda={1.30}$ and ${r_0=3.40}$~GeV$^{-1}$ (see solid lines and bullets in both figures). So this resonance, with a pole position of $(2439-i\times{3.5})$ MeV, almost decouples from the
only open OZIA MM channel \cite{EPJC32p493}, viz.\ $D^\ast\pi$, representing a quasi-bound state in the continuum (QBSC) \cite{PLB647p133}. Moreover, it is a good candidate for the \dc, though its width of roughly 7~MeV is somewhat too small and its mass 16 MeV too high. These minor discrepancies may be due to the neglect of the PS channels, with broad resonances in the final states, as suggested above.
\begin{figure}[h]
\centering
\resizebox{!}{400pt}{\includegraphics{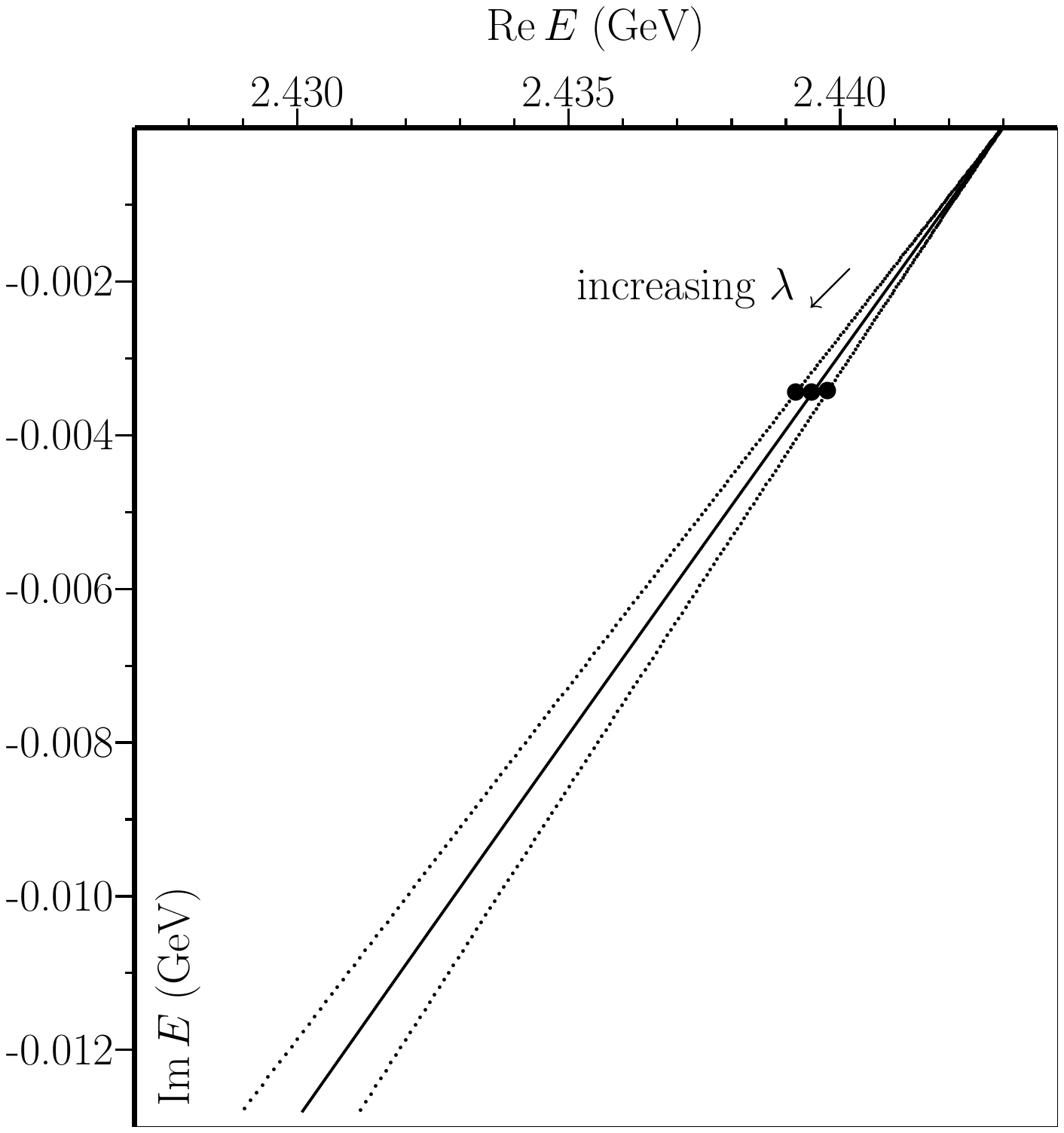}}
\mbox{} \\[-5.5cm]
\caption{\label{donehard}\dc\ pole trajectories as a function of $\lambda$, for $r_0=$ 3.3--3.5 GeV$^{-1}$ (left to right). Solid curve and bullets correspond to $r_0=$ 3.40~GeV$^{-1}$ and $\lambda=1.30$, respectively.}
\end{figure}
Nevertheless, these encouraging results might be partly due to a fortuitous choice of the parameters $\lambda$ and {$r_0$.} Therefore, we now check the $c\bar{s}$ system, {thereby scaling $r_0$ and $\lambda$} with the square root of the reduced quark
mass { (see Ref.~\cite{EPJC32p493}, Eq.~(13))}, so as to respect flavor independence of our equations, {which yields the $c\bar{s}$ values $r_0=3.12$~GeV$^{-1}$ and $\lambda=1.19$.} The ensuing $c\bar{s}$ pole trajectories are depicted in Fig.~\ref{dsonesofthard}, but
\begin{figure}[h]
\centering
\resizebox{!}{400pt}{\includegraphics{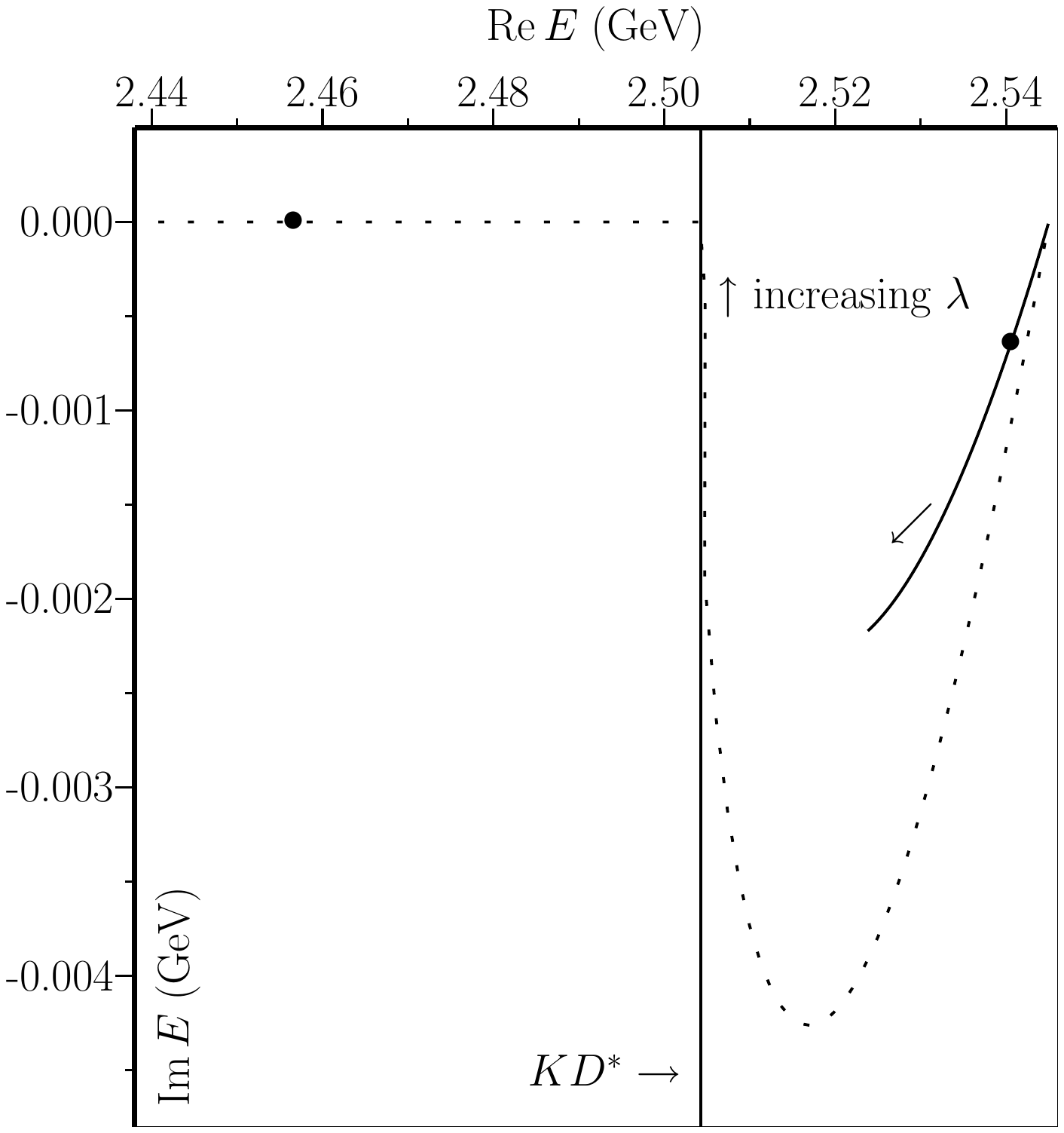}}
\mbox{} \\[-5.5cm]
\caption{\label{dsonesofthard}\dsd\ (dashed) and \dsc\ (solid) pole trajectories as a
function of $\lambda$, for $r_0=$ 3.12 GeV$^{-1}$. Bullets correspond to
$\lambda=1.19$; vertical line shows $D^\ast K$ threshold.}
\end{figure}
now for ${r_0=3.12}$~GeV$^{-1}$ only. Thus, for
$\lambda={1.19}$, the strongly
coupling state comes out at 2452 MeV, i.e., only 7.5 MeV below
the \dsd\ mass, with a vanishing width, as the pole ends up below the lowest
OZIA channel. As for the $c\bar{s}$ QBSC, it indeed shifts very little from
the bare state, settling at $(2540 - i\times0.7)$~MeV, i.e., only 5 MeV
above the \dsc\ mass, and having a width fully compatible with experiment
\cite{JPG37p075021}.\\
Besides the above ground-state AV charmed mesons, the present model of course
also predicts higher recurrences of these resonances. However, due caution is 
necessary so as to account for the most relevant open and closed decay channels
at the relevant energy scales. Now, the first radially excited HO levels of the
\tpo/\spo\ $c\bar{n}$ and $c\bar{s}$ states lie at 2823~MeV and 2925~MeV,
respectively,
which allows the corresponding resonances to be reasonably described by the
channels included in Tables~\ref{cn}, \ref{cs}. Thus, we find again 4 poles,
tabulated in Table~\ref{recurrences}, together with those of the ground-state
\begin{table}[h]
\centering
\begin{tabular}{c|c|l}
Quark Content & Radial Excitation & Pole in MeV \\[0.5ex] 
\hline&& \\[-9pt]
$c\bar{q}$  & $0$ & $2439 - i\times3.5$     \\[1mm]
$c\bar{q}$  & $0$ & $2430 - i\times191$     \\[1mm]
$c\bar{s}$  & $0$ & $2540 - i\times0.7$     \\[1mm]
$c\bar{s}$  & $0$ & $2452 - i\times0.0$     \\[1mm]
$c\bar{q}$  & $1$ & $2814 - i\times7.8$     \\[1mm]
$c\bar{q}$  & $1$ & $2754 - i\times47.2$    \\[1mm]
$c\bar{s}$  & $1$ & $2915 - i\times6.7$     \\[1mm]
$c\bar{s}$  & $1$ & $2862 - i\times25.7$    \\[1mm]
\end{tabular}
\caption{\label{recurrences}Poles of ground-state ($n\!=\!0$) and first radially-excited ($n\!=\!1$) AV charmed mesons. Parameters: $\lambda=1.30$ (1.19) and $r_0=$ 3.40 (3.12) GeV$^{-1}$, for $c\bar{q}$ ($c\bar{s}$) states.}
\end{table}
AV charmed mesons. For the radially excited states, we observe a similar
pattern as for the ground states, namely two poles that remain close
to the bare HO levels, whereas two other poles shift considerably. Note,
however, that the difference is not as dramatic as in the $n=0$ case.
This may be due to the fact that several decay channels are open now. As
for a possible observation of the here predicted $2\,P_1$ states, no
experimental candidates have been reported so far. Namely, in the nearby
$c\bar{q}$ mass region, the two listed \cite{JPG37p075021} resonances $D(2600)$
and $D(2750)$ \cite{JPG37p075021} both decay to $D^\ast\pi$ and $D\pi$, which
excludes an AV assignment.\\
Concerning the $c\bar{s}$ sector, the only listed \cite{JPG37p075021} state
around 2.8--2.9~GeV is the $D^\ast_{sJ}(2860)$
\cite{PRD80p092003},
with natural parity and so not an
AV, decaying to $D^\ast K$ and $DK$, which makes it a good candidate for the
\ttpt\ state, possibly overlapped by the \ttpz\ \cite{PRD81p118101}. Note that
the lower of our two predicted $2\,P_1$ resonances also practically coincides
with the $D^\ast_{sJ}(2860)$, both in mass and width. This may be a further
indication that the $D^\ast_{sJ}(2860)$ structure corresponds to more than one
resonance only.\\
To conclude this section, we study --- for the $c\bar{q}$ system --- the
dependence of the lowest-lying poles on the number of included quark-antiquark
and MM channels. In Table~\ref{poletests}, besides the \dc\ and \dd\
\begin{table}[h]
\centering
\begin{tabular}{c|c|cc}
$c\bar{q}$ channels & MM channels & Pole 1 (MeV) & Pole 2 (MeV)
\\[0.5ex] 
\hline&&& \\[-9pt]
\tpo+\spo  & $20$ & $2430 - i\times191$      & $2439 - i\times3$  \\[1mm]
\tpo+\spo  & $2$  & $2402 - i\times36$\,\,\, & $2441 - i\times1$  \\[1mm]
\tpo+\spo  & $1$  & $2431 - i\times39$\,\,\, &       -            \\[1mm]
\tpo       & $20$ & $2409 - i\times65$\,\,\, &       -            \\[1mm]
\spo       & $20$ & $2425 - i\times96$\,\,\, &       -            \\[1mm]
\end{tabular}
\caption{\label{poletests}Poles of AV $c\bar{q}$ mesons, for different sets of
included channels. Parameters: $\lambda=1.30$, $r_0=3.40$~GeV$^{-1}$.}
\end{table}
poles resulting from the full calculation, with the 20 MM channels from
Table~\ref{cn}, we first give the pole positions for the cases that only 2
($D^\ast\pi$, $L=0,2$) or 1 ($D^\ast\pi$, $L=0$) MM channels are included.
The last two poles then correspond to calculations with the full 20 MM
channels but only one quark-antiquark channel, viz.\ \tpo\ or \spo. Notice
that only one pole is found when the number of quark-antiquark or MM channels
is equal to 1. This confirms our above conjecture that the number of poles
for each bare HO level is given by min$(N_{q\bar{q}},N_{MM}$).

\section{\label{couplings}Three-meson couplings}
The ground-state couplings in Tables~\ref{cn} and \ref{cs} are obtained by multiplying the isospin recouplings given in Table~\ref{isospin} with the $J^{PC}$ couplings in Table~\ref{jpc}, for an OZIA process $M_A\to M_B+M_C$, Sec.~\ref{RSE}. For clarity, we represent here all couplings by rational numbers. Note that $\eta_n$ and $\eta_s$ in Table~\ref{isospin} stand for the pseudoscalar $I=0$ states $(u\bar{u}+d\bar{d})/\sqrt{2}$ and $s\bar{s}$, respectively. Then, we get the couplings to the physical $\eta$ and $\eta^\prime$ mesons by applying a mixing angle --- in the flavor basis, see Eq.~\eqref{pmixa} --- of $41.2^\circ$, as in Ref.~\cite{PRL97p202001}. For the $\omega$ and $\phi$ we assume ideal mixing.
\begin{table}[h]
\centering
\begin{tabular}{c|c|c|c}
$\;\;\;M_A\;\;\;$ & $\;\;\;M_B\;\;\;$ & $\;\;\;M_C\;\;\;$ &
$\;\;\;g_I^2\;\;\;$ \\[0.5ex]
\hline&&&\\[-9pt]
$D_{s1}$ & $D_s,D_s^\ast$ & $\eta_s,\phi$ & $1/3$ \\[1mm]
$D_{s1}$ & $D,D^\ast$     & $K,K^\ast$    & $2/3$ \\[2mm]
$D_{1}$  & $D_s,D_s^\ast$ & $K,K^\ast$      & $1/3$ \\[1mm]
$D_{1}$  & $D,D^\ast$     & $\pi,\rho$      & $1/2$ \\[1mm]
$D_{1}$  & $D,D^\ast$     & $\eta_n,\omega$ & $1/6$ \\[1mm]
\end{tabular}
\caption{\label{isospin}Squared isospin recouplings for the 3-meson process $M_A\to M_B+M_C$, with $M_A=c\bar{s}$ or $c\bar{q}$.}
\end{table}
\begin{table}[h]
\centering
\begin{tabular}{c|c|c|c|c|c}
$J^{PC}(M_A)$ & $J^{PC}(M_B)$ & $J^{PC}(M_C)$ & $L_{M_B M_C}$ & $S_{M_B M_C}$ &
$g^2_{(n=0)}$
\\[0.5ex]
\hline &&&&& \\[-9pt]
$1^{++}$ & $0^{-+}$ & $1^{--}$ & $0$ & $1$ & $1/18$\\[1mm]
$1^{++}$ & $0^{-+}$ & $1^{--}$ & $2$ & $1$ & $5/72$\\[1mm]
$1^{++}$ & $1^{--}$ & $1^{--}$ & $0$ & $1$ & $0$\\[1mm]
$1^{++}$ & $1^{--}$ & $1^{--}$ & $2$ & $2$ & $5/24$\\[1mm]
$1^{+-}$ & $0^{-+}$ & $1^{--}$ & $0$ & $1$ & $1/36$\\[1mm]
$1^{+-}$ & $0^{-+}$ & $1^{--}$ & $2$ & $1$ & $5/36$\\[1mm]
$1^{+-}$ & $1^{--}$ & $1^{--}$ & $0$ & $1$ & $1/36$\\[1mm]
$1^{+-}$ & $1^{--}$ & $1^{--}$ & $2$ & $1$ & $5/36$\\[1mm]
\end{tabular}
\caption{\label{jpc}Squared ground-state coupling constants for the 3-meson process $M_A\to M_B+M_C$, with $J^{PC}(M_A)=1^{+\pm}$, and $M_A$, $M_B$ belonging to the lowest pseudoscalar or vector nonet.}
\end{table}

\section{Summary and conclusions}
\label{summary}
In the foregoing, we have managed to rather accurately reproduce the masses and widths of the \dc, \dd, \dsc, and \dsd\ with only 2 free parameters, one of which is already constrained by previous model calculations, as well as by reasonable estimates for the size of these mesons. Crucial is the approximate decoupling from the continuum of one combination of \tpo\ and \spo\ components, which amounts to a mixing angle close to $35^\circ$. Namely, if we express a QBSC as $|\mbox{QBSC}\rangle=-\sin\theta\, $\tpo$\rangle+\cos\theta\,|$\spo$\rangle$, it decouples from the $L\!=\!0$ $D^\ast\pi$ channel (for $c\bar{q}$) or $D^\ast K$ channel (for $c\bar{s}$), if $\theta=\arccos\sqrt{2/3}\approx35.26^\circ$ (see Tables~\ref{cn}, \ref{cs}). Inclusion of the other, practically all closed, channels apparently changes the picture only slightly in our formalism. This result is in full agreement with the findings in Ref.~\cite{PRD77p074017}. However, in the present approach this particular mixing \cite{1105.6025} comes 
out as a completely dynamical result, and is not chosen by us beforehand. Moreover, the bare-mass degeneracy of \tpo\ and \spo\ states is adequately lifted via the decay couplings in Tables~\ref{cn} and \ref{cs}, dispensing with the usual $\vec{S}\cdot\vec{L}$
splitting. Also note that the occurrence of (approximate) bound states in the continuum for AV charmed mesons had already been conjectured by two of us \cite{EPJC32p493}, based on more general arguments.\\
The puzzling discrepancy between the AV-V mass splittings in the $c\bar{q}$ and $c\bar{s}$ sectors is resolved in our calculation by dynamical, nonperturbative coupled-channel effects. A similar phenomenon we have observed before \cite{PRL91p012003} for the
$D_0^\ast$(2300--2400) \cite{JPG37p075021} resonance, and may be related to an effective Adler-type zero \cite{PR397p257,AIPCP756p360} in the $D^\ast\pi$ and $D\pi$ channels in the AV and scalar $c\bar{n}$ cases, respectively, owing to the small pion mass.\\
Summarizing, we have reproduced the whole pattern of masses and widths of the AV charmed mesons dynamically, by coupling the most important open and closed two-meson channels to bare $c\bar{q}$ and $c\bar{s}$ states containing both \tpo\ and \spo\ components. The dynamics of the coupled-channel equations straightforwardly leads to one pair of strongly shifted states and another pair of QBSCs. Ironically, the state that shifts most in mass, namely the \dsd, ends up as the narrowest resonance. This 
emphasizes the necessity \cite{PTP125p581} to deal with unquenched meson spectroscopy in a fully nonperturbative framework.\\
One might argue that these conclusions will depend on the specific model employed. Admittedly, our numerical results could change somewhat if slightly different bare masses for the AV charmed mesons were chosen,  non-$S$-wave decay channels were included as well, or a different scheme was used to calculate the decreasing couplings of the higher recurrences. Nevertheless, we are convinced the bulk of our results will not change, most notably the appearance of QBSCs and the large shifts of their partner states, as the almost inevitable consequence of exact nonperturbative coupled-channel dynamics.

\chapter{The charmonium $X(3872)$\label{chp5}}
\thispagestyle{empty}
\vspace*{-1.2cm}{\normalsize S. Coito, G. Rupp, and E. van Beveren, {\it EPJC} {\bf71}, 1762 (2011).}\\[1cm]
The $X(3872)$ charmonium-like state was discovered in 2003 by the Belle Collaboration \cite{PRL91p262001}, as a $\pi^+\pi^- J\!/\!\psi$ enhancement in the decay $B^{\pm} \rightarrow K^{\pm} \pi^+\pi^- J\!/\!\psi$. The same structure was then observed, again in $\pi^+\pi^- J\!/\!\psi$, by CDF II \cite{PRL93p072001}, D0 \cite{PRL93p162002}, and BABAR \cite{PRD71p071103}. Moreover, CDF \cite{PRL96p102002} showed that the $\pi^+\pi^-$ mass distribution favors decays via a $\rho^0$ resonance, implying positive $C$-parity for the $X(3872)$. The $X(3872)$ has also been observed in the $\bar{D}^0D^0\pi^0$ and $\bar{D}^{\ast0}D^{\ast0}$ channels, by Belle \cite{PRL97p162002} and BABAR \cite{PRD77p011102}, respectively. CDF \cite{PRL103p152001} measured the $X(3872)$ mass with even higher precision, viz.\ $3871.61\pm0.16\pm0.19$~MeV, with a width fixed at $1.34\pm0.64$~MeV, while BABAR \cite{PRD82p011101} presented evidence for the long-awaited $\omega J\!/\!\psi$ decay mode (also see Ref.\ \cite{
HEPEX0505037}), and a surprising preference for the $2^{-+}$ assignment. At last, LHCb unequivocally determined the quantum numbers $J^{PC}=1^{++}$, based on angular correlations in $B^\pm\to X(3872)K^+$ decays \cite{PRL110p222001}. The $X(3872)$ resonance is listed in the 2012 PDG tables \cite{PRD86p010001}, with a mass of $3871.68\pm0.17$ MeV, a width $<\!1.2$ MeV.\\
On the theoretical side, the first to foresee a narrow $1^{++}$ state
close to the $DD^\ast$ threshold was T\"{o}rnqvist \cite{ZPC61p525}, arguing on the basis of strongly attractive one-pion exchange for $S$-wave meson-meson systems, which he called deusons. For further molecular descriptions and studies, see Ref.~\cite{molecular}, as well as the reviews by Swanson \cite{PR429p243} and Klempt \& Zaitsev \cite{PR454p1}. In Ref.~\cite{exotics}, a few exotic model descriptions can be found, such as a hybrid or a tetraquark; also see the reviews \cite{PR429p243,PR454p1}. For further reading, we  recommend the very instructive analyses by Bugg \cite{PRD71p016006} and Kalashnikova \& Nefediev \cite{PRD80p074004}. Much more in the spirit of our own calculation is the coupled-channel analysis
by Danilkin and Simonov \cite{PRD81p074027}, which studies resonances and level shifts of conventional charmonium states due to the most important open and closed decay channels. We shall come back to their results below.\\
According to the PDG 2010 \cite{JPG37p075021}, the $X(3872)$ could be either a $1^{++}$ or a $2^{-+}$ state, which implied $2\,{}^{3\!}P_1$ or $1\,{}^{1\!}D_2$, as other radial excitations would be much too far off (see e.g.\ Ref.~\cite{PRD32p189}). In the present chapter, we study the $1^{++}$ scenario, despite the conclusion by BABAR \cite{PRD82p011101}, the last result by the time this study was performed, from the $\omega J\!/\!\psi$ mode, that $2^{-+}$ was more likely. Indeed, the latter assignment appeared to be at odds with radiative-transition data \cite{1007.4541}. For a further discussion of electromagnetic decays, see e.g.\ the molecular description of Ref.~\cite{PRD79p094013}. But more importantly, in all charmonium models we know of, the $1\,{}^{1\!}D_2$ $c\bar{c}$ state lies well below 3.872~GeV, i.e., in the range 3.79--3.84~GeV (see e.g.\ Ref.~\cite{1011.6124}). Our own \em bare \em \/$1\,{}^{1\!}D_2$ state comes out at 3.79~GeV, just as the corresponding single-channel state in Ref.~\cite{
PRD81p074027}. Now, the crucial point is that loops from closed meson-meson channels are \em always \em \/attractive \cite{ZPC19p275}. Hence, since $DD^\ast$\footnote{Henceforth, we omit the bar in $D\overline{D}^*$, for notational simplicity.} at 3.872--3.880~GeV is the lowest OZIA channel that couples to a $1\,{}^{1\!}D_2$ $c\bar{c}$ state, the coupled-channel mass shift will inexorably be further downwards (also see Ref.\ \cite{1011.6124}).\\
The present sections aims to show that the mass and width of the $X(3872)$, as well as the corresponding observed amplitudes in the $D^0D^{\ast0}$, $\rho^0J\!/\!\psi$, and $\omega J\!/\!\psi$ channels, are compatible with a description in terms of a regular \ttpo\ charmonium state, though mass-shifted and unitarized via open and closed decay channels.

\section{The RSE applied to charmonium $1^{++}$}
Sticking to the $1^{++}$ scenario, we employ again the RSE in order to couple one $c\bar{c}$ channel,
with $l_c=1$, to several OZIA pseudo\-scalar-vector (PV) and vector-vector
(VV) channels, just as in our preliminary study \cite{1005.2486} of the
$X(3872)$. However, we now also couple the 
OZI-suppressed (OZIS) $\rho^0\jpsi$ and $\omega\jpsi$ channels,
to account for the bulk of the observed $\pi^+\pi^-\jpsi$
and $\pi^+\pi^-\pi^0\jpsi$ decays, respectively.
Although
the former channel is
isospin breaking 
as well,
the extreme closeness of its
central threshold at 3872.4~MeV to the $X(3872)$ structure makes it absolutely
nonnegligible, despite a very small expected coupling. A complication, though,
is the large $\rho$ width, which does not allow the $\rho^0\jpsi$ channel to
be described through a sharp threshold. 
Effects in the $X(3872)$ from nonzero $\rho$ and $\omega$ widths 
were already estimated in Ref.~\cite{PRD76p034007}.
We tackle this problem by taking a complex mass for the $\rho$, from its pole
position \cite{EPJA44p425}, and then apply the novel, empirical yet
rigorous, unitarization procedure to the $S$-matrix, derived in
Subsec.~\ref{redsm}. The analyticity and causality implications of
complex masses in asymptotic states were already studied a long time ago
\cite{NPB12p281}.\\
For consistency, we apply the same procedure to the $\omega$ meson, despite
the fact that its width is a factor 17.5 smaller than that of the
$\rho$. Nevertheless, the $\omega$ width of about 8.5~MeV is very close
to the energy difference between the $\omega\jpsi$ and $D^0D^{\ast0}$
thresholds, and therefore not negligible. Finally, we shall neglect the
unknown small ($<2.1$ MeV \cite{JPG37p075021}) $D^{\ast0}$ width
\cite{PRD80p074004}, because of the relatively large error bars on the
$D^0D^{\ast0}$ data, though this width may have some influence on the 
precise $X(3872)$ pole position.
Nevertheless, reasonable estimates of the $D^{\ast0}$ width yield values
clearly smaller than 100~keV \cite{PRD76p094028}, so that its effect should
be largely negligible as compared to that of the $\rho$ and $\omega$ widths.\\
Let us now proceed with our RSE calculation of a bare $2\,{}^{3\!}P_1$
(with $n\!=\!1$, $J\!=\!1$, $L\!=\!1$, $S\!=\!1$) $c\bar{c}$ state,  coupled
to a number of MM channels. The resulting closed-form $T$-matrix is given, onec mores, in Sec.~\ref{RSE}. In Table~\ref{MMpp} we list the
\begin{table}[h]
\centering
\begin{tabular}{l|c|c|c}
Channel & $\left(g^i_{(l_c=1,n=0)}\right)^2$ & $\;L\;$ &
Threshold (MeV) \\[2pt] 
\hline &&& \\[-9pt]
$\rho^0 J/\psi$	& variable  & 0 & $3872.406 - i\,74.7$\\
$\omega J/\psi$	& variable  & 0 & $3879.566 - i\,4.25$\\
\hline &&& \\[-9pt]
$D^0D^{\ast0}$	& 1/54 & 0 & 3871.81\\ 
$D^0D^{\ast0}$ 	& 5/216 & 2 & 3871.81\\ 
$D^{\pm}D^{*\mp}$ & 1/54 & 0 & 3879.84\\
$D^{\pm}D^{*\mp}$ & 5/216 & 2 & 3879.84\\
$D^{*}D^{*}$ & 5/36 & 2 & 4017.24\\ 
$D_s^{\pm}D_s^{*\mp}$& 1/54 & 0 & 4080.77\\ 
$D_s^{\pm}D_s^{*\mp}$& 5/216 & 2 & 4080.77\\
\end{tabular}
\caption{\label{MMpp}Included meson-meson channels, with thresholds and ground-state
couplings. For simplicity, we omit the bars over the anti-charm mesons;
also note that $D^\ast D^\ast$ stands for the corresponding mass-averaged
charged and uncharged channels.}
\end{table}
considered PV and VV channels, including
$\rho^0\jpsi$ and $\omega\jpsi$.
Besides
the latter two OZIS channels and the also observed OZIA $D^0D^{\ast0}$
channel, we furthermore account for the OZIA PV and VV channels
$D^\pm D^{\ast\mp}$, $D^\ast D^\ast$, and $D_s D_s^\ast$, whose influence
on the $X(3872)$ pole position is not negligible, in spite of
being closed channels. The $D_s^\ast D_s^\ast$ channel, with threshold
about 350~MeV above the $X(3872)$ mass, we do not include.\\
The relative couplings of the OZIA channels have been computed as discussed in Sec.~\ref{RSE}.\\
Couplings calculated in the latter scheme for ground-state mesons generally
coincide with the usual recouplings of spin, isospin, and orbital angular
momentum. Moreover, for excited stat\-es the formalism yields clear
predictions as well, contrary to other appoaches.
In Table~\ref{MMpp}, the squares of the ground-state
($n\!=\!0$) couplings are given, which have to be multiplied by $(n+1)/4^n$
for the $S$-wave PV channels, and by $(2n/5+1)/4^n$ for the others, so as to
obtain the couplings in the RSE sum of Eq.~(\ref{rsep}). Also note that the
two (closed) $D^\ast D^\ast$ channels have been lumped together, with their
average threshold value and sum of squared couplings. As for the
couplings
of the
$\rho^0\jpsi$ and $\omega\jpsi$ channels,
the formalism of Ref.~\cite{ZPC21p291}, based on
OZIA decay via ${}^{3\!}P_0$ quark-pair creation, cannot make any prediction.
However, we know from experiment that the
couplings of OZIS channels are considerably smaller than those of OZIA
channels. Moreover, isospin-breaking channels are even further suppressed.
Thus, in the following we shall employ the values
$g_{\rho^0\jpsi}=0.07\times g_{D^0D^{\ast0}}$ and
$g_{\omega\jpsi}=0.21\times g_{D^0D^{\ast0}}$, which correspond to effective
relative strengths of 0.49\% and 4.41\%, respectively, which seem reasonable
to us. These values may also be compared to the corresponding relative
probabilites of about 0.65\% ($\approx0.006/0.92$) and 4.5\%
($\approx0.041/0.92$), respectively, employed in Ref.~\cite{PRD79p094013}.
Furthermore, we shall also test coupling values twice as large,
namely $g_{\rho^0\jpsi}=0.14\times g_{D^0D^{\ast0}}$ and
$g_{\omega\jpsi}=0.42\times g_{D^0D^{\ast0}}$.
Note that our coupling for the isospin-breaking channel $\rho^0\jpsi$ is
also in rough agreement with estimates from the rate of the observed
\cite{JPG37p075021} isospin-violating $\omega\to\pi^+\pi^-$ decay, which
amounts to about 1.5\% of the total width.
Another difference
between OZIA and OZIS channels is the average distance $r_i$ (see
Eqs.~(\ref{tmatp},\ref{omegap})) at which a light $q\bar{q}$ pair is created
before decay, which in the OZIA case we believe to take place in the core
region and in the OZIS case more in the periphery. Thus, we
employ
a larger
value for $r_1\equiv r_{\rho^0\jpsi}$
$=r_{\omega\jpsi}$
than for $r_0$, the single radius used
for all OZIA channels. Concretely, we take $r_0=2$~GeV$^{-1}\simeq0.4$~fm
and
$r_1=3$~GeV$^{-1}\simeq0.6$~fm,
while we also test the case $r_1=r_0$.\\
For the bare $c\bar{c}$ energy levels $E_n^{(l_c)}$ in the RSE sum of
Eq.~(\ref{rsep}), we take the equidistant harmonic oscillator \eqref{hop} with the parameters in Eq.~\eqref{ctepar}, as previously. The only parameter we adjust freely is
the overall coupling constant $\lambda$ in Eqs.~(\ref{tmatp},\ref{omegap}), which
is tuned to move the bare \ttpo\ state from 3979~MeV down to 
the $D^0D^{\ast0}$ threshold, requiring a $\lambda$ value of the order
of 3, i.e., not far from the values used in e.g.\ Chpater \ref{chp3} and Ref.~\cite{PLB641p265}. At the
same time, the bare \otpo\ state shifts from 3599 MeV down to 
about 3.55 GeV, though depending quite sensitively on the precise form of the
used subthreshold suppression of closed channels Chpater \ref{chp3}. Anyhow,
for the purpose of the present study, an accurate reproduction of the
$\chi_{c1}(1P)$ mass of 3511 MeV is not very relevant.

\section{$X(3872)$ poles and amplitudes vs.\ data \label{x3940}}
In Table~\ref{polespp}, we give some pole positions in the vicinity of the
\begin{table}[h]
\centering
\begin{tabular}{c|c|c|c|c}
Label & $\lambda$ & $\tilde{g}_{\rho^0\jpsi}$
& $\tilde{g}_{\omega\jpsi}$ & Pole (MeV)\\[1pt] 
\hline &&&& \\[-9pt]
\emph{1} & 3.028 & 0.07 & 0.21 & $3872.30 - i\,0.71$\\[1mm]
\emph{2} & 3.066 & 0.07 & 0.21 & $3871.83 - i\,0.40$\\[1mm]
\emph{3} & 3.083 & 0.07 & 0.21 & $3871.56 - i\,0.11$\\[1mm] 
\emph{a} & 2.981 & 0.14 & 0.42 & $3872.30 - i\,0.75$\\[1mm]
\emph{b} & 3.017 & 0.14 & 0.42 & $3871.82 - i\,0.48$\\[1mm]
\emph{c} & 3.033 & 0.14 & 0.42 & $3871.57 - i\,0.28$\\[1mm] 
\end{tabular}
\caption{\label{polespp}Pole positions of the dots and stars in Fig.~\ref{trajectories}.
In all cases, $r_1=3.0$ GeV$^{-1}$. Note that the OZIS couplings
$\tilde{g}_{\rho^0\jpsi}$ and $\tilde{g}_{\omega\jpsi}$ are given relative
to the coupling of the OZIA $D^0D^{\ast0}$ channel.}
\end{table}
$D^0D^{\ast0}$ and $\rho^0\jpsi$ thresholds, with the chosen  values
of $\lambda$ and $r_1$. In Fig.~\ref{trajectories}, third-sheet pole
trajectories
\begin{figure}[htb]
\begin{center}
\resizebox{!}{400pt}{\includegraphics{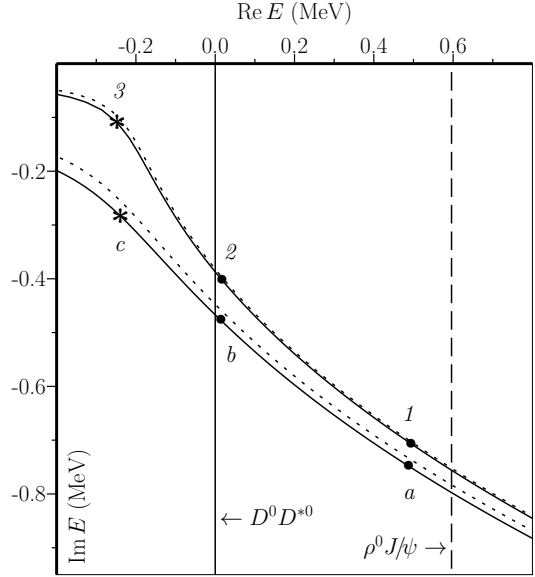}}
\mbox{} \\[-175pt]
\caption{Pole trajectories for $r_1$=3.0 GeV$^{-1}$ (solid curves) and $r_1$=2.0 GeV$^{-1}$ (dotted curves); $g_{\rho^0\jpsi}/g_{D^0D^{\ast0}}=0.07$ and $g_{\omega\jpsi}/g_{D^0D^{\ast0}}=0.21$ (upper two curves), $g_{\rho^0\jpsi}/g_{D^0D^{\ast0}}=0.14$ and $g_{\omega\jpsi}/g_{D^0D^{\ast0}}=0.42$ (lower two curves). Note that the CM energy $E$ is relative to the $D^0D^{\ast0}$ threshold in all figures. Also see Table~\ref{polespp}.}
\label{trajectories}
\end{center}
\end{figure}
in the complex energy plane (relative to the $D^0D^{\ast0}$ threshold)
are plotted, as a function of $\lambda$,
with the pole positions of Table~\ref{polespp} marked by
bullets and stars.
The solid curves represent the case $r_1=3.0$~GeV$^{-1}$, while the dotted
ones stand for $r_1=2.0$~GeV$^{-1}$, showing little sensitivity to the
precise decay radius.
Figure~\ref{trajectories} shows that the $X(3872)$ resonance pole may come out
below the $D^0D^{\ast0}$ threshold
with a nonvanishing width, which is moreover of the right order of magnitude,
viz.\ $<\!1$~MeV.
The recent CDF \cite{PRL103p152001} mass determination of the $X(3872)$ might
suggest that the
pole positions '3' or `c' (see Table~\ref{polespp} and Fig.~\ref{trajectories})
are favored.
However, one should realize that the differences amount to mere fractions of
an MeV,
while experimental uncertainties are at least of the same order.

Now we compare the corresponding $D^0D^{\ast0}$ amplitudes to Belle
\cite{0810.0358} data , for the six cases labeled `1, 2, 3', and 'a, b, c' in
Table~\ref{polespp} and Fig.~\ref{trajectories}. The results are depicted
in Figs.~\ref{DD123} and \ref{DD123}, respectively. Note that
\begin{figure}[h]
\centering
\begin{tabular}{cc}
\resizebox{!}{300pt}{\includegraphics{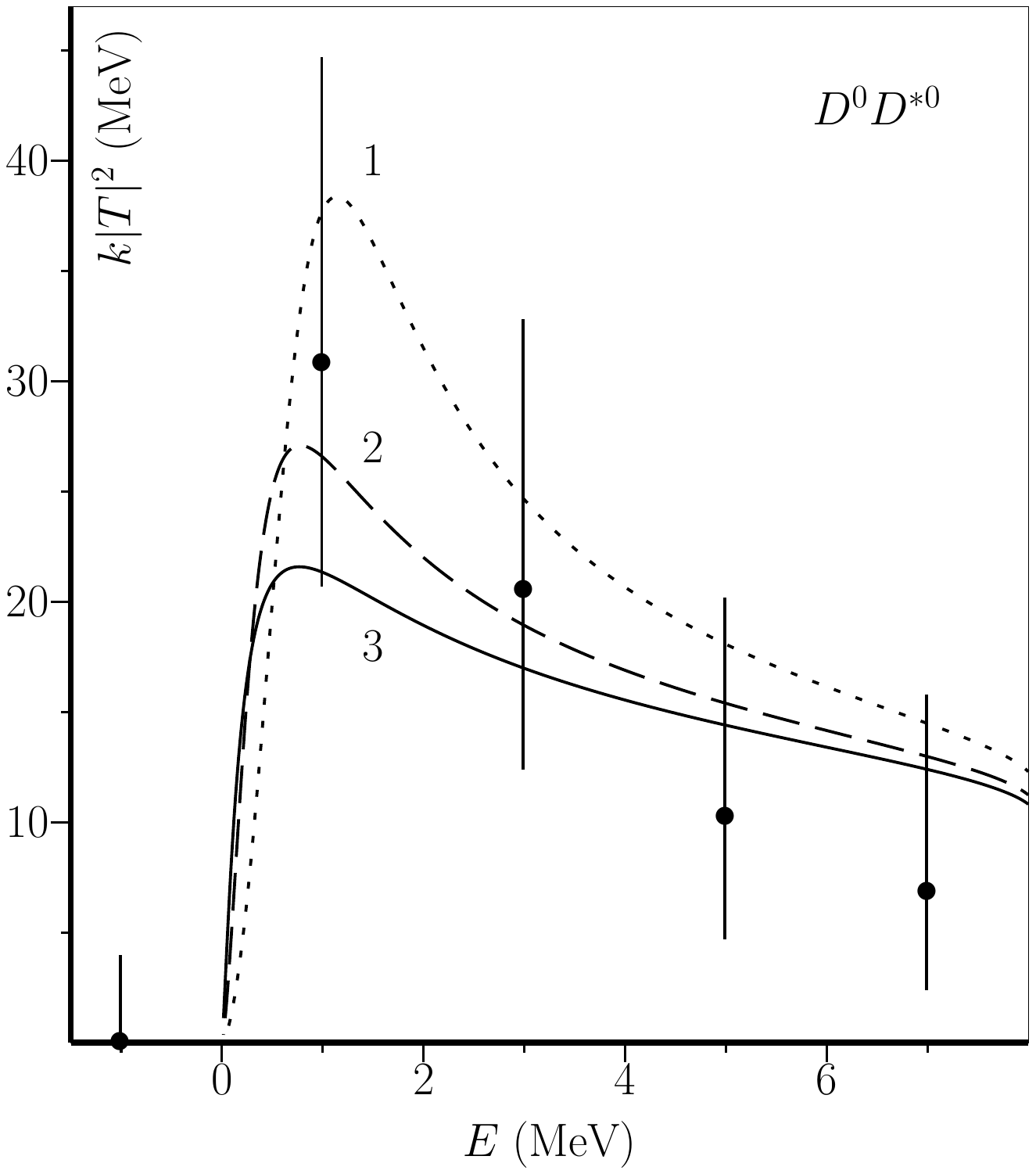}}
&
\hspace*{-34pt}
\resizebox{!}{300pt}{\includegraphics{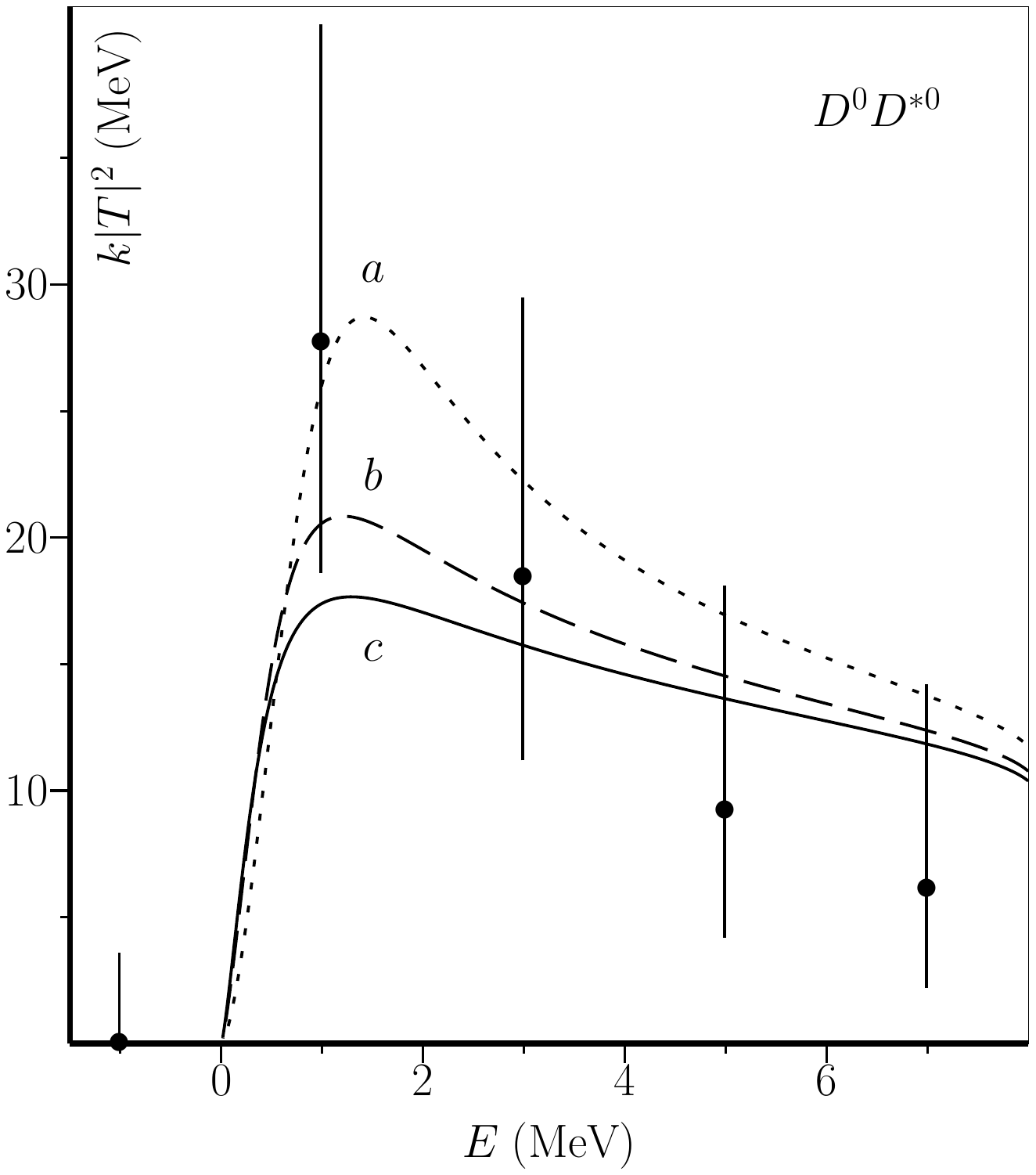}}
\end{tabular}
\mbox{ } \\[-130pt]
\caption{\label{DD123}$D^0D^{\ast0}$ elastic amplitude for poles 1, 2, 3 in Table~\ref{polespp} and Fig.~\ref{trajectories}; arbitrarily normalized data are from Ref.~\cite{0810.0358}. Elastic $T$-matrix elements follow from Eqs.~(\ref{tmatp}--\ref{rsep}); $k$ is on-shell relative momentum. (b)for poles a, b, c.}
\end{figure}
we allow for an arbitrary normalization of the data, which is inevitable as
we are dealing with production data, 
which cannot be directly
compared with our scattering amplitudes,
also because of the finite experimental mass bins.
From these figures we see that the best agreement with data is
obtained in case `2', though 5 out of the 6 curves pass through all error
bars. Nevertheless, in view of the large errors, one should be very cautious
in drawing definite conclusions on the precise pole position as well as the
preferred OZIS couplings $g_{\rho^0\jpsi}$ and $g_{\omega\jpsi}$.\\
Next we show, in Fig.~\ref{tsqrhoomeg123}, the elastic amplitudes in the
$\rho^0\jpsi$ and $\omega\jpsi$ channels, corresponding to the pole positions
1, 2, 3, i.e., for the smaller values of the OZIS couplings. We see that
both amplitudes are very
sensitive to the precise pole position, which
\begin{figure}[h]
\centering
\begin{tabular}{cc}
\resizebox{!}{300pt}{\includegraphics{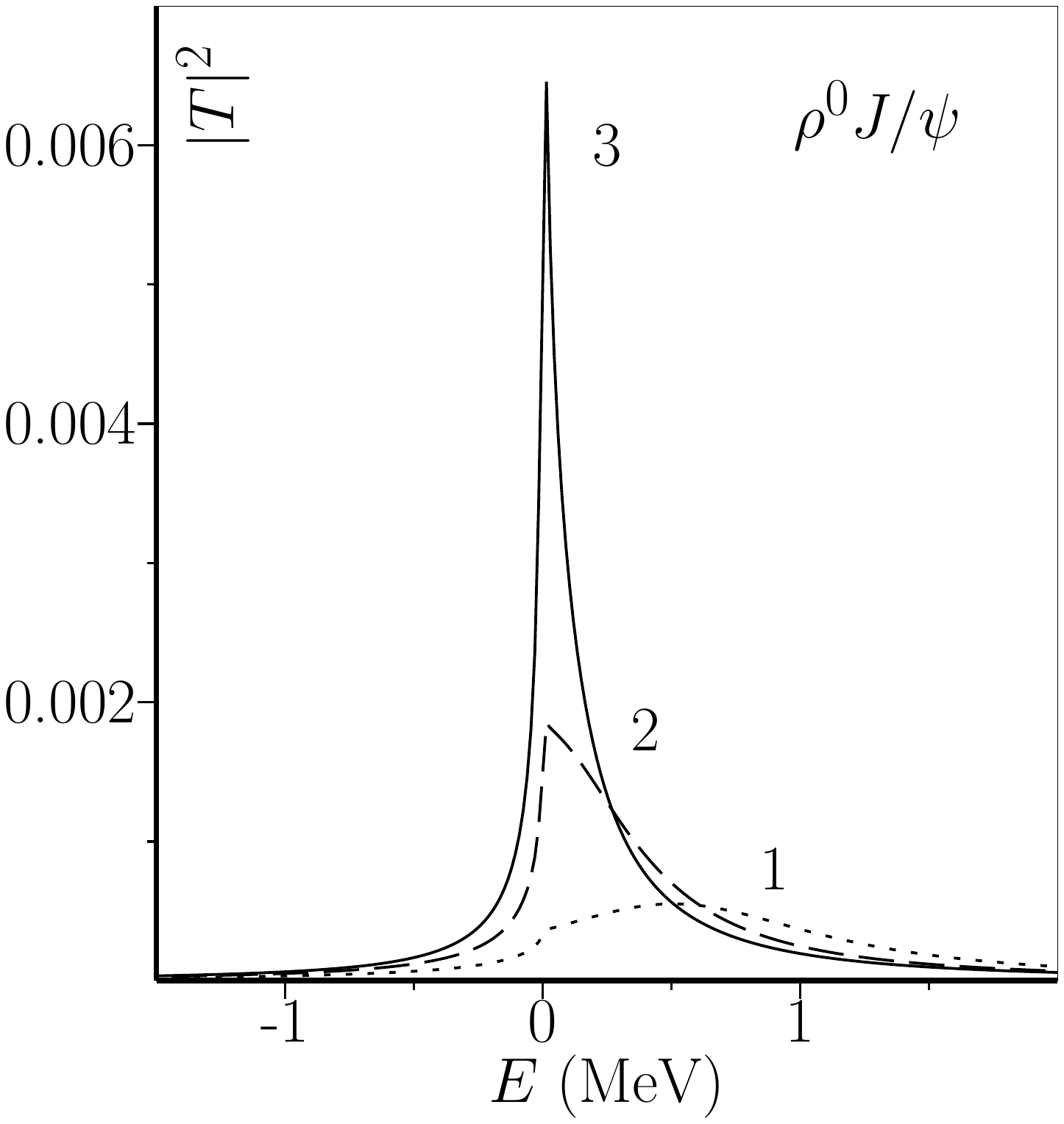}}
&
\hspace*{-34pt}
\resizebox{!}{300pt}{\includegraphics{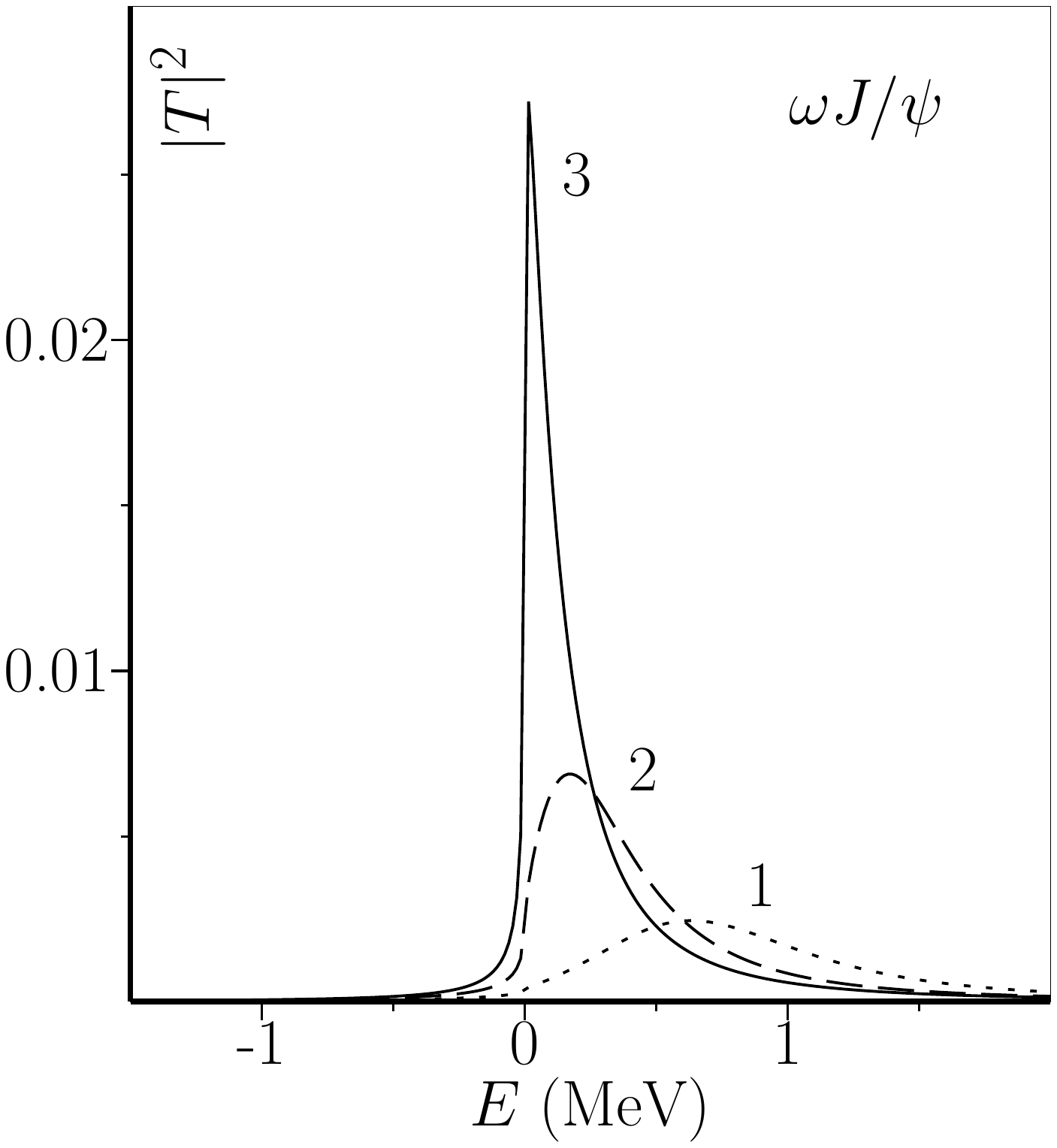}}
\end{tabular}
\mbox{ } \\[-130pt]
\caption{\label{tsqrhoomeg123}$\rho^0\jpsi$ (left) and $\omega\jpsi$ (right) elastic amplitudes for poles 1, 2, 3. Also see Fig.~\ref{trajectories} and Table~\ref{polespp}.}
\end{figure}
is logical, as the
OZIS channels couple
much more weakly to $c\bar{c}$ than $D^0D^{\ast0}$, so that the latter channel
will strongly deplete the former
ones,
as soon as it acquires some phase space. This is in line with our
analysis in e.g.\ Ref.~\cite{1005.1010}. Also note the strongly
cusp-like structure of the amplitude in the cases
2 and 3 for $\rho^0\jpsi$, and 3 for $\omega\jpsi$,
which is a manifestation of the depletion due to the opening of the
$D^0D^{\ast0}$ channel. Such a cusp makes the experimental determination
of the $X(3872)$ width very difficult.\\
In Fig.~\ref{tsqomegrho23} we take a closer look at the $\omega\jpsi$ and
$\rho^0\jpsi$ amplitudes, in particular how they compare to one another. Now,
the effective strength of the $\omega\jpsi$ elastic $T$-matrix element is
9 times that of $\rho^0\jpsi$, as its coupling has been chosen 3 times as
large (see Table~\ref{polespp} and Eqs.~(\ref{tmatp}--\ref{rsep})). For the
corresponding square amplitudes plotted in Fig.~\ref{tsqomegrho23}, this
amounts to a factor as large as 81. However, the central $\omega\jpsi$
threshold lies more than 7~MeV above that of $\rho^0\jpsi$, while the full
$\omega$ width is only 8.49~MeV. On the other hand, the central $\rho^0\jpsi$
threshold lies much closer to $D^0D^{\ast0}$, while the large physical $\rho$
width strongly boosts the associated amplitude, as demonstrated below. 
Qualitative arguments in agreement with our calculation were already presented
in Ref.~\cite{PRD80p014003}.
These effects make
the maximum $\omega\jpsi$ square amplitude to be only a factor 3.5--4 larger
than that of $\rho^0\jpsi$, both in case 2 and 3, as can be read off from
Fig.~\ref{tsqomegrho23}. Moreover, at the precise energy of the respective
pole position, the two amplitudes are almost equal in size. Therefore, the
observed branching ratio
$\mathcal{B}(X(3872)\to\omega\jpsi)/
\mathcal{B}(X(3872)\to\pi^+\pi^-\jpsi)\sim1$ \cite{PRD82p011101,HEPEX0505037}
is compatible with the present model calculation.
\begin{figure}
\centering
\begin{tabular}{cc}
\resizebox{!}{300pt}{\includegraphics{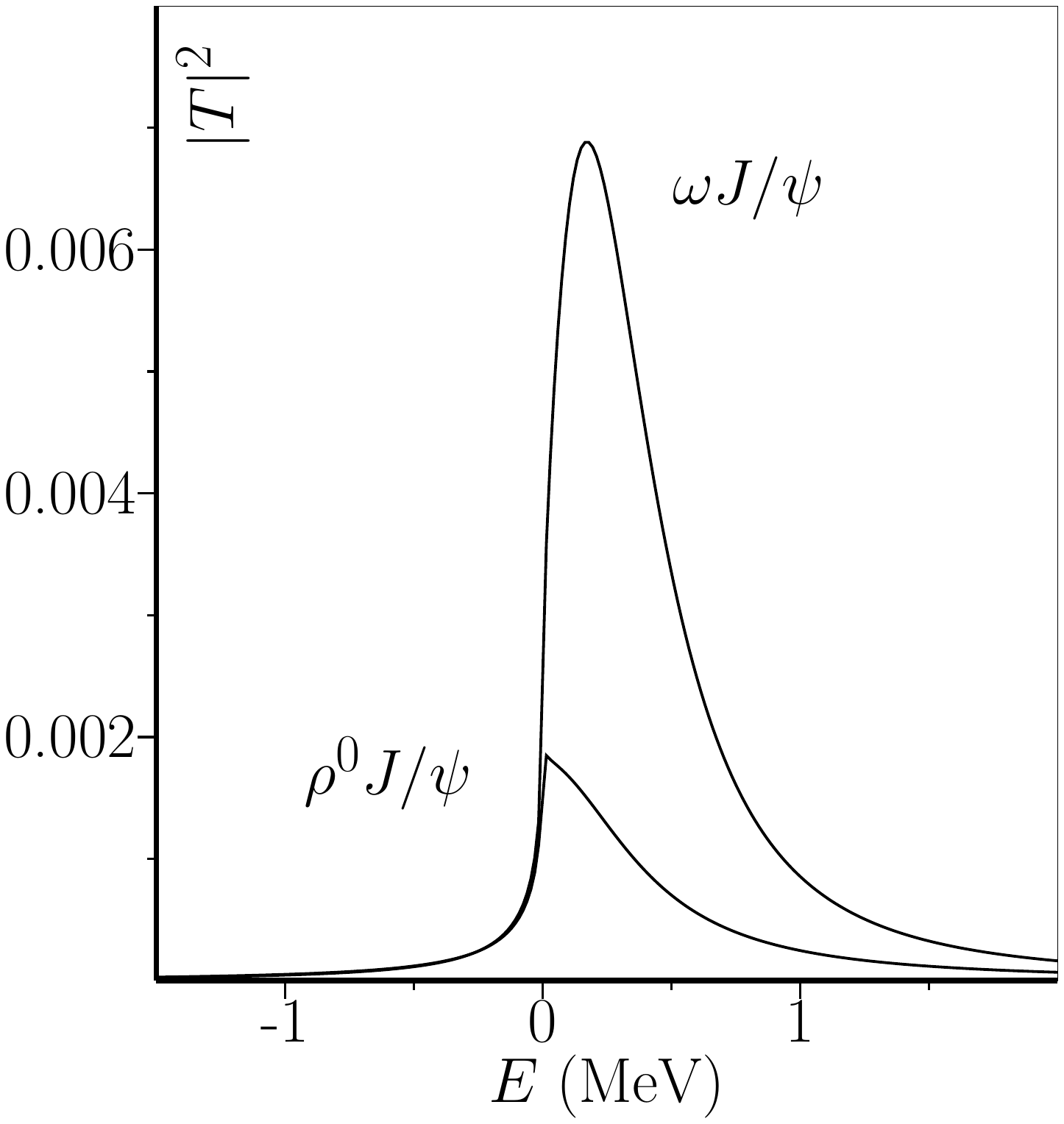}}
&
\hspace*{-34pt}
\resizebox{!}{300pt}{\includegraphics{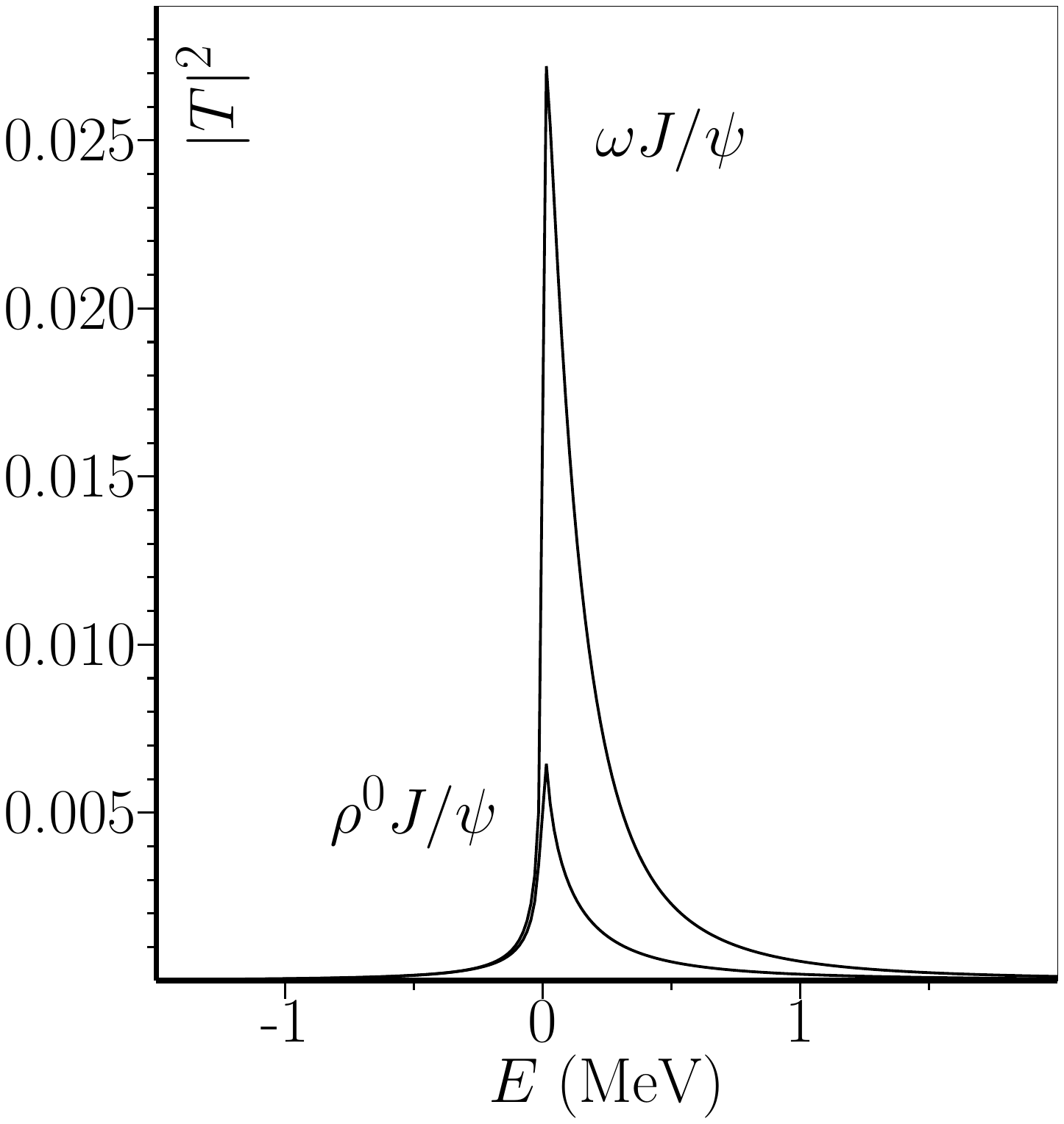}}
\end{tabular}
\mbox{ } \\[-110pt]
\caption{\label{tsqomegrho23}
$\rho^0\jpsi$ and $\omega\jpsi$ elastic amplitudes, for
poles 2 (left) and 3 (right).
Also see
Fig.~\ref{trajectories} and
Table~\ref{polespp}.}
\end{figure}
Finally, in
order to study the effect of using a complex mass for the $\rho^0$ in the
$\rho^0\jpsi$ channel, we vary the $\rho$ width from 0\% to
100\%
of its PDG \cite{JPG37p075021} value and plot the corresponding amplitudes in
Fig.~\ref{rhowidth}. We see that the maximum $|T|^2$ increases by
almost 3 orders of magnitude
when going from the 0\% case (dotted curve in left-hand plot) to
the
100\%
case (solid curve in right-hand plot).
Furthermore, the 0\% curve only starts out at the central $\rho^0\jpsi$
threshold, of course.
Thus, it becomes clear that no
realistic description of the $\rho^0\jpsi$ channel is possible without
smearing out somehow its threshold, so that its influence kicks in
before the $D^0D^{\ast0}$ channel opens and depletes the signal.
Naturally, similar conclusions apply in principle to the $\omega\jpsi$
channel, though there the effects are less pronounced because of the small
$\omega$ width and the somewhat higher threshold.
These results show that
our unitarization procedure for complex masses in the asymptotic states
performs as expected
in accounting for thresholds involving resonances.
\begin{figure}
\centering
\begin{tabular}{cc}
\resizebox{!}{300pt}{\includegraphics{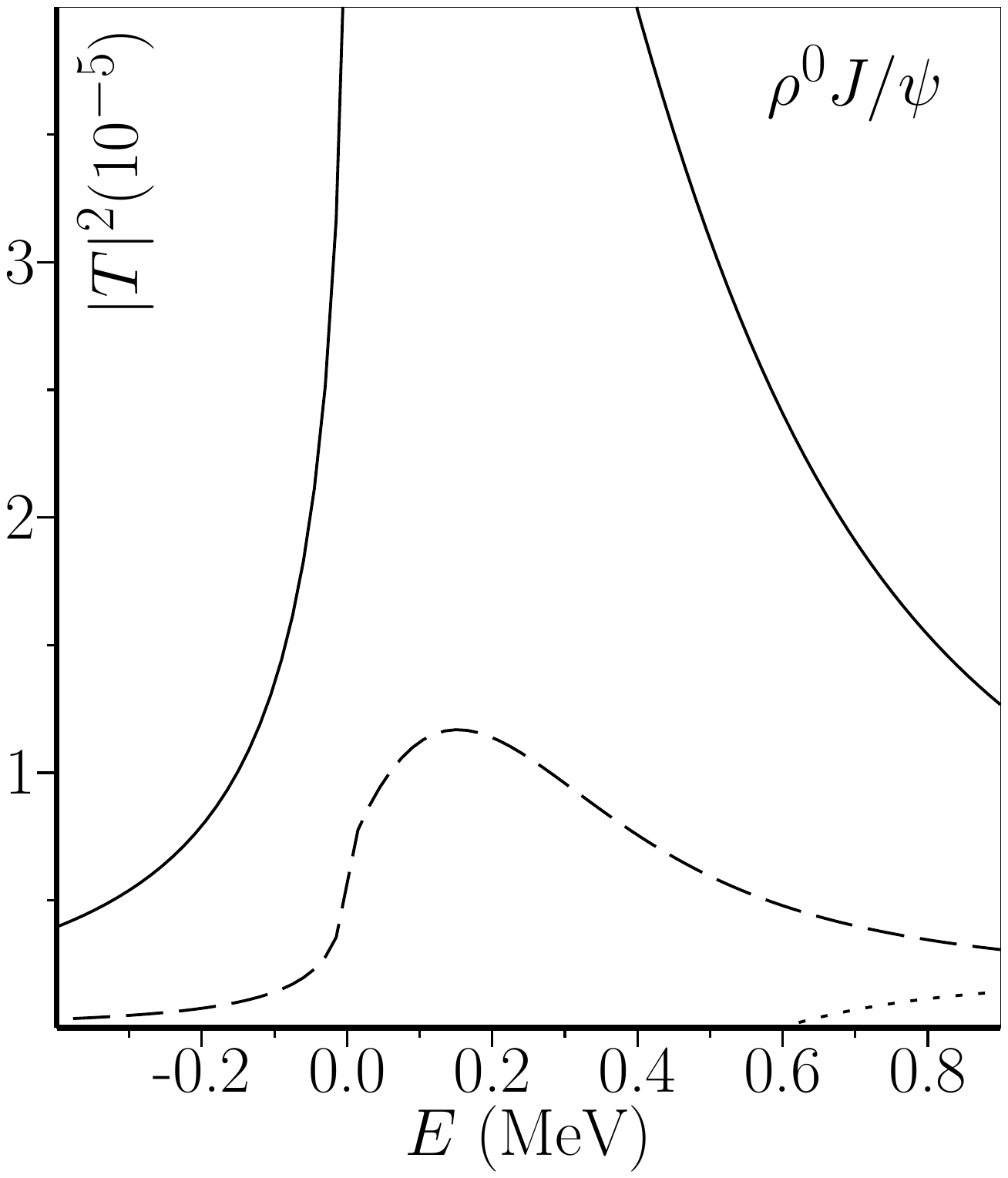}}
&
\hspace*{-34pt}
\resizebox{!}{300pt}{\includegraphics{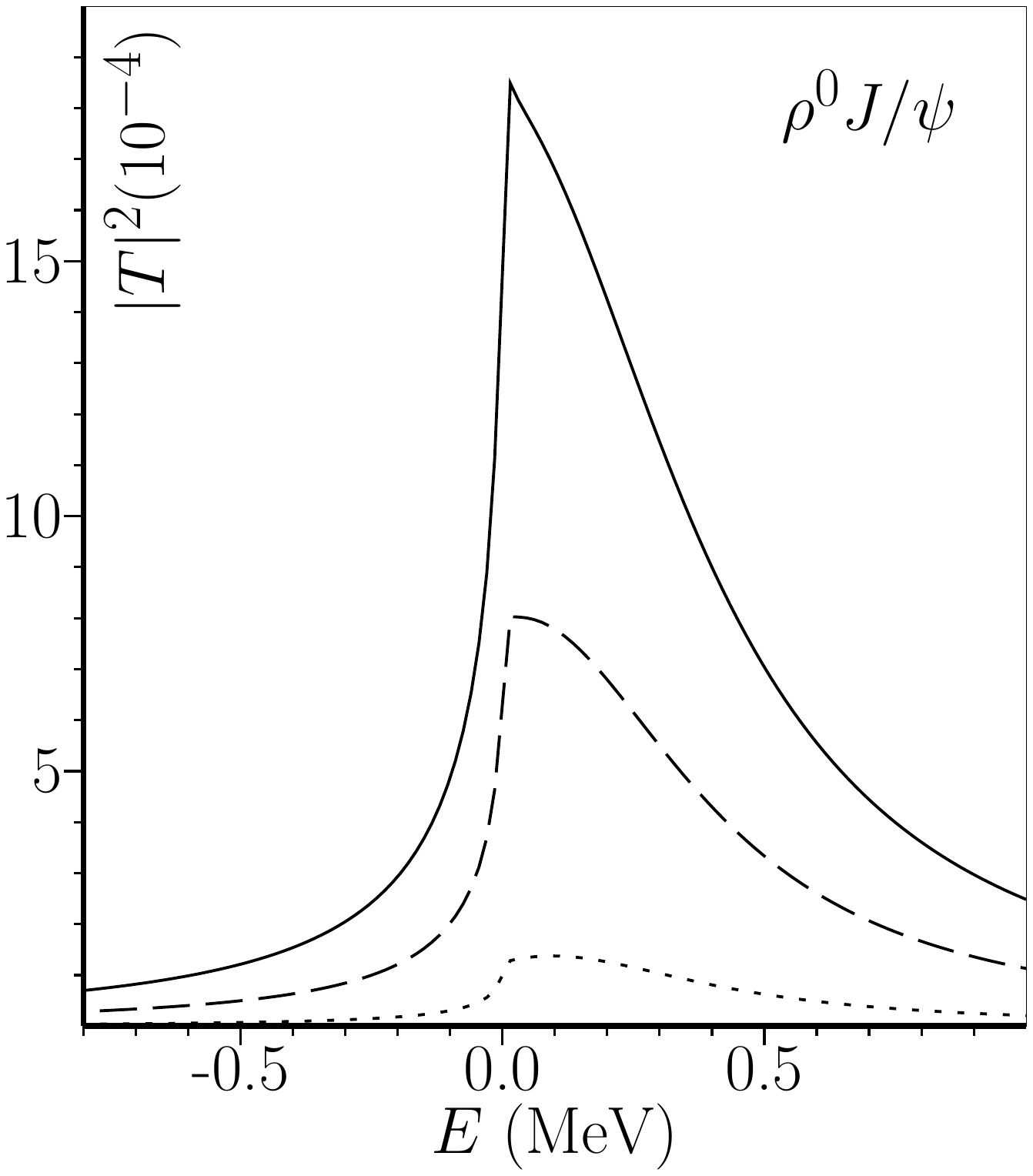}}
\end{tabular}
\mbox{ } \\[-140pt]
\caption{\label{rhowidth} $\rho^0\jpsi$
elastic
amplitude for reduced $\rho^0$ width.
Left: 0\% (dots), 1\% (dashes), 5\% (full); right: 10\% (dots),
50\% (dashes),
100\%
(full). 
Studied case: pole
2
in Table~\ref{polespp}.
Also see Fig.~\ref{trajectories}.}
\end{figure}
To conclude our discussion, we should mention that our results are
qualitatively in agreement with those of Danilkin \& Simonov
\cite{PRD81p074027}, in the sense that a single resonance pole originating
from the $2\,{}^{3\!}P_1$ $c\bar{c}$ state is capable of describing the
$X(3872)$ data. However, we disagree with their conclusions on the 
$2\,{}^{3\!}P_0$ state. In an earlier, single-channel description
\cite{PRD74p037501}, we found a resonance at 3946~MeV with a width of 58~MeV,
and we do not believe a detailed multichannel calculation will change these
values dramatically. Thus, the listed $X(3945)$ \cite{JPG37p075021} resonance,
with mass 3916 MeV, width 40 MeV, and positive $C$-parity, appears to be a good
candidate. As for the $2\,{}^{1\!}P_1$ state, the $X(3940)$ \cite{JPG37p075021}
resonance, with mass 3942 MeV, width 37 MeV, and principal decay mode $DD^\ast$,
seems the obvious choice. With the old $Z(3930)$ meanwhile identified as the
$2\,{}^{3\!}P_2$ ($\chi_{c2}(2P)$ \cite{JPG37p075021}) state, we
might
so understand all 4 charmonium states in the range 3.87--3.95~MeV.

\section{Summary and conclusions}
Summarizing, we have investigated the $1^{++}$ charmonium scenario for the 
$X(3872)$ resonance, by analyzing in detail the influence of the
$D^0D^{\ast0}$,
$\rho^0\jpsi$, and $\omega\jpsi$
channels on pole positions and amplitudes.
In order to describe the latter OZIS
channels
in a realistic way, we have
used
complex masses for the $\rho^0$ and $\omega$,
and then restored unitarity of the
$S$-matrix by a new
and rigorous algebraic procedure, 
albeit physically heuristic.
It is true that the redefined $S$-matrix may have some unusual analyticity
properties \cite{NPB12p281}, but in our amplitudes no sign was found of any
nearby spurious singularities.
Moreover,
the behaviour of the $\rho^0\jpsi$
amplitude as a function of the $\rho^0$ width gives us confidence in our
approach. Concretely, we have shown that our scenario is compatible with the
$D^0D^{\ast0}$ and $\pi^+\pi^-\jpsi$ data, with a single resonance pole
on top of
or slightly below the $D^0D^{\ast0}$ threshold.
Moreover, our treatment of the $\rho^0\jpsi$ and $\omega\jpsi$ channels
has proven compatible with the observed branching ratio of these decays.\\
Thus, the data do not
seem to require a molecular or tetraquark interpretation of the $X(3872)$,
also in view of so far unobserved \cite{PRD71p031501} charged partner states.\\
Nevertheless, only further improved measurements and theoretical calculations
will in the end allow to draw a definitive conclusion on the scenario
preferred by nature.\\
In conclusion, we must stress that the $X(3872)$, whatever its
assignment, is an extraordinary structure, because of its coincidence ---
to an accuracy of less than 1 MeV --- with the central thresholds of the
principal decay modes. This circumstance is at the same time a blessing and a
curse. To start with the latter, no model can ambition to quantitatively 
describe the $X(3872)$ with present-day state-of-the-art in strong
interactions, while experiment will have an extremely hard time to reduce the
bin sizes to less than 1 MeV and simultaneously keep statistics sufficiently
high. On the other hand, with a strengthened effort of both theory and
experiment, a wealth of knowledge on charmonium spectroscopy and strong decay
--- OZI-allowed as well as OZI-suppressed --- may be gathered by further
studying this peculiar resonance.

\chapter{$X(3872)$ is not a true molecule \label{chp6}}
\thispagestyle{empty}
\vspace*{-1.2cm}{\normalsize S. Coito, G. Rupp, and E. van Beveren, {\it EPJC} {\bf73}, 2531 (2013).}\\[1cm]
On the theory side, the discussion about the nature of $X(3872)$ continues
most vivid. Even before the recent result of LHCb \cite{PRL110p222001},
most model builders described the state as an axial vector. For instance, model
calculations of semi-inclusive $B\to\eta_{c2}+X$ processes \cite{PRD85p034032}
as well as electromagnetic $\eta_{c2}$ decays \cite{1205.5725} have been shown
to disfavor the $2^{-+}$ scenario. The same conclusion was reached in a
tetraquark description of $X(3872)$ \cite{CTP57p1033}, while pion exchange in
a molecular picture would be repulsive in this case \cite{PLB590p209} and so
inhibitive of a bound state. Finally, unquenching a \osdt\ $c\bar{c}$ state 
by including meson-meson loops could only further lower the bare mass, which
lies in the range 3.79--3.84~GeV for all quenched quark models we know of,
thus making a $2^{-+}$ charmonium resonance at 3.872~GeV very unlikely. For further information and more references concerning
$X(3872)$, see e.g.\ a recent review \cite{PPNP67p390}, as well as our previous
coupled-channel analysis in Chapter \ref{chp5}.\\
The first suggestion of possible meson-meson molecules bound by pion exchange,
in particular a $DD^*$ state with quantum numbers $1^{++}$ or $0^{-+}$, was due to T\"{o}rnqvist
\cite{PLB590p209}.
With the discovery of $X(3872)$ just below the $D^0D^{*0}$ threshold,
this idea was revived, of course. In the present chapter, we intend to study
the issue, not from T\"ornqvist's pion-exchange point of view, but rather as
regards its possible implications for models based on quark degrees of
freedom. In this context, it is worthwhile to quote from
Ref.~\cite{PRD76p094028}, in which a molecular interpretation is advocated
(also see Ref.~\cite{PRD69p074005}):
\begin{quote} \em
``Independent of the original mechanism for the resonance,
the strong coupling transforms the resonance into a
bound state just below the two-particle threshold if $a>0$
or into a virtual state just above the two-particle threshold if $a<0$.
If $a>0$, the bound state has a molecular structure,
with the particles having a large mean separation of order $a$.'' \em
\end{quote}
(Note that, here, $a$ is the $S$-wave scattering length.)
Also:
\begin{quote} \em
``In this case \em [$1^{++}$], \em
the measured mass $M_X$ implies unambiguously that $X$ must be either a
charm meson molecule or a virtual state of charm mesons.'' \em
\end{quote}
In face of these peremptory claims about the molecular picture, 
it is of utmost importance to study in detail the $X(3872)$ wave function
for a model in which the {\em mechanism} \/generating the meson is quark
confinement combined with strong decay. Thus, we employ the simplified,
coordinate-space version of of RSE, defined in Sec.~\ref{CSSM}, to describe $X(3872)$ as a unitarized and
mass-shifted \ttpo\ charmonium state. The model's exact solvability then
allows to obtain analytic expressions for the wave-function components,
and follow bound-state as well as resonance poles on different Riemann
sheets.

\section{The coupled $\mathitbf{c\bar{c}}\,$-$\mathitbf{D^0\!D^{*0}}$ system}
Now we apply the formalism to the coupled $c\bar{c}$-$D^0D^{*0}$ system.
The $c\bar{c}$ channel is assumed to be in a \tpo\ state, i.e., with $l_c=1$,
implying the $D^0D^{*0}$ channel to have $l_f=0$ or 2. Nevertheless, we shall
restrict ourselves here to the $S$-wave channel only, which will be strongly
dominant, especially near threshold. The fixed parameters are given in
Table~\ref{tab:param}, where the meson masses are from the
\begin{table}[h]
\centering
\begin{tabular}{c|c|c|c|c|c}
Param.\ & $\omega$&$m_c$&$m_{D^0}$&$m_{D^{*0}}$&$m_{D^0}\!+\!m_{D^{*0}}$\\
\hline &&&&&\mbox{ }\\[-9pt]
(MeV)&$190$&$1562$&$1864.86$&$2006.98$&$\bf{3871.84}$\\
\end{tabular}
\caption{\label{tab:param} Fixed parameters.}
\end{table}
PDG \cite{PRD86p010001}, while $\omega$ and the constituent
charm quark mass $m_c$ are defined in \eqref{ctepar}. Thus,
from Eq.~(\ref{hop}) we get the lowest two harmonic oscilator states at $E_0=3599$~MeV and
$E_1=3979$ MeV, respectively. The former should give rise --- after
unquenching --- to the \otpo\ charmonium state $\chi_{c1}(1P)$
\cite{PRD86p010001}, with mass $3511$ MeV, while the latter is the bare
\ttpo\ state, which cannot so easily be linked to resonances
in the PDG tables, though both $X(3940)$ and $X(3872)$ are possible
candidates, in view of their mass and dominant $DD^*$ decay mode
\cite{PRD86p010001}. However, $X(3940)$ may just as well be the, so far
unconfirmed, \tspo\ ($1^{+-}$) state $h_c(2P)$, cf.~Sec.~\ref{x3940}.\\
The two remaining parameters, viz.\ the string-breaking distance $a$
and the global coupling $g$, have to be adjusted to the experimental data.
Nevertheless, these parameters are not completely free, as they both have a
clear physical interpretation, albeit of an empirical nature. Thus, $a$ is
the average interquark separation at which \tpo\ quark-pair
creation/annihilation is supposed to take place, while $g$ is the overall
coupling strength for such processes. Note that we do not assume a particular
microscopic model for string breaking inspired by QCD, like e.g.\ in a
very recent paper \cite{1210.4674}. Still, the values of $a$ found in
the present work are in rough agreement with our prior model findings, and
even compatible \cite{0712.1771} with a lattice study of string breaking in
QCD \cite{PRD71p114513}. Concretely, we have been obtaining values of $a$
in the range 1--4 GeV$^{-1}$ (0.2--0.8~fm), logically dependent on quark
flavor, since the string-breaking distance will scale with the meson's size,
being smallest for bottomonium.
As for the coupling parameter $g$, its empirical value will depend on $a$,
but also on the set of included decay channels. In
realistic calculations, values of the order of 3 have been obtained (see
e.g.\ Table \ref{polespp}, where $g\equiv\lambda$).

\section{Poles}
The crucial test the present model must pass is its capability of generating
a pole near the $D^0D^{*0}$ threshold. Indeed, a dynamical pole is
found slightly below threshold for different combinations of the free
parameters $a$ and $g$, several of which are listed in Table~\ref{ag}.
Examples are here given of bound states, virtual bound states,
and below-threshold resonances, the latter ones only occurring for $S$-wave
thresholds as in our case. Note that poles of both virtual bound states and
resonances lie on the second Riemann sheet, i.e., the relative momentum has a
negative imaginary part. From this table we also observe that larger and larger
couplings are needed to generate a pole close to threshold when $a$ approaches
the value 3.5~GeV$^{-1}$. We shall see below that this is due to the nodal
structure of the bound-state wave function.\\
\begin{table}[h]
\begin{center}
\begin{tabular}{c|c|l|c}
$a$ (GeV$^{-1}$)& $g$  & pole (MeV) & type\\
\hline &&&\mbox{ } \\[-9pt]
$2.0$ & $1.149$ & $3871.84$&VBS\\[1mm]
$2.5$ & $1.371$ & $3871.84$&VBS\\[1mm]
$3.0$ & $2.142$ & $3871.84$&VBS\\[1mm]
$3.1$ & $2.503$ & $3871.84$&VBS\\[1mm]
$3.2$ & $2.531$ & $3871.84-i12.01$&resonance\\[1mm]
$3.3$ & $3.723$ & $3871.84-i\ 4.45$&resonance\\[1mm]
$3.4$ & $7.975$ & $3871.84-i\ 0.39$&resonance\\[1mm]
$3.5$ & $\infty$ &  & - \\
\hline &&&\mbox{ } \\[-9pt]
$2.0$ & $1.152$ & $3871.84$&BS\\[1mm]
$2.5$ & $1.373$ & $3871.84$&BS\\[1mm]
$3.0$ & $2.145$ & $3871.84$&BS\\[1mm]
$3.1$ & $2.507$ & $3871.84$&BS\\[1mm]
$3.2$ & $3.083$ & $3871.84$&BS\\[1mm]
$3.3$ & $4.194$ & $3871.84$&BS\\[1mm]
$3.4$ & $8.254$ & $3871.84$&BS\\[1mm]
$3.5$ & $\infty$ &  & - \\[1mm]
\end{tabular}
\caption{\label{ag} Bound states (BS), virtual bound states (VBS), and
resonances closest to threshold, for various $g$ and $a$ combinations.}
\end{center}
\end{table}
\begin{table}[h]
\centering
\begin{tabular}{c|c|c|c}
$a$ (GeV$^{-1}$) & $g$ & dynamical pole & confinement pole\\
\hline &&&\mbox{ } \\[-9pt]
2.0 & 1.172 & 3871.68 & $4030.50 - i 136.51$\\[1mm]
2.5 & 1.403 & 3871.68 & $4063.27 - i 124.07$\\[1mm]
3.0 & 2.204 & 3871.68 & $4101.48 - i\ 88.03$\\[1mm]
3.4 & 8.623 & 3871.68 & $4185.85 - i\ 20.63$\\[1mm]
\end{tabular}
\caption{\label{twopoles} Pole doubling: pairs of poles (in MeV) for some sets
of $a$ and $g$ values, chosen such that the dynamical pole settles at
the $X(3872)$ PDG \cite{PRD86p010001} mass.}
\end{table}
Although a dynamical pole shows up near the $D^0D^{*0}$ threshold, there still
should be a confinement pole connected to the first radial \tpo\ excitation at
3979~MeV. Well, we do find such a pole, for each entry in Table~\ref{ag}. In
Table~\ref{twopoles} a few cases are collected, with the parameters tuned to 
generate a dynamical pole at precisely the $X(3872)$ PDG \cite{PRD86p010001}
mass of 3871.68~MeV. Note, however, that the associated confinement pole is not
necessarily of physical relevance, since at the corresponding energy several
other strong decay channels are open, which no doubt will have a very
considerable influence and possibly even change the nature of both poles. As
a matter of fact, in Chapter \ref{chp5}, with all relevant
two-meson channels included, the $X(3872)$ resonance was found as a confinement
pole, whereas dynamical poles were only encountered very deep in the complex
energy plane, without any observable effect at real energies. So here we show
these results only to illustrate that pole doubling may occur when strongly
coupling $S$-wave thresholds are involved, as we have observed in the past 
in the case of e.g.\ the light scalar mesons \cite{ZPC30p615} and 
$D_{s0}^*(2317)$ \cite{PRL91p012003}. The issue of confinement vs.\ dynamical
poles will be further studied in Sec.~\ref{secdvsc}.\\
In order to better understand the dynamics of the different poles, we plot
in Fig.~\ref{traja2} pole trajectories in the complex energy plane as a function
of the coupling constant $g$, and for three different values of $a$. For
vanishing $g$, the dynamical pole acquires a negative infinite imaginary part
and so disappears in the continuum, whereas the confinement pole moves to the 
real energy level of the bare \ttpo\ state, i.e., 3979~MeV. 
\begin{figure}[H]
\centering
\begin{tabular}{c c }
\hspace*{-20pt}
\resizebox{!}{400pt}{\includegraphics{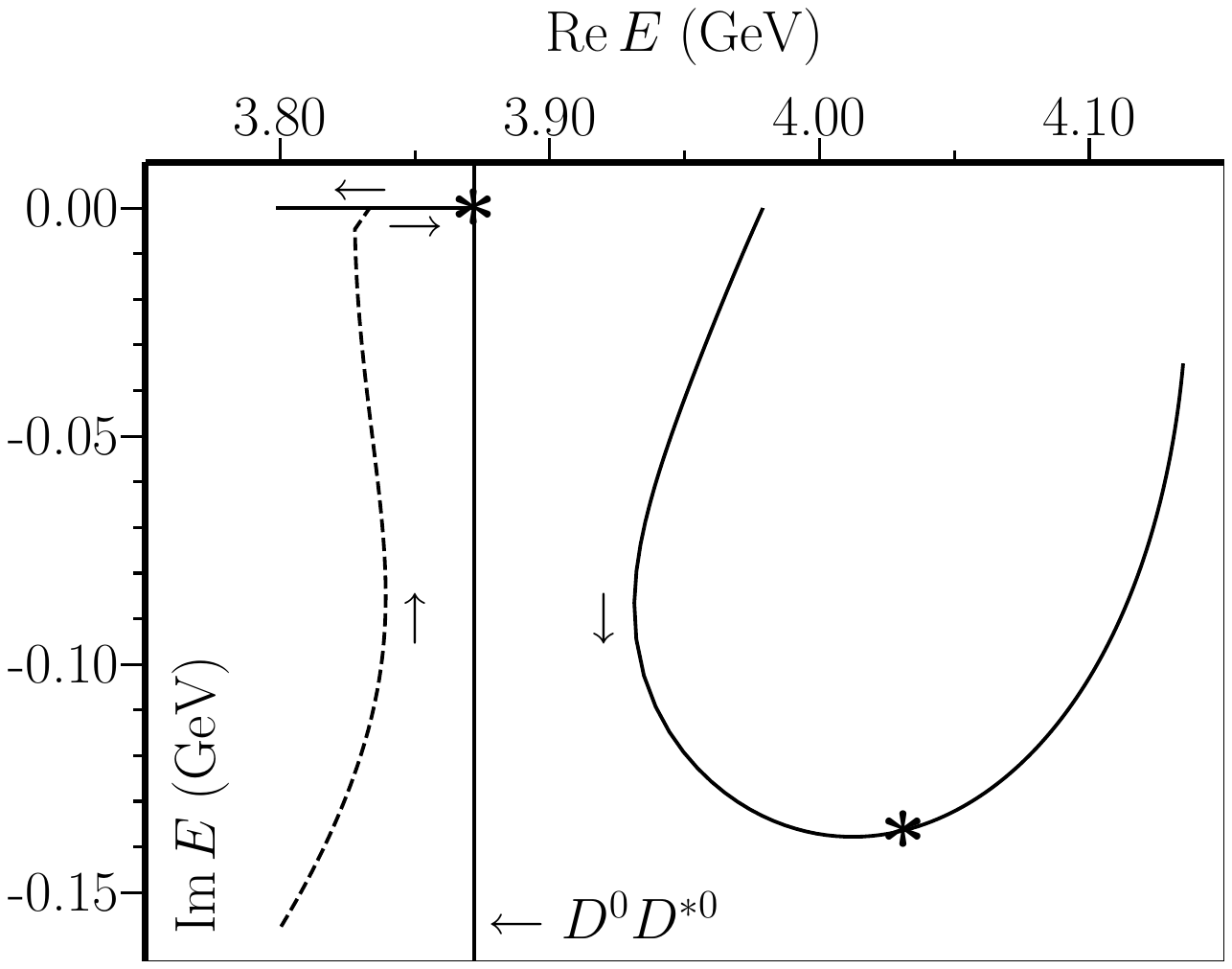}}\\
\mbox{} \\[-90mm]
\hspace*{-20pt}\resizebox{!}{400pt}{\includegraphics{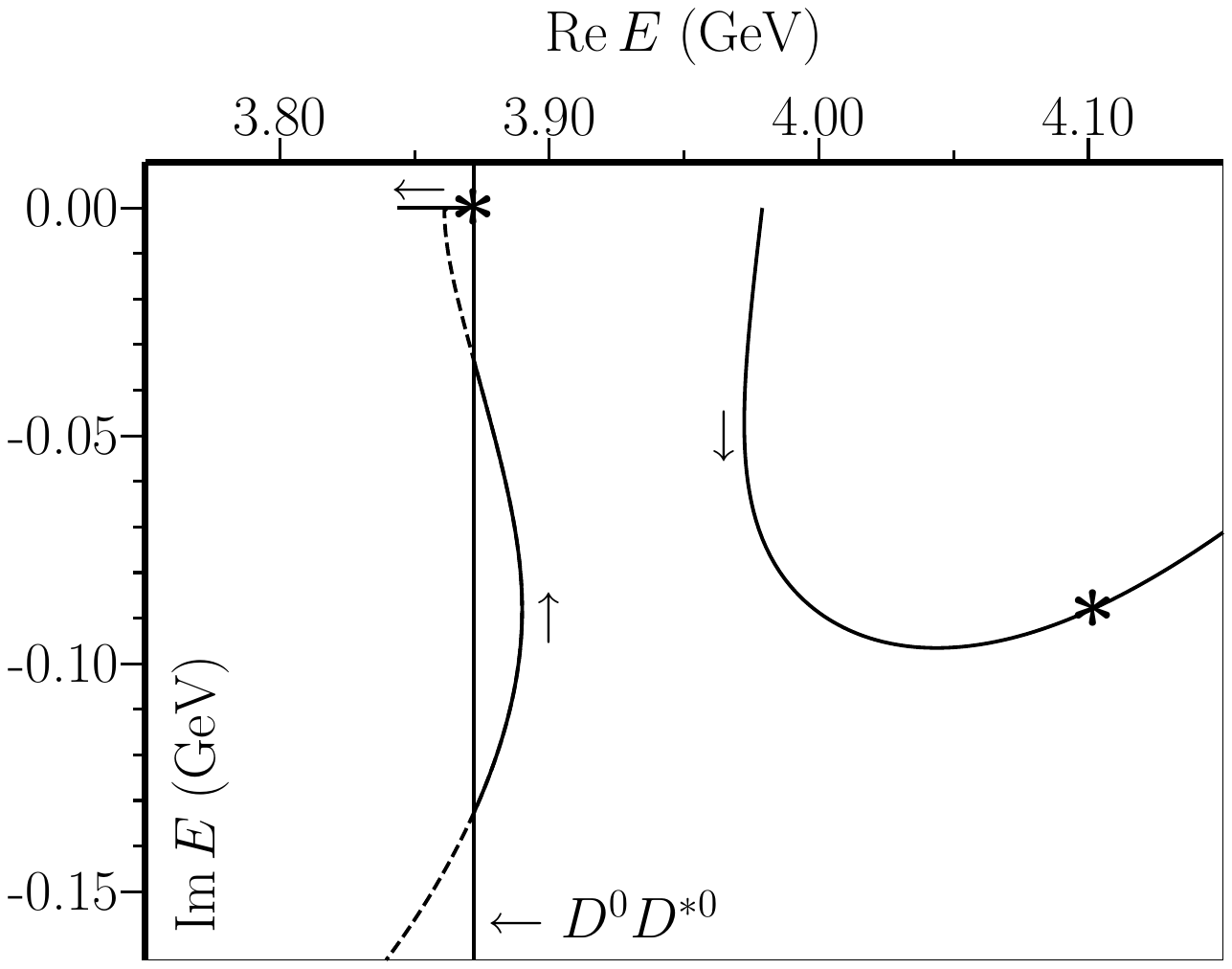}}\\
\mbox{} \\[-90mm]
\hspace*{-20pt}\resizebox{!}{400pt}{\includegraphics{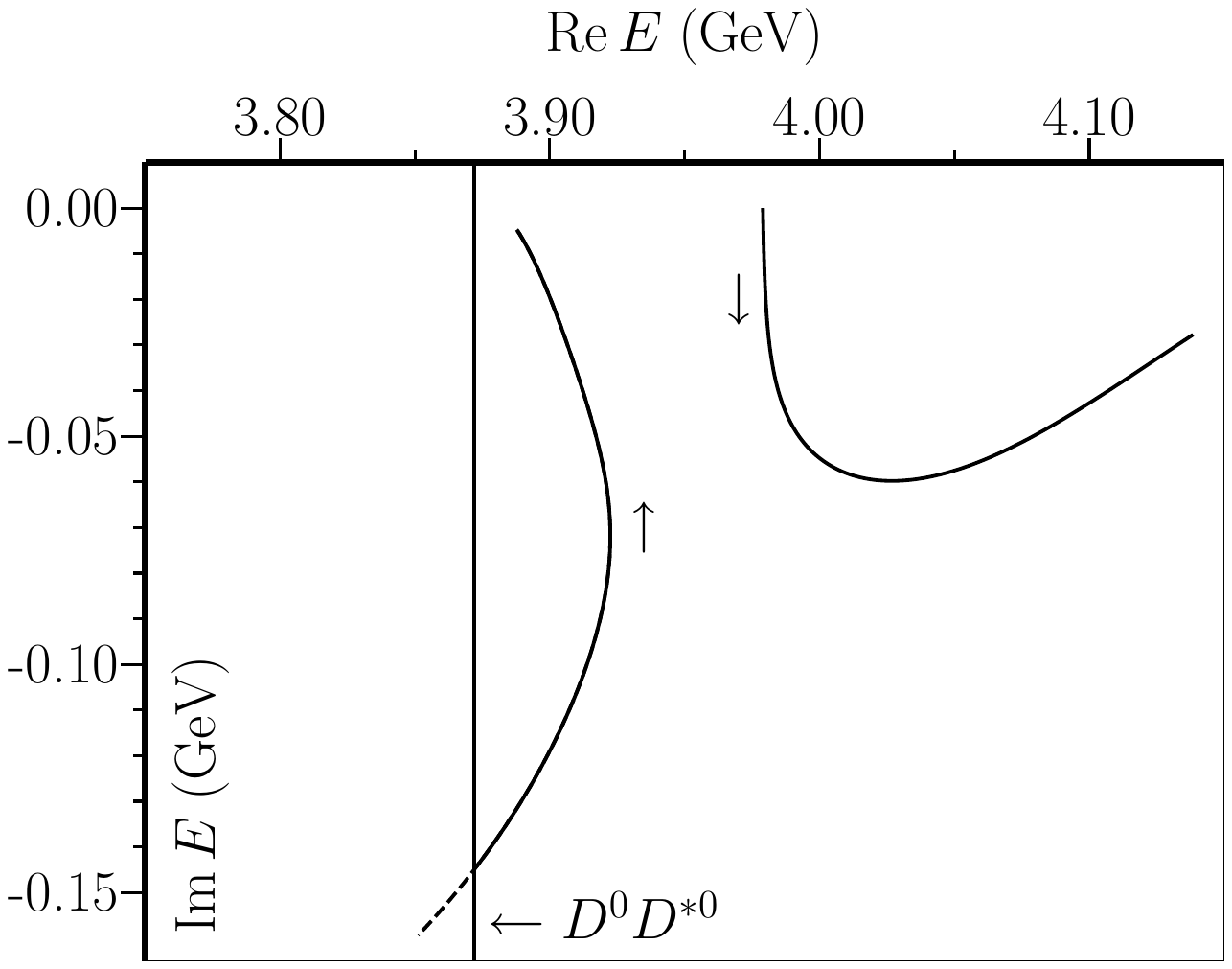}}\\
\mbox{} \\[-90mm]
\end{tabular}
\caption{\label{traja2}Pole trajectories of dynamical (left) and
confinement (right) poles as a function of $g$, for $a$=2.0 GeV$^{-1}$ (top),
3.0 GeV$^{-1}$ (middle), and 3.5 GeV$^{-1}$ (bottom), respectively.
In the last case, there is no bound state near threshold. Note: (i)
poles in Table.~\ref{twopoles} are here marked by
\boldmath$\ast$; (ii) arrows along curves indicate increasing $g$.}
\end{figure}
As $g$ increases,
and for both $a=2.0$ GeV$^{-1}$ and $a=3.0$ GeV$^{-1}$, the dynamical pole 
moves to the real axis below threshold, becoming first a virtual bound
state and then a genuine bound state. Note that, in the latter case, the
real part twice attains the $X(3872)$ mass even before the pole reaches
the real axis, but the corresponding imaginary parts are much too
large as compared with experiment \cite{PRD86p010001}, so only the bound
state can be considered physical. Finally, for $a=3.5$~GeV$^{-1}$ the pole
does never reach the real axis, which would require an infinite coupling.
For the other parameter sets listed in Table~\ref{ag}, we find intermediate
situations. Another feature we can observe for all trajectories is an initial
attraction and subsequent repulsion between the dynamical and the confinement
poles.
\section{Wave function}
\label{w-f}
Now we are in a position to study the $X(3872)$ bound-state wave function in
several situations. We choose two values for the string-breaking parameter, 
viz.\ $a=2.0$~GeV$^{-1}$ and $a=3.0$~GeV$^{-1}$. In Table \ref{binden} five
different binding energies (BEs) are chosen with respect to the $D^0D^{*0}$
channel, including the PDG \cite{PRD86p010001} value labeled by $X$. We have
computed and normalized (see Subsec.~\ref{appAp}) the two-component radial
wave function $R(r)$ for each  of the five cases. In
Fig.~\ref{fig:rwf} we depict the cases labeled by $A$, $X$ and $D$, the other
two representing intermediate situations. General features we immediately
observe are the typical $S$-wave behavior of the $D^0D^{*0}$ wave-function
component $R_f$, while the $c\bar{c}$ wave function $R_c$ is in a $P$ state,
the latter also having a node, as it is dominantly a first radial excitation.
Furthermore, $|R_f|$ is larger than $|R_c|$ in most situations, for
all $r$, except for unphysically large BEs (cf.\ plot $D$).
Nevertheless, the two components are of comparable size for intermediate $r$
values. Then, as the BE becomes smaller, the tail of $R_f$ grows
longer, as expected, whereas $R_c$ always becomes negligible for
distances larger than roughly 11--12 GeV$^{-1}$. Now, the increased $R_f$ tail
affects the normalization of \em both \em \/$R_c$ and $R_f$. Thus, the ratio
$|R_f(r)|/|R_c(r)|$ is quite robust for most $r$ values, as it does not
significantly change with the BE.
\begin{table}[h]
\centering
\begin{tabular}{c| c| c| c| c|c}
\multicolumn{2}{c|}{$a$ (GeV$^{-1}$)}   & \multicolumn{2}{c|}{2.0} &
\multicolumn{2}{c}{3.0}\\
\hline &&&&&\mbox{ } \\[-9pt] 
label& BE (MeV) & $g$     & pole    & $g$     & pole \\
\hline &&&&&\\[-9pt] 
$A$      & $\ 0.00$ & 1.152 & 3871.84 & 2.145 & 3871.84\\[1mm]
$B$      & $\ 0.10$ & 1.167 & 3871.74 & 2.191 & 3871.74\\[1mm]
$\mathbf{X}$ & $\mathbf{\ 0.16}$ & \bf{1.172} & \bf{3871.68} & \bf{2.204} & \bf{3871.68}\\[1mm]
$C$      & $\ 1.00$ & 1.207 & 3870.84 & 2.311 & 3870.84\\[1mm]
$D$      & $ 10.00$ & 1.373 & 3861.84 & 2.899 & 3861.84\\[1mm]
\end{tabular}
\caption{\label{binden}Five chosen binding energies (BE) in the $D^0D^{*0}$
channel, for two different $a$ values and the corresponding couplings $g$.}
\end{table}
\begin{figure}[H]
\centering
\begin{tabular}{c c}
\hspace*{-70pt}\resizebox{!}{530pt}{\includegraphics{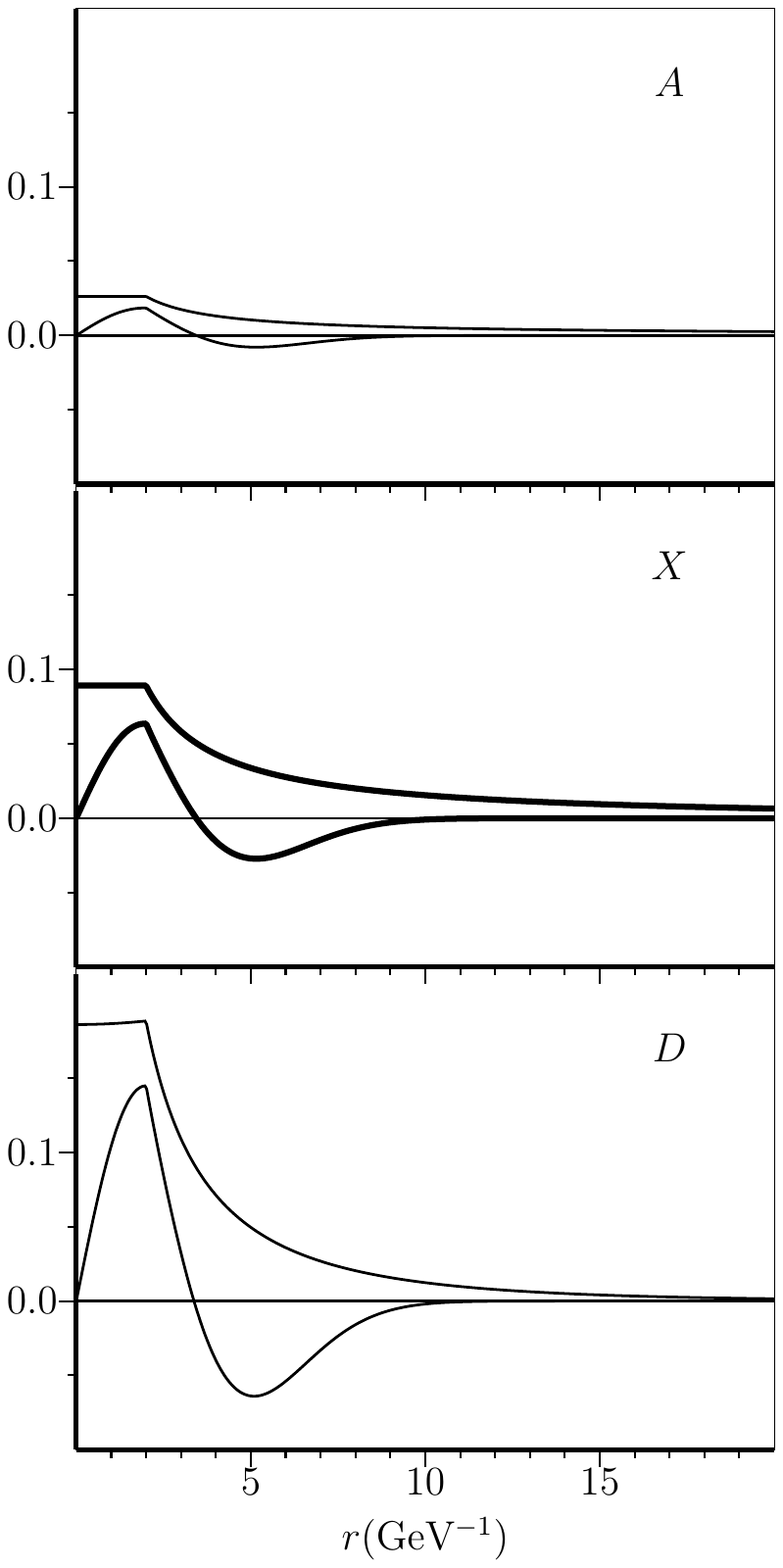}}&
\hspace*{-215pt}\resizebox{!}{530pt}{\includegraphics{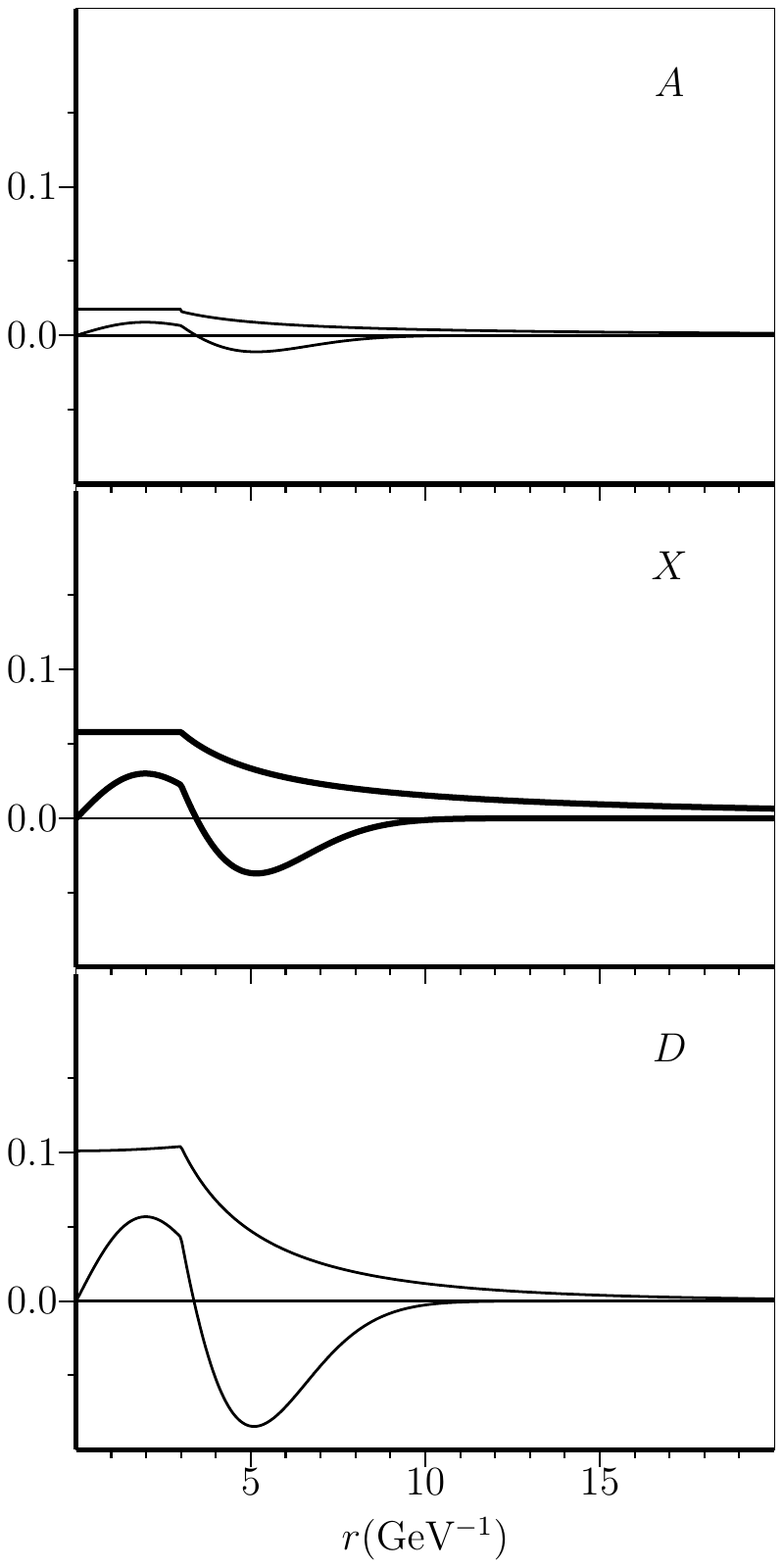}}\\
\mbox{} \\[-7.5cm]
\end{tabular}
\caption{\label{fig:rwf}Normalized two-component radial wave function $R(r)$
for three BEs, corresponding to labels $A,X,D$ in
Table~\ref{binden}, and two $a$ values. Upper curves: $R_f(r)$; lower curves:
$R_c(r)$. Left: $a=2$~GeV$^{-1}$; right: $a=3$~GeV$^{-1}$.}
\end{figure}

\section{Probabilities and r.m.s.\ radii}
\label{obs}
Having derived the $X(3872)$ wave function for several scenarios, we can
now straightforwardly compute the relative probabilities of the $c\bar{c}$
and $D^0D^{*0}$ components (see Subsec.~\ref{appAp}), with the results given
in Table~\ref{prob}, for the
\begin{table}
\centering
\begin{tabular}{ c|c|c|c|c|c|c}
$a$  & channel        & $A$ & $B$ & $\mathbf{X}$ & $C$ & $D$\\
\hline &&&&&&\mbox{ } \\[-9pt]
2.0 &$c\bar{c}$   & $\ 0.63$  & $\ 6.00$ & ${\bf \ 7.48}$ & $16.98$ & $39.68$\\[1mm]
2.0 &$D^0D^{*0}$  & $ 99.37$  & $ 94.00$ & $ 92.52$       & $83.02$ & $60.32$\\[2mm]
3.0 &$c\bar{c}$   & $\ 0.97$  & $\ 9.01$ & ${\bf 11.18}$  & $24.65$ & $55.54$\\[1mm]
3.0 &$D^0D^{*0}$  & $ 99.03$  & $ 90.99$ & $88.82$        & $75.35$ & $44.46$\\[1mm]
\end{tabular}
\caption{\label{prob} Probabilities (in \%) of the two wave-function
components, for the cases specified in Table~\ref{binden} ($a$ in GeV$^{-1}$).}
\end{table}
five BEs and two $a$ values from Table~\ref{binden}. Note that the
probability in the $D^0D^{*0}$ channel is only computed for normalization
purposes, since in a more realistic calculation at least the $D^\pm D^{*\mp}$
component would acquire a nonnegligible probability as well, as the
corresponding threshold lies only 8~MeV higher. Nevertheless, our
simplification is unlikely to have an appreciable effect on the $c\bar{c}$
probability and will only increase that of the $D^0D^{*0}$ component
accordingly. Also note that the $c\bar{c}$ probability includes all \tpo\
states, with the \ttpo\ being dominant, because the corresponding bare
eigenstate lies only 100~MeV higher.  However, also the \otpo\ state is
nonnegligible in the physical $X(3872)$ wave function. In the coupled-channel
approach of Ref.~\cite{PRD85p114002}, a \otpo\ admixture of about 15\% was
found.  Notice that --- inevitably --- unquenching not only mixes meson-meson
components into the total bound-state wave function, but also quark-antiquark
components of confinement states other than the one under consideration (also
see Ref.~\cite{ZPC19p275}). Here, for a BE of 0.16~MeV, corresponding to the
physical \cite{PRD86p010001} $X(3872)$, case $\mathbf{X}$ in Table~\ref{prob},
has a 7.48\% $c\bar{c}$ probability for $a=2.0$ GeV$^{-1}$ and 11.18\% for
$a=3.0$ GeV$^{-1}$. For smaller BEs, the $c\bar{c}$ probability decreases as
expected, because of the growing weight of the $D^0D^{*0}$ tail. On the other
hand, for a BE of 10~MeV and $a=3.0$~GeV$^{-1}$, the charmonium probability
becomes even larger than that of the meson-meson component. Now,
the experimental errors in the average mass of the $X(3872)$ and the
$D^0D^{*0}$ threshold allow for a maximum BE of 0.57~MeV, i.e., somewhere
in between cases $\mathbf{X}$ and $C$. This would then correspond to a
$c\bar{c}$ probability roughly midway in the range 7.48\%--16.98\%
($a=2.0$~GeV$^{-1}$) or 11.18\%--24.65\% ($a=3.0$~GeV$^{-1}$).
In the limiting case of zero binding, the $c\bar{c}$ probability
would eventually vanish. Also notice that, in all five cases of
Table~\ref{prob}, the $c\bar{c}$ probability rises by about 50\%
when $a$ is increased from 2.0 to 3.0~GeV$^{-1}$.  Nevertheless,
if we take $a=2.0$~GeV$^{-1}$ as in Chapter \ref{chp5},
we get a $c\bar{c}$ probability of 7.48\%, very close the 7\% found
in Refs.~\cite{PRD81p054023,1206.4877}.\\
Next we use the normalized wave functions and Eq.~(\ref{rmsp}) to compute
the $X(3872)$ r.m.s.\ radius for the five cases discussed before
(see Table~\ref{binden}), with the results presented in Table~\ref{tab:avr}. 
\begin{table}
\centering
\begin{tabular}{c| c| c| c| c| c| c}
$a$ (GeV$^{-1}$) & $a$ (fm) & $A$ & $B$ & $\mathbf{X}$ & $C$ & $D$\\[1mm]
\hline &&&&&&\mbox{ } \\[-9pt]
2.0 & 0.39 & 100.22 & 9.92 & \bf{7.82} & 3.10 & 1.15\\[1mm]
3.0 & 0.59 & 100.14 & 9.85 & \bf{7.76} & 3.05 & 1.23 \\[1mm]
\end{tabular}
\caption{\label{tab:avr} R.m.s.\ radii of the wave function, expressed in fm,
for the cases specified in Table~\ref{binden}.}
\end{table}
It is interesting to observe that the r.m.s.\ radius, which in principle
is an observable, is much less sensitive to the choice of $a$ than de
wave-function probabilities. Furthermore, the large to very large
r.m.s.\ radii in the various situations are hardly surprising, in view
of the small binding energies and the resulting very long tails of the
$D^0D^{*0}$ wave-function components (see Fig.~\ref{fig:rwf} above).\\
Using Eq.~(\ref{cotanp}), we now also evaluate the $S$-wave scattering length
\bdm
a_S=-\lim_{E\to0}\left[k(E)\cot\delta_0(E)\right]^{-1}.
\edm
 In case $\mathbf{X}$ and for $a=2.0$ GeV$^{-1}$ we thus find $a_S=11.55$ fm, which is
large yet of the expected order of magnitude for a BE of 0.16 MeV. For even
smaller BEs, the scattering length will further increase, roughly like
$\propto\!1/\sqrt{\mbox{BE}}$. Let us here quote from Ref.~\cite{PRD69p074005}:
\begin{quote} \em
``Low-energy universality implies that as the scattering length $a$ increases,
the probabilities for states other than $D^0\bar{D}^{*0}$ or
$\bar{D}^0D^{*0}$ decrease as $1/a$ \ldots''
\end{quote}
Indeed, we verify from our Table~\ref{prob} that  --- very roughly ---
the $c\bar{c}$ probability decreases as $\propto\!\sqrt{\mbox{BE}}$,
and so like $\propto\!1/a_S$.

\section{Stability of results and nature of poles}
\label{secdvsc}
In this section we are going to study the stability of our results, as well as
the nature of the found solutions. So let us
vary the two usually fixed parameters, viz.\ $\omega$ and $m_c$, in such a
way that the bare \otpo\ mass remains unaltered at 3599~MeV, whereas that of the
\ttpo\ changes as shown in Table.~\ref{tab:newpar}.
\begin{table}[h]
\centering
\begin{tabular}{c|c| c| c}
            &  $I$ &  standard   &  $II$\\[1mm]
\hline &&&\mbox{ } \\[-9pt]
$E_1$ (MeV) & 3954 & 3979 & 4079\\[1mm]
\hline &&&\mbox{ } \\[-9pt]
$m_c$ (MeV)       & 1577.63 & 1562  & 1499.5\\[1mm]
$\omega$ (MeV)    & 177.5   & 190   & 240\\[1mm]
$g$               & 1.034   & 1.172 & 1.572\\[1mm]
$c\bar{c}$ ($\%$) & 9.49 & $\mathbf{7.48}$ & 6.51 \\[1mm]
$r_{\mbox{\scriptsize r.m.s.}}$ (fm)    & 7.72 & 7.82 & 8.83\\[1mm]
\end{tabular}
\caption{\label{tab:newpar}Probability of $c\bar{c}$ component and $X(3872)$
r.m.s.\ radius for varying $\omega,m_c$, with bare $E_0$ fixed at 3599~MeV,
$X(3872)$ pole at 3871.68~MeV, and $a=2.0$~GeV$^{-1}$.}
\end{table}
Thus, in case $I$ $E_1$ is lowered by $25$~MeV, while in case $II$ it rises
by $100$ MeV. The trajectories for these two new situations are plotted in
Fig.~\ref{newtraj}.
\begin{figure}[h]
\centering
\begin{tabular}{c c}
\hspace*{-50pt}
\resizebox{!}{430pt}{\includegraphics{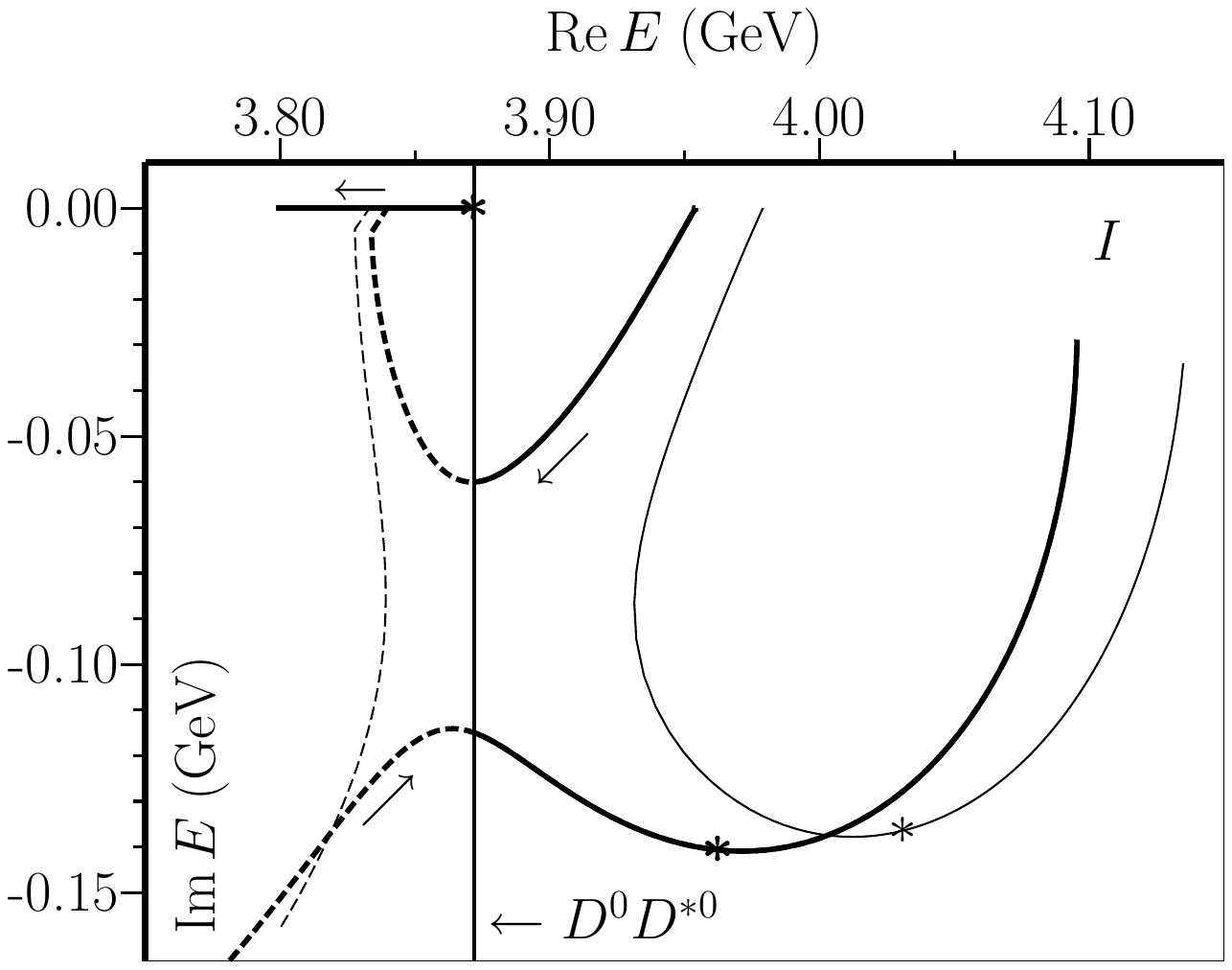}}&\hspace*{-125pt}\resizebox{!}{430pt}{\includegraphics{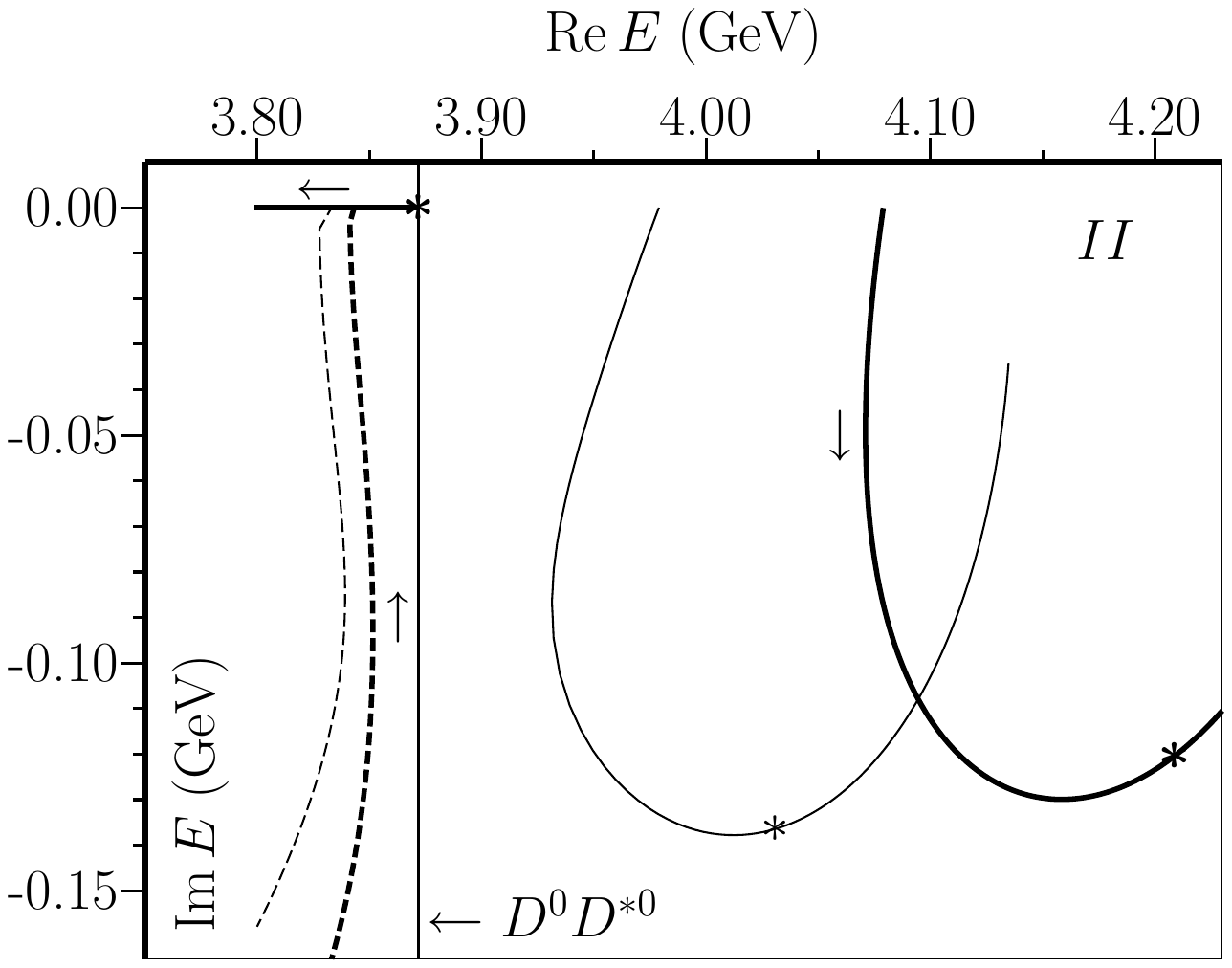}}\\
\mbox{} \\[-8.5cm]
\end{tabular}
\caption{\label{newtraj} Trajectories of dynamical and confinement poles.
The bold curves represent cases $I$ \/(top graph) and $II$ \/(bottom graph)
defined in Table~\ref{tab:newpar}, and the others the standard case of
Fig.~\ref{traja2}; the solid (dashed) lines stand for normal (below-threshold)
resonances. All trajectories lie on the second Riemann sheet. The pole
positions for the $g$ values in Table~\ref{tab:newpar} are marked by
$\mathbf{\ast}$.}
\end{figure}
For $I$ \/we observe that, just as in the standard case depicted
in Fig.~\ref{traja2}, two poles are found relatively close to the
real axis, of a dynamical and a confinement origin, respectively. However,
now it is the \ttpo\ confinement pole that moves steadily downwards
and settles on the real axis below threshold, whereas the dynamical
pole moves to higher energies and eventually approaches the real axis. So
the poles interchange their roles when going from the standard case to
case $I$. Nevertheless, the values of $g$ needed to get a bound state
at 3871.68~MeV are not very different in the two cases, viz.\ 1.172 vs.\ 1.034.
Such a behavior was already observed almost a decade ago, namely for
$D_{s0}^*(2317)$ \cite{PRD86p010001} charmed-strange meson. In a first,
two-channel model calculation \cite{PRL91p012003} the $D_{s0}^*(2317)$ showed
up as a dynamical resonance, settling below the $S$-wave $DK$ threshold, 
whereas the \otpz\ $c\bar{s}$ state turned out to move to higher energies,
with a large width, similarly to the standard $X(3872)$ case in
Figs.~\ref{traja2} and \ref{newtraj} above. However, in a more complete,
multichannel approach \cite{PRL97p202001} the situations got reversed,
just as in the present case $I$. Also in previous Chapter \ref{chp5}, with nine coupled channels, we reproduced the meson as a
confinement pole. What appears to happen in the present case $I$ \/is that
shifting the bare \ttpo\ state to somewhat lower energies is just enough to
deflect the confinement pole to the left and not the right
when approaching the continuum pole. Clearly, there will be an intermediate
situation for which the left/right deflection will hinge upon only marginal
changes in the parameters, but resulting in two completely different
trajectories. Therefore, identifying one pole as dynamical and the other
as linked to a confinement state is entirely arbitrary, the whole system being
dynamical because of unquenching. At the end of the day, the only thing that
really counts is where the poles end up for the final parameters. The
trajectories themselves are not observable and only serve as an illustration
how a coupled-channel model as the one employed here mimics the physical
situation. Suffice it to say that the lower pole, representing the $X(3872)$,
is quite stable with respect to variations in the parameters, owing to its
proximity to the only and most relevant OZI-allowed decay channel. The higher
pole, on the other hand, should not be taken at face value, since a more
realistic calculation should include other important decay channels,
such as $D^*D^*$, with threshold just above 4~GeV.\\
Concerning the other scenario with changed parameters, labeled $II$ \/in
Table~\ref{tab:newpar} and depicted in the lower graph of Fig.~\ref{newtraj},
we see that the trajectories do not change qualitatively when going from
the standard case to $II$. There is a displacement of the right-hand
branch, about 100~MeV to the right on average, in accordance with the same
shift of the bare \ttpo\ state. But the change in the lower, dynamical branch,
is much less significant, though the value of $g$ needed to produce a bound
state at 3871-68~MeV now increases to 1.572 (see Table~\ref{tab:newpar}).
We also notice from Fig.~\ref{newtraj} that the two pole-trajectory branches
hardly move towards one another, signaling less attraction between the
poles due to a larger initial separation.\\
Inspecting again Table~\ref{tab:newpar} as for the $c\bar{c}$ probability
in cases $I$ \/and $II$ \/compared to the standard situation, we observe an
increased value for case $I$ and a decreased one for $II$. This is logical,
since in case $I$ \/the bare \ttpo\ state lies closer to the $X(3872)$, whereas
in case $II$ \/it lies farther away. Nevertheless, the difference in $c\bar{c}$
probability between $I$ \/and $II$ \/is only about 3\%, i.e., less than the
variation with $a$ in the standard case. These comparisons lend further support
to the stability of our results.\\
Finally, in Fig.~\ref{nwf} we compare the wave function for case $II$ \/with the
\begin{figure}[h]
\centering
\resizebox{!}{400pt}{\includegraphics{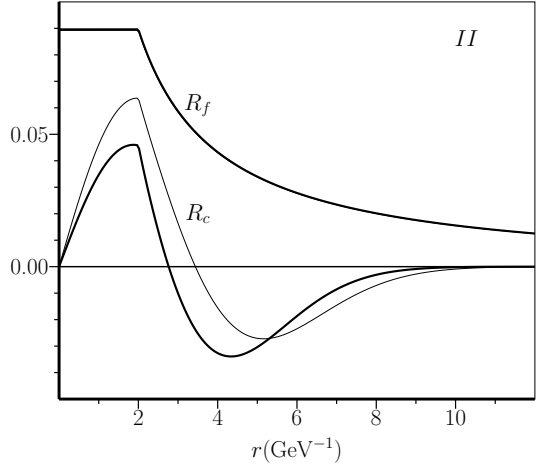}}\mbox{} \\[-6.5cm]
\caption{\label{nwf}Normalized two-component radial wave function $R(r)$, for cases $II$
\/and ''standard'', corresponding to parameters in Table~\ref{tab:newpar}. Bold
curves refer to case $II$, normal curve to $R_c$ for standard case. Note: $R_f$
is indistinguishable within graphical accuracy for the two cases.}
\end{figure}
standard one. We see there is no visible change in the $R_f$ component. 
As for $R_c$, the first maximum gets somewhat reduced, but the secondary,
negative bump even becomes a bit larger, owing to an inward shift of the node,
lying now at about 3~GeV$^{-1}$. Yet, also in case $II$ \/the $c\bar{c}$
component is still very significant, despite the large separation of more
than 200 MeV between the $X(3872)$ bound state and the bare \ttpo\ state.\\
From the latter and all previous results we may safely conclude that the
$c\bar{c}$ component of the $X(3872)$ wave function remains nonnegligible
in a variety of scenarios, being even of comparable size as the $D^0D^{*0}$
component in the inner region, save at very short distances.

\section{Summary and conclusions}  
\label{conc}
In the present chapter, we have employed a simple and solvable
Schr\"odinger model to study the  wave function of the $X(3872)$ meson, by
treating it as a coupled $c\bar{c}$-$D^0D^{*0}$ system with $J^{PC}=1^{++}$
quantum numbers. Transitions between the two channels are described
with the \tpz\ mechanism, through string breaking at a sharp distance $a$.
The exact solutions to the equations allow us to easily study the
trajectories of $\mathcal{S}$-matrix poles as a function of the decay
coupling constant $g$. Thus, a dynamical pole is found, becoming a 
bound state just below the $D^0D^{*0}$ threshold, for different
string-breaking distances $a$, and an appropriate coupling $g$. On the other
hand, the pole arising from the bare \ttpo\ confinement state moves to 
higher energies and acquires a large imaginary part. However, the latter pole
may not be very relevant physically, because of neglected additional 
meson-meson channels which will become important in that energy
region.\\
As for the $X(3872)$ radial wave function, the $c\bar{c}$ component $R_c$
turns out to be of significant size as compared to the $D^0D^{*0}$ component
$R_f$, especially for intermediate $r$ values. Moreover, even for other trial
BEs, the global shape of $R_c$ and its relative magnitude vis-\`{a}-vis $R_f$
in the central region is remarkably stable. But the corresponding $c\bar{c}$
probability is relatively low, due to the very long tail of the $D^0D^{*0}$
wave function at small binding. These results are along the lines of the
analysis based on general arguments presented in Ref.~\cite{PPNP61p455}.
Quantitatively, for the average
\cite{PRD86p010001} $X(3872)$ binding of 0.16 MeV, a $c\bar{c}$ probability
of 7.5--11.2\% is found, for $a$ in the  range 0.4--0.6~fm, which is compatible
with other recent approaches \cite{PRD81p054023,1206.4877}.
The corresponding r.m.s.\ radius turns out to be quite stable at about 7.8~fm,
for the latter range of $a$ values, while the $S$-wave scattering length of
11.6~fm, for $a\approx0.4$~fm, is in agreement with expectations for a BE of
0.16 MeV. \\
Finally, we have studied the nature of the $\mathcal{S}$-matrix pole giving 
rise to $X(3872)$, by varying some of the otherwise fixed parameters. Thus,
a drastic modification of pole trajectories is observed, for relatively
small parameter variations, making the $X(3872)$ pole transform from a
dynamical pole into one directly connected to the \ttpo\ bare confinement
state. However, the corresponding changes in the $c\bar{c}$ probability and
r.m.s.\ radius, as well as the coupling $g$ needed to reproduce $X(3872)$,
are quite modest.\\
In conclusion, we should revisit the claims about $X(3872)$ made in
Ref.~\cite{PRD76p094028}, quoted in the Introduction above, namely about the
inevitability of $X(3872)$ being a charm-meson molecule or virtual
state, independently of the mechanism generating the state. Now, it
is true that our analysis has confirmed some of the quantitative
predictions in Ref.~\cite{PRD76p094028}, viz.\ concerning the vanishing
probability of wave-function components other than $D^0D^{*0}$ as the
BE approaches zero, and the related behavior of the $D^0D^{*0}$ scattering
length. However, we have also shown that the $c\bar{c}$ component is certainly
not negligible and quite stable, in a variety of scenarios. Especially in
electromagnetic processes, the prominence of this component for relatively
small as well as intermediate $r$ values will no doubt result in a significant
contribution to the amplitudes. Moreover, as already mentioned above, the very
unquenching of a \ttpo\ $c\bar{c}$ state will not only introduce meson-meson
components into the wave function, but also a contribution of the \otpo\
$c\bar{c}$ state, which can change predictions of electromagnetic transition
rates very considerably \cite{PRD85p114002}. We intend to study such processes
for $X(3872)$ in future work, on the basis of a model as the one used in the
present paper, by employing the formalism developed and successfully applied
in Ref.~\cite{PRD44p2803}. However, in a detailed and predictive calculation
of electromagnetic $X(3872)$ decays, the inclusion of the charged
$D^\pm D^{*\mp}$ channel will be indispensable \cite{PRD81p014029}.\\
For all these reasons, we do not consider $X(3872)$
a charm-meson molecule, but rather a very strongly unitarized charmonium
state. As a matter of fact, we do not believe \em any \em \/non-exotic mesonic
resonance --- whatever its origin --- qualifies as a true meson-meson molecule,
simply because such a state will inexorably mix with the nearest $q\bar{q}$
states having the same quantum numbers. Indeed, we have demonstrated above
that, even with a bare $c\bar{c}$ state 200~MeV higher in mass, the resulting
$c\bar{c}$ component in the wave function is still appreciable. So let us
conclude this discussion by quoting and fully endorsing the following statement
from Ref.~\cite{PRD76p094028}:
\begin{quote} \em
``Any model of the $X(3872)$ that does not take into account its
strong coupling to charm meson scattering states should not be taken seriously.''
\em
\end{quote}

\backmatter
\chapter{Summary and Conclusions}
\thispagestyle{empty}
\onehalfspacing
The RSE coupled-channel model defined in Sec.~\ref{RSE} was applied to three different meson sectors, namely the isoscalar vector $\phi$, the axial-vectors - pseudovectors with open-charm $D_1$ and $D_{s1}$, and the charmonium-like axial-vector $X(3872)$. A  harmonic-oscillator (HO) confining potential was used with fixed parameters for frequency and constituent quark masses (c.q.m.), Eq.~\eqref{ctepar}, values defined in Ref.~\cite{PRD27p1527}. The frequency of 190 MeV has been used in all RSE applications, and the c.q.m.\ are in agreement with other spectroscopy models. Only two free parameters are left, the dimensionless global coupling $\lambda$ and the ``string-breaking" distance $a$. For the present RSE applications their range is between 2 and 4 GeV$^{-1}$, and between 1.2 and 4, respectively for $a$ and $\lambda$. We see these values are pretty close, considering a likely c.q.m.\ dependence of the parameters, as well as the set of included decay channels.\\    
This work may shed light on the excited $J^{PC}=1^{--}$ $\phi$ states and on the classification of the $\phi(2170)$ resonance, originally denoted $X(2175)$. Among the model's $S$-matrix poles, there are good candidates for observed resonances, as well as other ones that should exist according to the quark model. Besides the expected resonances as unitarized confinement states, a dynamical resonance pole is found at $(2186-i246)$~MeV. The huge width makes its interpretation as the $\phi(2170)$ somewhat dubious. On the other hand, the calculated resonances originating in the confinement spectrum are generally too narrow. We consider the inclusion of sharp thresholds only as the main problem of our description. Including the widths of final-state resonances could probably {\em increase} \/the widths of the now too narrow excited $\phi$ resonances stemming from the confinement spectrum. Since a simple substitution of the here used real masses by the true complex masses destroys the manifest 
unitarity of the $S$-matrix, to account for the nonvanishing widths of mesons in the
coupled channels is a very difficult problem. The formalism described in Subsec.~\ref{redsm} was developed {\it a posteriori} and has not been applied to the $\phi$ vectors.\\
We dynamically reproduced the whole pattern of masses and widths of the axial-vector - pseudovector charmed mesons, by coupling the most important open and closed two-meson channels to bare $J^P=1^+$ $c\bar{n}$ and $c\bar{s}$ states containing both \tpo\ and \spo\ components. The coupling to two-meson channels dynamically mixes and lifts the mass degeneracy of the spectroscopic \tpo\ and \spo\ states, as an alternative to the usual spin-orbit splitting. Of the two resulting $S$-matrix poles in either case, one stays very close to the energy of the bare state, as a quasi-bound state in the continuum, whereas the
other shifts considerably. This is in agreement with the experimental observation that the \dc\ and \dsc\ have much smaller widths than one would naively expect. Predictions for pole positions of radially excited axial-vector charmed mesons are presented.\\
The nature of the $X(3872)$ enhancement was analyzed in the framework of the RSE, by studying it as a regular $J^{PC}=1^{++}$ charmonium state, though strongly influenced and shifted by
open-charm decay channels. The observed but OZI-suppressed $\rho^0J\!/\!\psi$ and $\omega J\!/\!\psi$ channels were coupled as well, but effectively smeared out by using complex $\rho^0$ and $\omega$ masses, in order to account for their physical widths, followed by a rigorous algebraic procedure to restore unitarity. A very delicate interplay between the $D^0D^{\ast0}$, $\rho^0J\!/\!\psi$, and $\omega J\!/\!\psi$ channels was observed. The data clearly suggest that the $X(3872)$ is a very narrow axial-vector $c\bar{c}$ resonance, with a pole at or slightly below the $D^0D^{\ast0}$ threshold.\\
A solvable coordinate-space model was employed to study the $c\bar{c}$ component of the $X(3872)$ wave function, by coupling a confined \tpo\ $c\bar{c}$ state to the almost unbound $S$-wave $D^0\overline{D}^{*0}$ channel via the \tpz\ mechanism. The two-component wave function was calculated for different values of the binding energy and the transition radius $a$, always resulting in a significant $c\bar{c}$ component. However, the long tail of the $D^0\overline{D}^{*0}$ wave function, in the case of small binding, strongly limits the $c\bar{c}$ probability, which roughly lies in the range 7--11\%, for the average experimental binding energy of 0.16~MeV and $a$ between 2 and 3~GeV$^{-1}$. Furthermore, a reasonable value of 7.8~fm was obtained for the $X(3872)$ r.m.s.\ radius at the latter binding energy, as well as an $S$-wave $D^0\overline{D}^{*0}$ scattering length of 11.6~fm. Finally, the $S$-matrix pole trajectories as a function of coupling constant show that $X(3872)$ can be generated either 
as a dynamical pole or as one connected to the bare $c\bar{c}$ confinement spectrum, depending on details of the model. From these results we conclude that $X(3872)$ is not a genuine meson-meson molecule, nor actually any other mesonic system with non-exotic quantum numbers, due to inevitable mixing with the corresponding quark-antiquark states.\\
All the studied resonances are controversial, with $X(2175)$, alias $\phi(2170)$, and $X(3872)$ often thought to be exotics, while the open-charm axials exhibit an unexpected pattern of masses and widths. Within the unquenched and unitarized RSE model we are able to reproduce all the main features of these resonances as observed in experiments. Furthermore, we have shown, with a simplified Schr\"odinger model, the importance of a charmonium component in $X(3872)$. The relevance of ``dressing" the bare states is well known in spectroscopy, and so we consider the $X(3872)$ is indeed the $\chi_{c1}(2P)$ resonance, strongly influenced by the $D^0D^{*0}$ threshold, and strongly deviated from its bare energy due to the unquenching.\\
\begin{center}
$\therefore$
\end{center}
Future research work following the present thesis will include the development of unquenched models, such as the RSE but not only, to approach enigmatic resonances. In particular, we plan to clarify the nature of mesonic structures observed in experimental data, and help to disentangle peaks from resonance poles and nonresonant enhancements due to threshold openings. Also, we aim to distinguish between intrinsic resonances, directly linked to a confinement spectrum, and dynamically generated ones. Another goal is the development of a unitarization scheme for quenched tetraquark states. At last, the ultimate purpose of meson spectroscopy is to understand confinement and decay mechanisms, i.e., the strong interaction itself.  
\clearpage
\thispagestyle{empty}

\end{document}